\documentclass[iop]{emulateapj}

\bibpunct[;]{(}{)}{;}{a}{}{,}
\newcommand{\kms}{{\ensuremath{{\rm km~s}^{-1}}}}
\newcommand{\etal}{et~al.\/}
\newcommand{\msun}{\ensuremath{M_\sun}}

\newcommand{\dmt}[2]{\ensuremath{{#1}\degr\,{#2}\arcmin}}
\newcommand{\hmt}[2]{\ensuremath{{#1}^h\,{#2}^m}}
\newcommand{\dms}[3]{\ensuremath{{#1}\ {#2}\ {#3}}}
\newcommand{\hms}[3]{\ensuremath{{#1}\ {#2}\ {#3}}}
\newcommand{\sqdeg}{square degree}
\newcommand{\sqdegs}{square degrees}

\newcommand{\ujy}{\ensuremath{\mu}Jy}

\newcommand{\mjypbm}{{\rm mJy~beam}\ensuremath{^{-1}}}

\newcommand{\ie}{i.\,e.\,}
\newcommand{\eg}{e.\,g.\,}

\newcommand{\esh}{erg s\ensuremath{^{-1}} Hz\ensuremath{^{-1}}}

\newcommand{\chisq}{\ensuremath{\chi^2}}
\newcommand{\rchisq}{\ensuremath{\chi^2_\nu}}

\newcommand{\uv}{\ensuremath{(u,v)}}

\newcommand{\paperi}{Paper I}
\newcommand{\paperii}{Paper II}
\providecommand{\pasa}{PASA}
\providecommand{\nar}{NewAR}

\shorttitle{PiGSS III}
\shortauthors{Croft \etal}

\begin{document}
\title{The Allen Telescope Array Pi GHz Sky Survey -- III: The ELAIS-N1, Coma, and Lockman Hole Fields}
\author{Steve Croft\altaffilmark{1}, Geoffrey C.\ Bower\altaffilmark{1}, and David Whysong\altaffilmark{1}}
\altaffiltext{1}{University of California, Berkeley, Astronomy Dept., B-20 Hearst Field Annex \#3411, Berkeley, CA 94720, USA }
\tabletypesize{\scriptsize}

\begin{abstract}

We present results from a total of 459 repeated 3.1\,GHz radio continuum observations (of which 379 were used in a search for transient sources) of the ELAIS-N1, Coma, Lockman Hole, and NOAO Deep Wide Field Survey fields as part of the Pi GHz Sky Survey (PiGSS). The observations were taken approximately once per day between 2009 May and 2011 April. Each image covers 11.8~\sqdegs\ and has 100\arcsec\ FWHM resolution. Deep images for each of the four fields have rms noise between 180 and 310\,\ujy, and the corresponding catalogs contain $\sim 200$ sources in each field. Typically 40 -- 50 of these sources are detected in each single-epoch image. This represents one of the shortest cadence, largest area, multi-epoch surveys undertaken at these frequencies.

We compare the catalogs generated from the combined images to those from individual epochs, and from monthly averages, as well as to legacy surveys. We undertake a search for transients, with particular emphasis on excluding false positive sources. We find no confirmed transients,  defined here as sources that can be shown to have varied by at least a factor 10. However, we find one source which brightened in a single-epoch image to at least six times the upper limit from the corresponding deep image. We also find a source associated with a $z = 0.6$ quasar which appears to have brightened by a factor $\sim 3$ in one of our deep images, when compared to catalogs from legacy surveys.

We place new upper limits on the number of transients brighter than 10\,mJy: fewer than 0.08 transients deg$^{-2}$ with characteristic timescales of months to years; fewer than 0.02 deg$^{-2}$ with timescales of months; and fewer than 0.009 deg$^{-2}$ with timescales of days. We also plot upper limits as a function of flux density for transients on the same timescales.

\end{abstract}

\keywords{catalogs --- methods: data analysis --- radio continuum: galaxies --- radio continuum: general --- surveys}

\section{Introduction}

\subsection{Recent Results from Radio Transient Surveys}

The past few years have seen an increase in multi-epoch surveys specifically designed to probe the population of radio transients, as well as transient searches undertaken using archival data \citep[for a recent summary, see][]{fender:11}. Transients can arise from processes taking place in a variety of progenitors \citep{lazio:09}, with typical timescales dictated by the characteristic processes causing the transient emission, and the sizes of the emitting regions. Radio transients are generally classified as short (characteristic timescales $\lesssim 1$\,s) and long (timescales $\gtrsim 1$\,s), where the shorter timescales typically correspond to coherent emission processes and the longer timescales to incoherent emission \citep{cordes:04}. The timescales probed correspond to characteristic processes and size scales in the emitting objects, and so two surveys with differing temporal coverage will see a different mix of objects from each other.

Of particular interest for the current generation of radio sky surveys are tidal disruption events (TDEs) --- flares connected to sudden increases in accretion rate caused when stars are tidally shredded as they pass close to black holes with mass $\lesssim 10^8$\,\msun\ \citep{rees:88}. Such events may be the dominant contributor to transient source counts in the radio for surveys focusing on long duration transients \citep{frail:12,bower:tde}. Recently around 10 TDE candidates have been detected \citep[\eg,][]{gezari:09,svv:11b} by high energy surveys, of which some have been detected in radio-follow-up \citep[\eg,][]{zauderer:11,cenko:12}. However, no TDEs have yet been unambiguously detected in blind radio surveys. 

Other classes of sources that are expected to dominate blind transient surveys are radio supernovae, gamma ray burst afterglows, radio counterparts to gravitational wave sources such as neutron star mergers, flares from low-mass stars and brown dwarfs, and refraction and scintillation of extragalactic sources due to clumpiness and turbulence in our own Galaxy's interstellar medium \citep{frail:12,bower:11}. Some of the aforementioned source classes have been detected in follow-up of triggers at other wavelengths, but until recently, radio surveys have not had the required throughput to detect them blindly in significant numbers. Some progenitors may only be detectable in the radio, however, and in other cases radio variability may be more extreme than variability at other wavelengths, making these sources easier to find. Blind surveys also have the potential to serendipitously detect new classes of objects.

Some transient candidates have been detected in blind radio surveys, though --- for example, those reported by \citet{bower:07}, a large fraction of which remained undetected in deep optical \citep{bower:07,ofek:10} and X-ray \citep{croft:11} follow-up. A recent reanalysis of the \citeauthor{bower:07}\ data by \citet{frail:12}, however, suggests that some of these transients may be data artifacts, and the significance of some of the others may have been overestimated (Section~\ref{sec:falsepos}). Many other radio transient surveys may also have been somewhat overoptimistic in their claims of detections, particularly of sources seen in a single epoch and never again, and we urge appropriate skepticism. In this paper we discuss some of the safeguards we have implemented to ensure that any transients we detect have a high probability of being real.

\subsection{Limitations of Current Surveys}

Survey design involves trade-offs between a number of factors, including frequency, bandwidth, frequency resolution, spatial resolution, limiting sensitivity, sky area, number of epochs, cadence (the time between epochs), integration time at each epoch, and Galactic latitude. The number of transients detected by a survey depends on the lightcurves, radio spectra, luminosity functions, and distribution in Galactic latitude and in distance, for a mix of different classes of progenitor, so comparing results from one survey to another is not trivial. 

For example, a survey with many epochs with a cadence of days might see different objects to a shallow large-area survey with fewer epochs with a timescale of months, even if the effective area covered (the number of pointings, multiplied by the area covered by each pointing) is similar. Large area surveys with contiguous coverage have the advantage that sensitivity is approximately uniform across most of the resulting mosaic, whereas surveys with small numbers of pointings are more affected by primary beam falloff at the mosaic edges. Small area surveys with many repeated epochs, however, have the advantage of allowing a deep comparison image to be constructed from all of the data.
Different authors also define ``transient'' in different ways, and as a result, measurements of transient surface densities (sometimes referred to as ``two-epoch rates'' of transients per unit effective area) from different surveys may not be directly comparable.

The sensitivity of a given survey is particularly critical in influencing the number of transients seen; a transient seen in one survey may be classified as a variable source by a more sensitive survey if detected in additional epochs. 

We suggest that a sensible threshold for a source to be considered a transient is that it brightens by a factor of at least ten\label{sec:fac10}. \label{sec:factor10}The distinction between transient and variable sources is always somewhat arbitrary, but we pick this threshold to distinguish between sources that would be detected in quiescence with a modest increase in sensitivity (\eg, variable AGNs) and those that brighten more dramatically due to some explosive event (e.g. SNe), very significant changes in accretion rate (\eg, TDEs), and the like. As shown in \citet{bower:11}, and in Section~\ref{sec:maxfracmod}, a few percent of sources exhibit apparent variability of a factor of a few, but almost none of these ``normal'' variables exhibit variations of an order of magnitude or more in flux. The choice of a factor 10 for transient detection naturally limits the sensitivity of any two-epoch survey to around $30\sigma$ (for a source to be undetected at $\lesssim 3\sigma$ in one image, and to change in flux density by a factor of 10 or more on its detection in the other image).

It should also be noted that explosive events may occur within galaxies that are themselves radio sources (for example, radio galaxies). If the spatial resolution is not sufficient to distinguish between the two sources of radio emission, then the explosive event will not be characterized as a transient, by our definition, unless it is at least nine times brighter than the emission from the host galaxy. This will also tend to reduce the amount of variability seen by surveys like ours (with relatively low spatial resolution) where extended non-variable emission is unresolved from variable sources such as radio galaxy cores \citep[\eg,][]{duric:89}.

The distinction between transient and variable sources aside (which is of most importance in comparing transient statistics between surveys) it is of course still of interest to study sources which exhibit large changes in flux density, and we remark upon some of these extreme variables, as well as analyzing variability statistics for large numbers of sources, in Section~\ref{sec:variability}.

If we consider a multi-epoch survey with a detection threshold of $6\sigma_{single}$ for a transient in an individual epoch, and a $3\sigma_{deep}$ upper limit in a deep comparison image made from all observations of the same field (with $\sigma_{deep} = \sigma_{single} / \sqrt{n}$ for $n$ epochs), we see that to detect transients close to the threshold (where $6\sigma_{single} \geq 10 \times 3 \sigma_{deep}$) requires 25 or more epochs. Such repeated observations of the same field (as opposed to using the allocated observing time to increase the area of the survey while maintaining shallower depth) also have the advantage of allowing us to extract multi-epoch lightcurves for sources of interest, and to make images from subsets of the data with different timescales. Repeated observations of the same field also allow us to disentangle intrinsic variability from telescope calibration errors. However, these requirements must be balanced against the fact that quadrupling the area covered by a survey at the expense of doubling the detection threshold should result in an increase of the number of transients detected of $\sim 40\%$ (naively assuming an $S^{-3/2}$ dependence of rates on flux density, $S$). The tradeoff between depth and area must also be balanced with the desire to obtain adequate sampling for the timescales of interest.

Some surveys \citep[\eg][]{atats} use a deep comparison field that was taken a long time (compared to the timescales of interest) before the survey data. For surveys which use their own data to generate a deep field, there is an additional complication. Transients that are bright, or those that are detected in more than a few of the individual images, will also contribute significant flux to the deep image, potentially resulting in a detection even when no long-term counterpart is present.

\subsection{False Positives}\label{sec:falsepos}

The reassessment of the \citet{bower:07} transients by \citet{frail:12}; and of the \citet{levinson:02} transients by \citet{galyam:06} and \citet{ofek:10}; as well as questions raised by \citet{bs:11} about transients such as those seen by \citet{lorimer:07}; and about the \citet{matsumura:09} transients by \citet{atatsii}, illustrate that events detected only in a single epoch have a significant chance of being spurious. In fact many current transient searches may be operating in an uncomfortable regime where the number of false positive detections is larger than the number of true transients.

Spurious sources in imaging searches can arise due to amplification of noise by overly large primary beam gain corrections; from various image artifacts including sidelobes of bright sources and corrupted data; or simply due to thresholds that are overly optimistic. The latter can occur because the large area surveyed means that even rare events are seen occasionally; because underestimates of the background noise cause the significance of detections to be overestimated; or because the assumption of Gaussian noise is not valid (which it may not be in the presence of artifacts as well as non-linear imaging algorithms such as CLEAN). If the primary beam is well-known, relatively conservative gain cutoffs are easy to implement; in this paper we discard sources where the primary beam correction is greater than a factor 5 (Section~\ref{sec:biggain}). 

To have the best chance of finding real transients, we must carefully choose thresholds to lower the number of false positives. One way is to choose a very conservative threshold, perhaps 8 -- $10\sigma$, in individual epoch images (and look for counterparts at any flux density in the deep image). However, the assumed $S^{-3/2}$ dependence of rates on flux density, $S$, means that high thresholds also sacrifice a lot of sensitivity to potential transients.

A better method (Section~\ref{sec:dualimageimp}) is to split the \uv\ data in half and create two separate images, which are then searched simultaneously for candidates appearing in the same place. Noise peaks will tend not to coincide in two images, and since sidelobes and other image defects also often do not coincide, this method can also reduce the number of false positives due to artifacts.

The data can be halved in a variety of ways --- in time, frequency, or by taking subsets of baselines, for example. Then the catalog can be generated with a lower threshold in both images ($\sim 4.2\sigma$, in our case). This will result in relatively large numbers of spurious sources, but any source not appearing in both images is then discarded, resulting in a comparatively clean catalog with a $\sim 6\sigma$ threshold. For the number of independent beams searched, the probability of a single source appearing in our catalogs above this threshold due to random Gaussian noise is 0.7\%; as noted above, however, artifacts or other imaging defects can still easily result in the presence of spurious detections above this threshold.

The observing strategy used for PiGSS (two correlators delivering adjacent frequency bands) lends itself naturally to the two-image approach, but it could be applied by other telescopes with a single correlator too, by dividing the data up in some way before imaging. Dividing in frequency may of course miss transients with very steep spectral indices, and dividing in time reduces sensitivity to rapid variability, but in such cases more than one method of halving the data could be used, and the catalogs checked against each other.\label{sec:dualimageint} We discuss our implementation of this method in Section~\ref{sec:dualimageimp}.

\section{Transient Surveys with the Allen Telescope Array}

The Allen Telescope Array \citep[ATA;][]{welch:09}, is a 42-element interferometer, located at Hat Creek Radio Observatory in Northern California. Each element is a
6.1-m dish with a broadband feed that receives radiation from 0.5 to 11\,GHz. Two independently-tunable correlators select bands of 104\,MHz width, each consisting of 1024 channels, from anywhere in this frequency range. The maximum baseline is 300\,m, yielding resolution $\sim 100$\arcsec\ at $\sim 3$\,GHz. Typical system equivalent flux densities at $\sim 3$\,GHz for antennas that are performing as expected are $\sim 10^4$\,Jy.

\citet{atats} and \citet{atatsii} presented results from a pilot program, the ATA Twenty-centimeter Survey (ATATS). They compared the resulting catalogs to those from the NRAO VLA Sky Survey \citep[NVSS;][]{condon:nvss}, a survey also at 1.4\,GHz undertaken with the VLA between 1993 and 1997. \citet{bower:10}, hereafter \paperi, introduced a larger ATA survey, the Pi GHz Sky Survey (PiGSS). PiGSS covered $\sim 2500$ \sqdegs\ in two epochs, and an additional $\sim 2500$ \sqdegs\ in a single epoch, in two simultaneous 100\,MHz bands centered at 3.04 and 3.14\,GHz (Croft \etal~in prep.). As well as the large field, PiGSS also incorporates four smaller ($\sim 12$ \sqdeg) fields observed with a cadence of days. \paperi\ presents results from a deep image made from 75 epochs of observation of one of these fields (centered on the location of the NOAO Deep Wide Field Survey, NDWFS). \citet{bower:11}, hereafter \paperii, compares catalogs generated from the individual epochs of this field (and monthly averages of them) to each other, and to other radio \citep{condon:nvss,first,gb6,devries:02}, infrared \citep{sdwfs,sdwfs2}, and X-ray \citep{kenter:05} catalogs.

In this paper, we present results from the other three 12 \sqdeg\ deep fields, centered on the ELAIS-N1, Lockman Hole, and Coma fields (Table~\ref{tab:fields}), as well as a partial reanalysis of the NDWFS results from \paperii\ to conform with updated analysis techniques used on the other three fields. The ELAIS-N1 and Lockman fields were selected due to the availability of large amounts of deep imaging and spectroscopic coverage from other telescopes. The Coma field also has extensive multi-wavelength coverage, and in addition the wide-field of the ATA enabled our survey to include a large number of low-redshift galaxies in the Coma cluster, as opposed to the higher-redshift galaxies in the other fields. All four fields are at high Galactic latitude ($b > 50\degr$) and as such primarily target extragalactic transients; results from another recent ATA survey primarily targeting Galactic transients are reported by \citet{williams:12}. Most existing transient surveys have focussed on frequencies in the 1 -- 2\,GHz range, with typical cadences rather longer than used for our observations, and so PiGSS provides a unique view of the transient and variable radio sky.

We place a particular emphasis here on techniques that will be scalable to the next generation of imaging surveys, where processing and analysis must take place with an increasing degree of automation due to the very large data volumes involved. Although PiGSS is still (just) at a scale where manual processing is possible, we focused on developing routines to reject bad data both before and after the imaging stage, and on understanding our results in the context of other surveys, both past and future.
    
\begin{deluxetable*}{lllllllll}
\tablewidth{0pt}
\tabletypesize{\scriptsize}
\tablecaption{\label{tab:fields} PiGSS Deep Fields}
\tablehead {
\colhead{Field} & 
\colhead{RA} &
\colhead{Dec} &
\colhead{Date Range} &
\multicolumn{4}{c}{Number of Epochs} &
\colhead{Deep Field}\\
 & 
\colhead{(J2000)} &
\colhead{(J2000)} &
 &
\colhead{Total} & 
\colhead{Deep\tablenotemark{a}} &
\colhead{Transient\tablenotemark{b}} &
\colhead{Variable\tablenotemark{c}} &
\colhead{rms (\ujy)} 
}
\startdata
NDWFS\tablenotemark{d} & \hms{14}{32}{00} & \dms{34}{16}{00} & 2009 May 20 -- Sep 29 & 96 & 78 & 67 & 66 & 250 \\
Lockman & \hms{10}{52}{00} & \dms{57}{38}{00} & 2009 Dec 2 -- 2010 Jul 7 & 132 & 87 & 110 & 101 & 250\\ 
ELAIS-N1 & \hms{16}{12}{42} & \dms{53}{48}{31} & 2010 May 30 -- 2011 Jan 20 & 179 & 154 & 165 & 145 &180 \\
Coma & \hms{13}{00}{43} & \dms{27}{11}{13} & 2011 Jan 24 -- Apr 7  & 52 & 43 & 37 & 33 & 310 \\
\enddata
\tablenotetext{a}{Median number of epochs for each field with rms flux density $< 3.5$\,\mjypbm\ in images made from individual {\em pointings}. These data were used to generate the deep field images, except for the NDWFS deep field where selection was performed differently (see \paperi)}
\tablenotetext{b}{Number of epochs with 90\%\ completeness better than 50\,mJy at both frequencies, that were used in the transient search (Section~\ref{sec:transients}).}
\tablenotetext{c}{Number of epochs with 90\%\ completeness better than 50\,mJy at both frequencies, and Post Imaging Calibration gain within 0.1 of the median for that field (Section~\ref{sec:medianpic}). These were the epochs used to determine variability statistics (Section~\ref{sec:variability}).}
\tablenotetext{d}{Reanalysis of data presented in \paperi\ and \paperii.}
\end{deluxetable*}

Throughout this paper, we use J2000 coordinates.

\section{Data Acquisition and Reduction}

The PiGSS observations reported here were undertaken approximately daily between 2009 December and 2011 April, in campaigns lasting several months for each field (Table~\ref{tab:fields}). We attempted to observe all seven pointings for each field for three minutes, three times, distributed in hour angle so as to optimize \uv\ coverage, during the 4 -- 12\,hr allocated to each PiGSS epoch. As in \paperi, pointings were separated by 47\arcmin, the voltage FWHM at 3.14\,GHz, to provide approximately uniform sensitivity in the central region of the field. Data were taken in two 100\,MHz bands centered at 3.04 and 3.14\,GHz, each covering 1024 channels. One of the bright calibrators 3C\,286, 3C\,48, or 3C\,147 was observed hourly.

\subsection{Flagging}

Bad data arose from a variety of sources: radio frequency interference (RFI); clock synchronization problems; Walsh system problems; temperature related problems for individual boards (particularly radio frequency converter boards) causing spectral corruption; antennas that were offline or had problems with feeds; and other issues, known and unknown. Most problems except RFI were relatively infrequent and typically affected a single epoch of data. RFI occurred in essentially all datasets, both as narrow-band spikes (relatively easy to excise) and wide-band RFI affecting a larger number of frequency channels. Identifying and excising (``flagging'') these bad data in an efficient, automatic manner is important for large surveys like ours where the large amount of data makes pipeline processing essential. Flagging was performed by MIRIAD \citep{sault:95} tasks wrapped in our own custom Perl pipeline (``autoflag'') which expands and improves upon the earlier RAPID / ARTIS \citep{keating:10} ATA data reduction software. 

The autoflag software computes the median and variance of the real and imaginary values for each channel / polarization / baseline combination in each single-epoch dataset as a function of time. Channels where more than 60\%\ of the visibilities deviate by more than four standard deviations from the median values, as well as their immediate neighboring channels, are flagged as suspected RFI. Additionally, individual visibilities that are more than $10\sigma$ from the median amplitude are flagged. To flag wide-band RFI, spectra are binned (using a range of bin sizes) and bins with excessive variance (three or more times the median) are flagged, on a baseline by baseline basis. Where exactly one quarter of the channels in spectra associated with a particular antenna show large variance, spectral corruption is suspected and the channels are flagged on all baselines associated with that antenna. Baselines are also flagged if calibrator phases show excessive ($> 1$ radian) variability.

Baselines shorter than 110 wavelengths (which tended to be subject to solar interference), along with 100 channels at each edge of the band (due to filter roll-off) and the central DC channel were also flagged. The visibility retention (the percentage of cross-correlations for each dataset remaining after all flagging) ranged from 1\%\ to 78\%. 93\%\ of datasets had retention of at least 65\%. Fields with poor retention often showed obvious defects after imaging, or failed to image entirely; such fields were excluded from further analysis.

\subsection{Calibration and Imaging}

Where multiple calibrators were used, the primary calibrator was taken to be the one with the most unflagged data, and the reference antenna was taken to be the one with the most unflagged visibilities in the primary calibrator data. Initial calibration was performed using MFCAL with flux densities from \citet{baars:77}, and then amplitudes which deviated by more than 4\,Jy from the nominal flux density, and phases which deviated by more than 20\degr, were flagged in the calibrator data. A second round of MFCAL was performed, bandpass solutions were copied to the other calibrators (if present), and MFCAL was run on these calibrators to obtain a gain solution. The gain and bandpass solutions were copied to the target data from the primary calibrator, and the gain solutions for the other calibrators (if present) were merged into the target data. No polarization calibration was performed.

At each pointing, images of $1024 \times 1024$ pixels, where each pixel is $15 \times 15\arcsec$, were produced, using natural weighting and multi-frequency synthesis. In order to produce mosaics, the `mosaic' option was used with the INVERT command, and the `offset' keyword used to set the reference coordinate for all seven pointings to a single position at the center of the mosaic. If these steps are not performed, MIRIAD LINMOS does not correctly handle the sky projection when mosaics are produced, resulting in offsets between ATA positions and the standard reference frame. This was the cause of the offsets between ATA and NVSS positions seen in \paperi, which are corrected here in our reanalysis.

An initial CLEAN was performed down to the first negative component, and sources in the resulting map were found using the MIRIAD source finder, SFIND. The output image from SFIND was convolved with the dirty beam, and pixels below $10\sigma$ were blanked. This image was then used as a mask for a second round of CLEAN (on the image which had already been CLEANed down to the first negative component), which continued until a cutoff of 5\,mJy or 5000 iterations was reached. The two stage CLEAN ameliorates non-linear behavior in CLEAN, and essentially performs more iterations on sources brighter than $10\sigma$ in the initial image.

If sources brighter than 500\,mJy were present in the field, amplitude self-calibration was performed, and then the image was inverted and CLEANed again. The final image was restored using a $100\arcsec \times 100\arcsec$ Gaussian beam.

\begin{figure*}[htp]
\centering
\includegraphics[width=0.45\linewidth,draft=false]{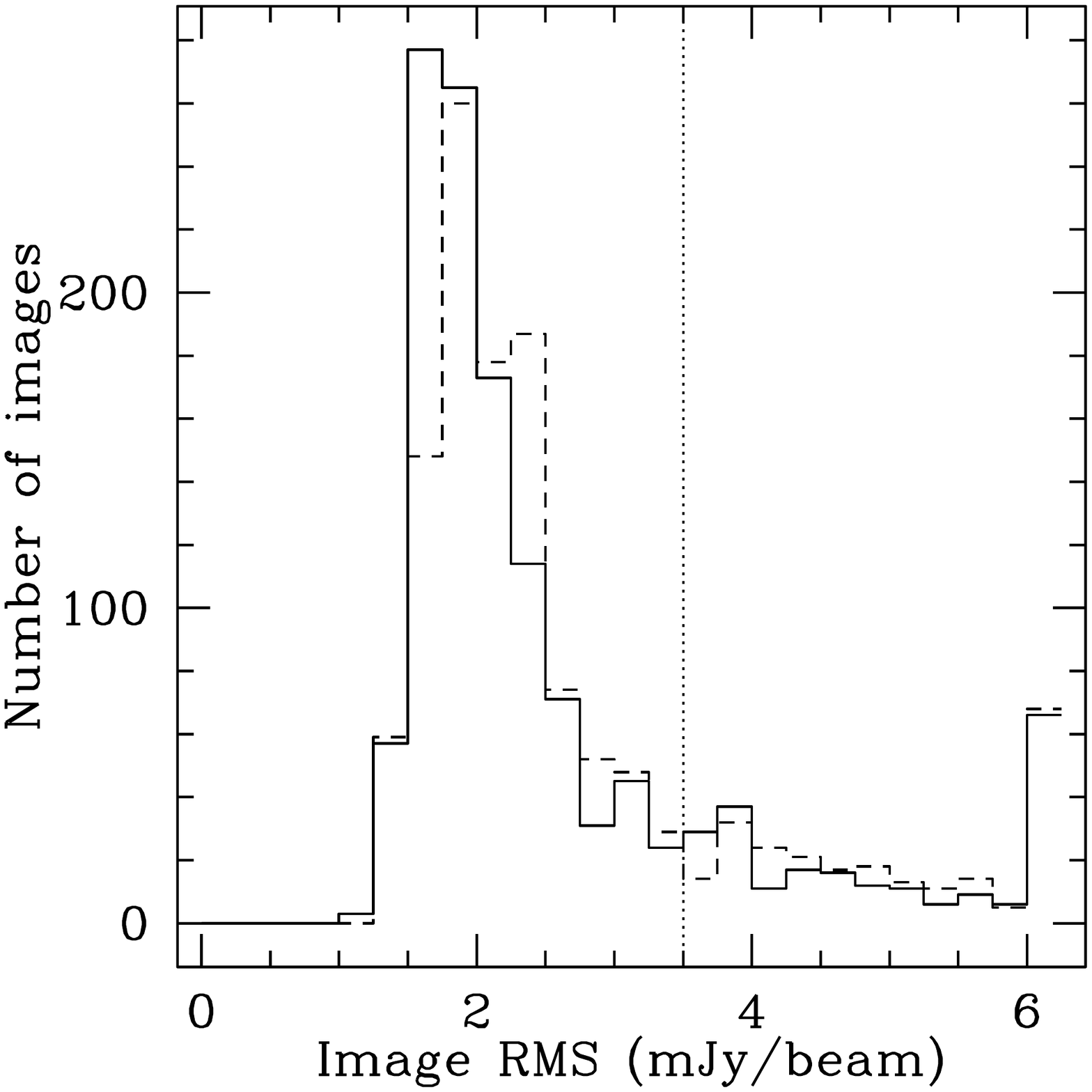}\hspace{0.03\linewidth}%
\includegraphics[width=0.45\linewidth,draft=false]{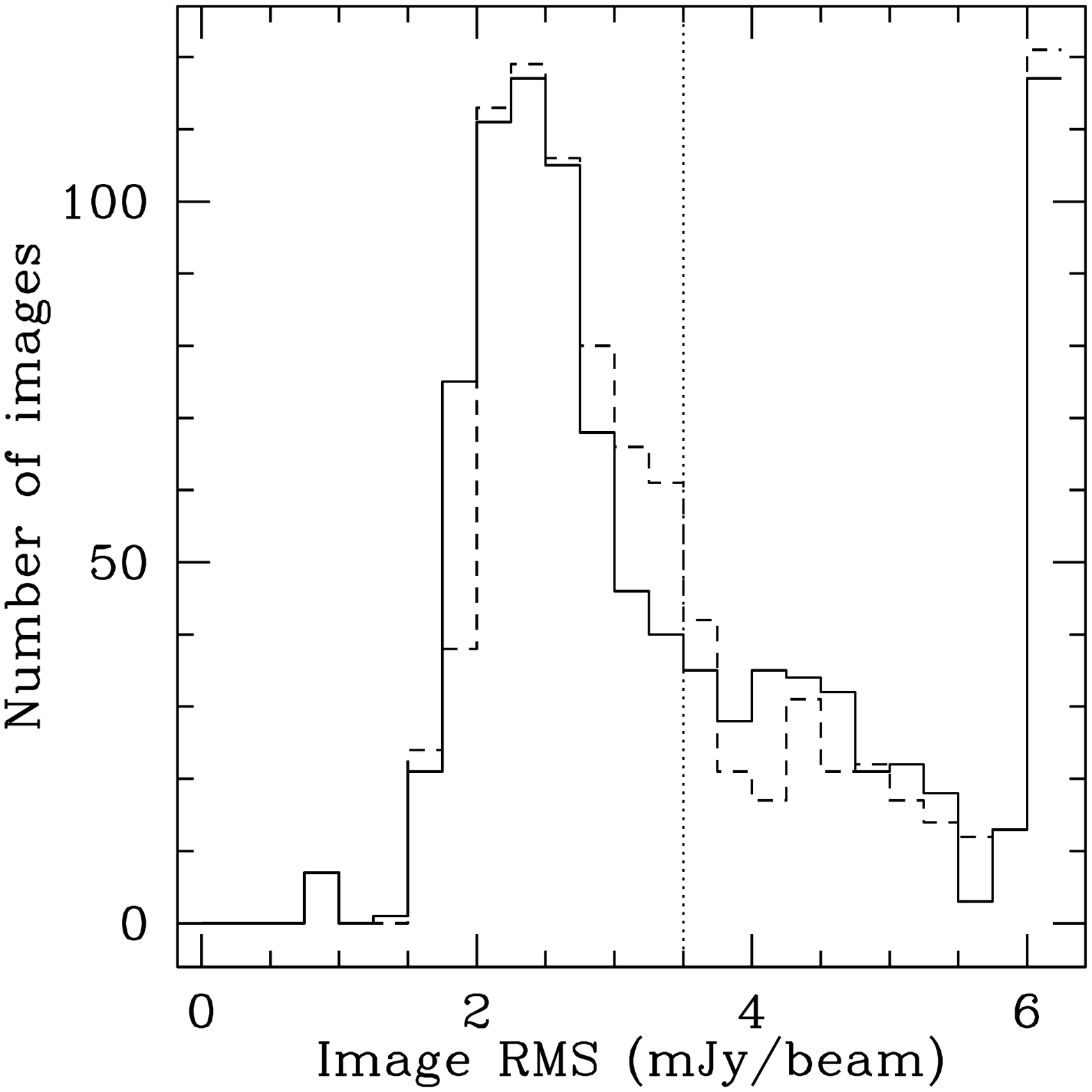}
\includegraphics[width=0.45\linewidth,draft=false]{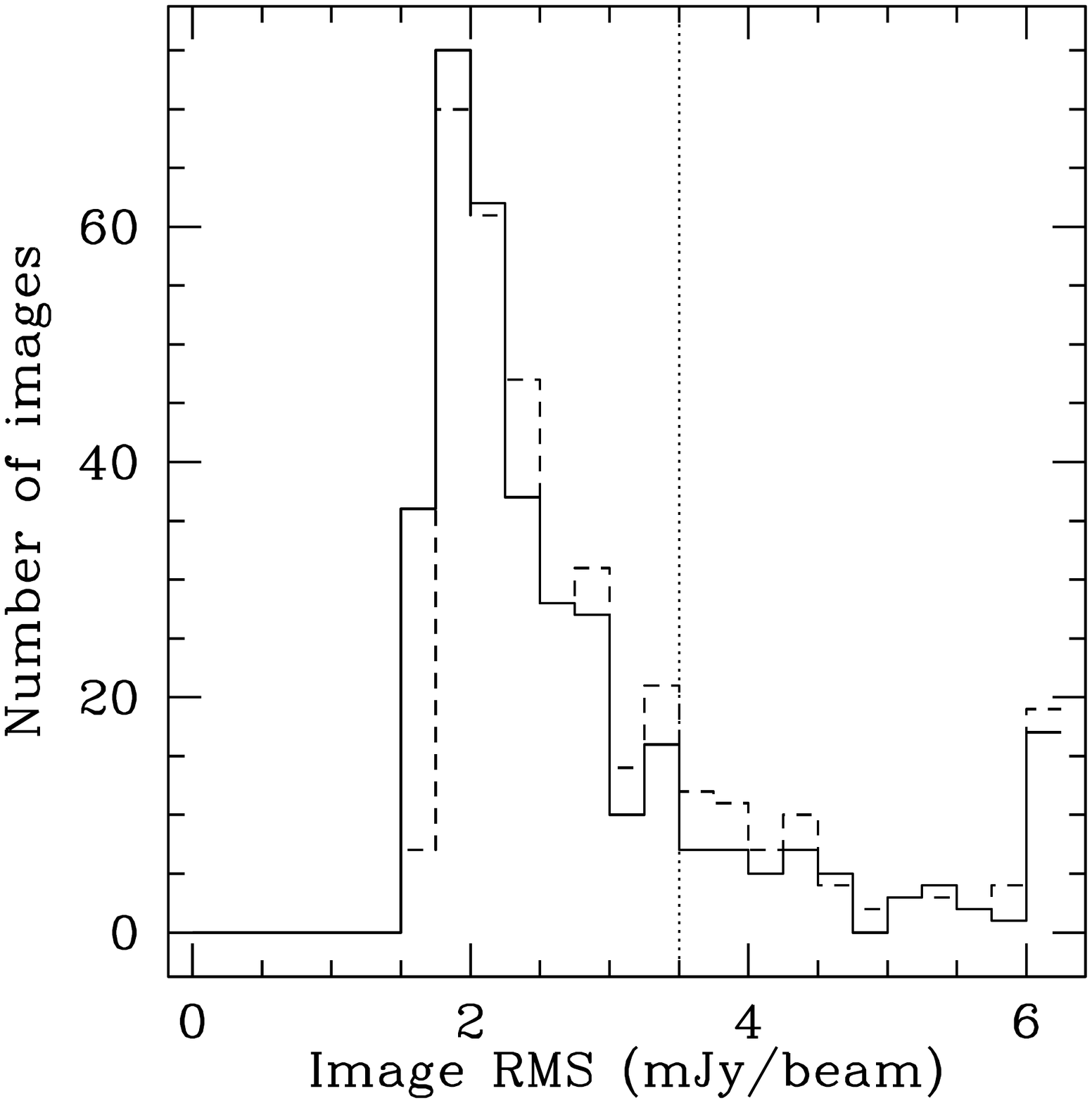}\hspace{0.03\linewidth}%
\includegraphics[width=0.45\linewidth,draft=false]{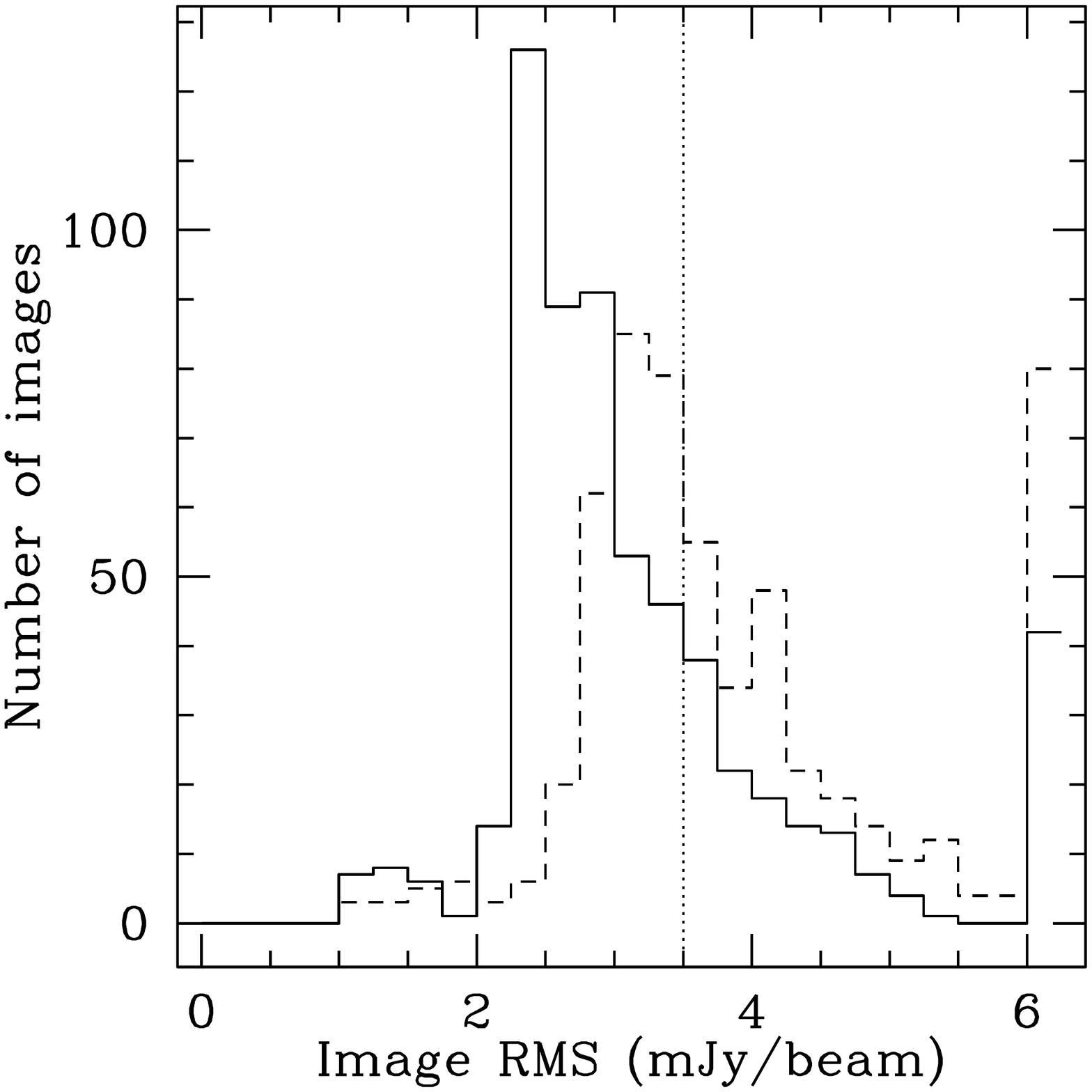}
\caption{\label{fig:rmshist}
Flux density rms for images of all individual {\em pointings} at 3040 (solid line) and 3140\,MHz (dashed line) for ELAIS-N1, Lockman, Coma, and NDWFS. The dotted lines show the 3.5\,\mjypbm\ cut used to select data to construct the deep images.
}
\end{figure*}

\begin{figure*}[htp]
\centering
\includegraphics[width=0.49\linewidth,draft=false]{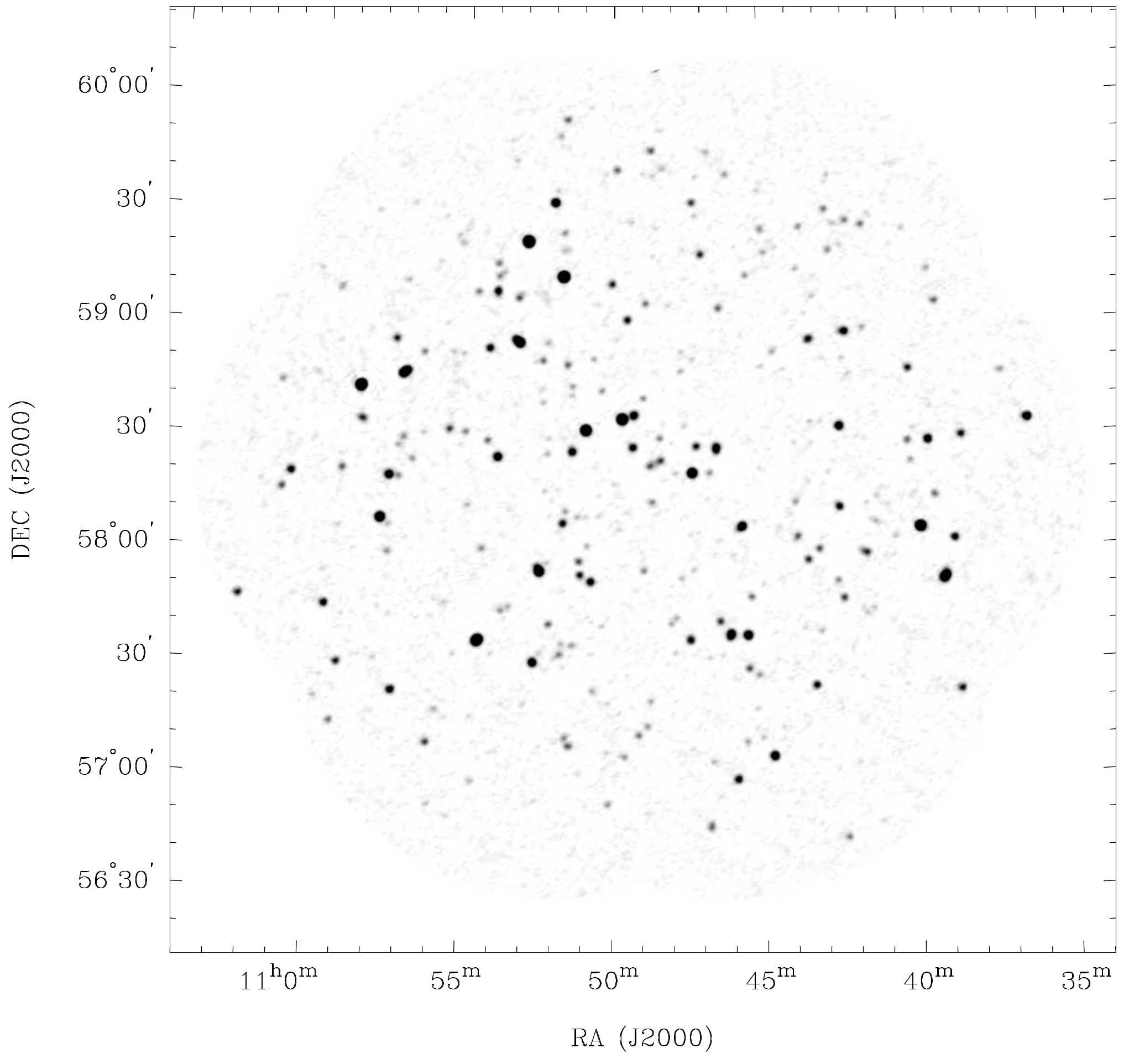}%
\includegraphics[width=0.49\linewidth,draft=false]{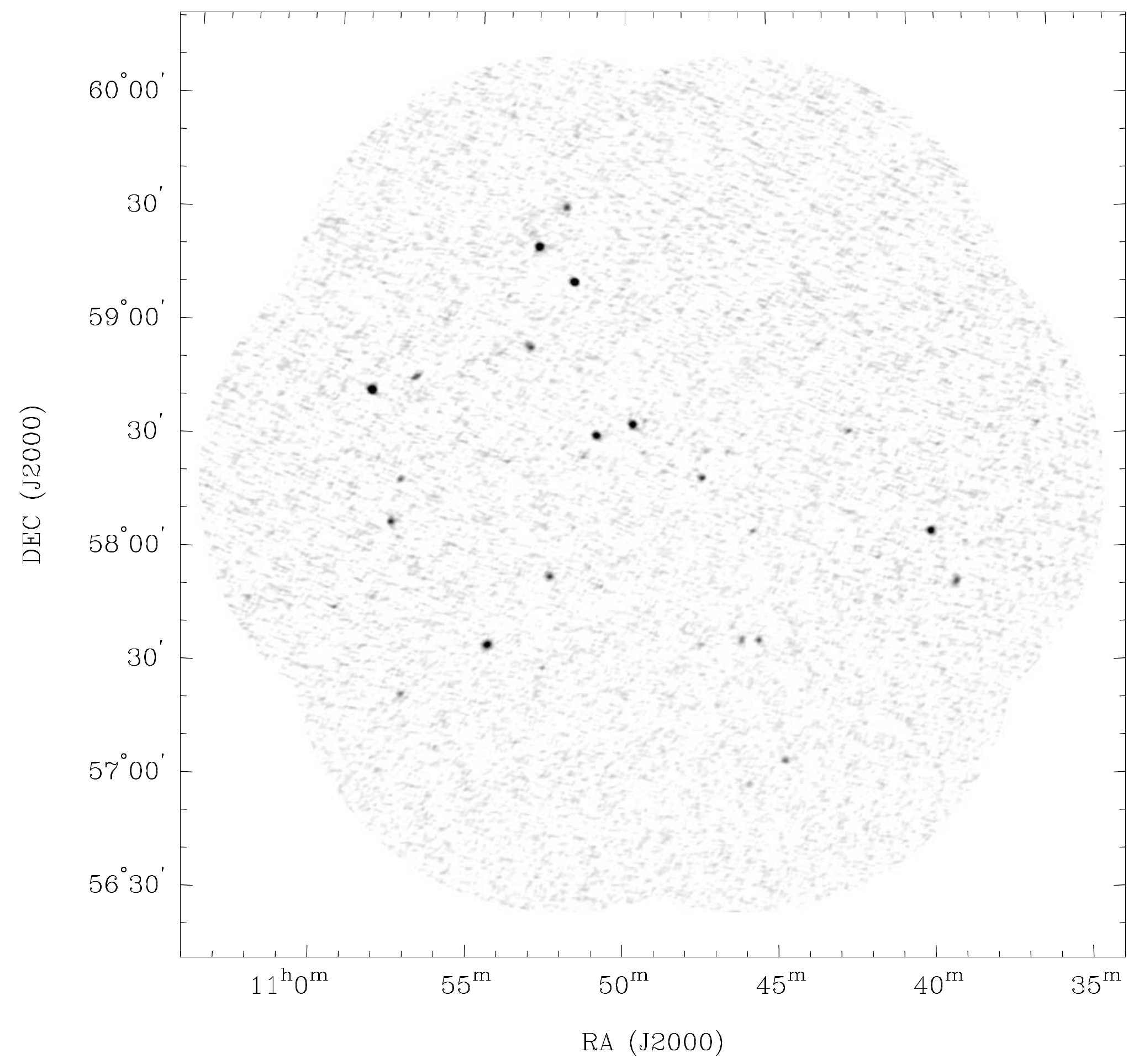}
\caption{\label{fig:lockmandeep}
{\em Left:}
Lockman deep field, made with data at both 3040 and 3140\,MHz.
The greyscale runs from 0 to 10\,\mjypbm.
\label{fig:lockmantyp}
{\em Right:}
Typical epoch (90\%\ completeness 15.6\,mJy) of the Lockman field at 3040\,MHz.
The greyscale runs from 0 to 50\,\mjypbm.
}
\end{figure*}

\begin{figure*}[htp]
\centering
\includegraphics[width=0.49\linewidth,draft=false]{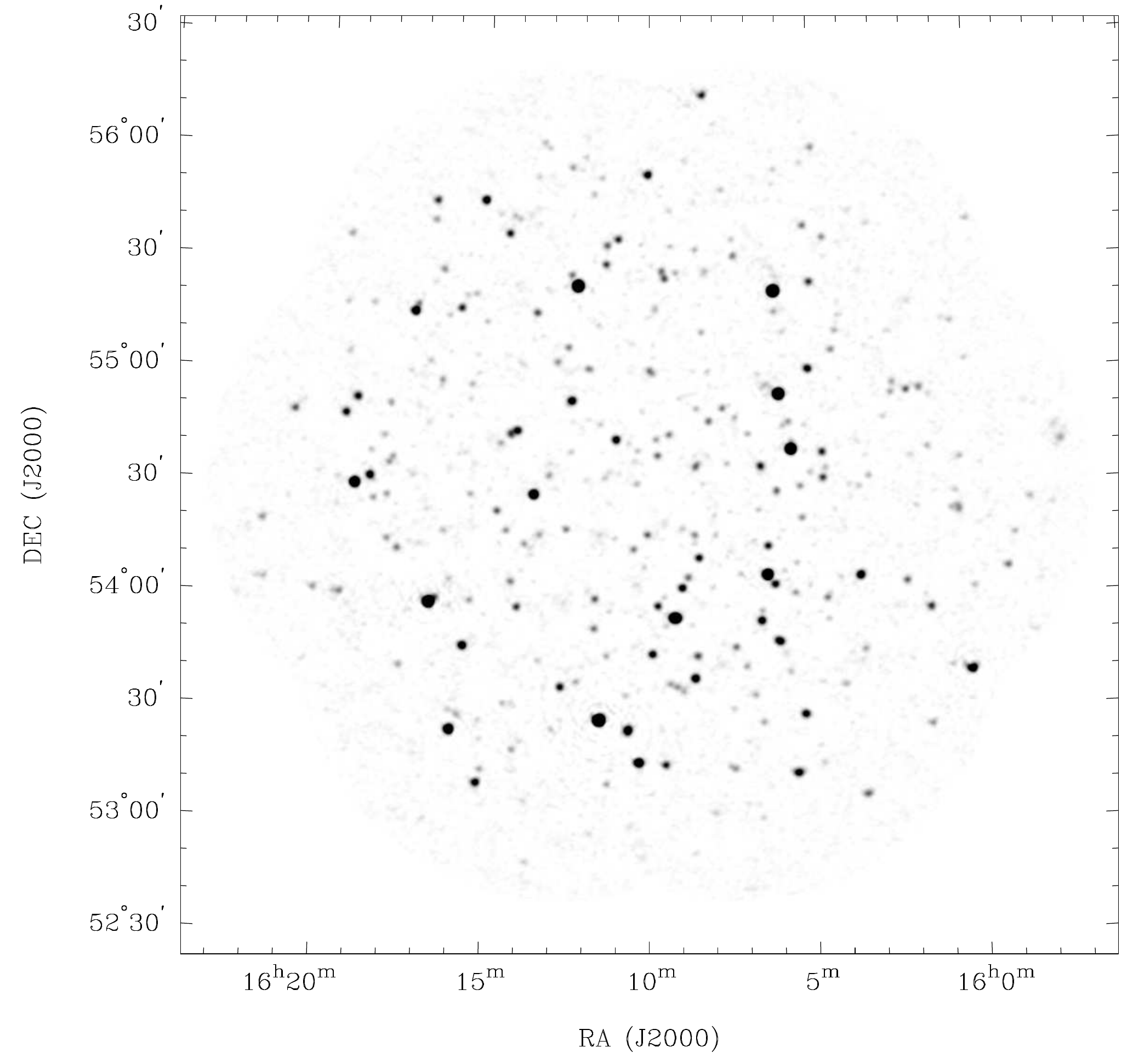}%
\includegraphics[width=0.49\linewidth,draft=false]{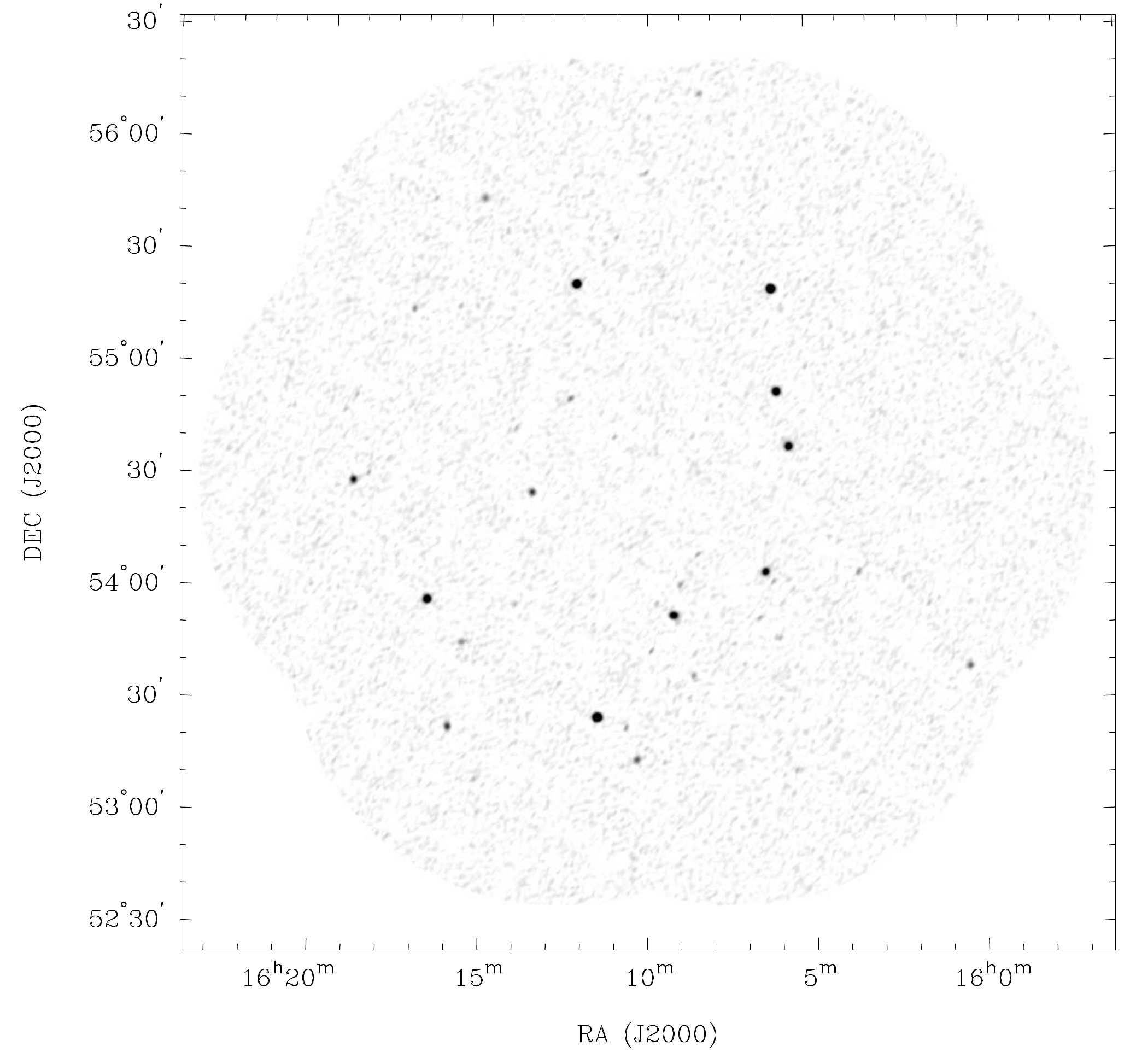}
\caption{\label{fig:elaisdeep}
{\em Left:}
ELAIS-N1 deep field, made with data at both 3040 and 3140\,MHz.
The greyscale runs from 0 to 10\,\mjypbm.
\label{fig:elaistyp}
{\em Right:}
Typical epoch (90\%\ completeness 10.0\,mJy) of the ELAIS-N1 field at 3040\,MHz.
The greyscale runs from 0 to 50\,\mjypbm.
}
\end{figure*}

\begin{figure*}[htp]
\centering
\includegraphics[width=0.49\linewidth,draft=false]{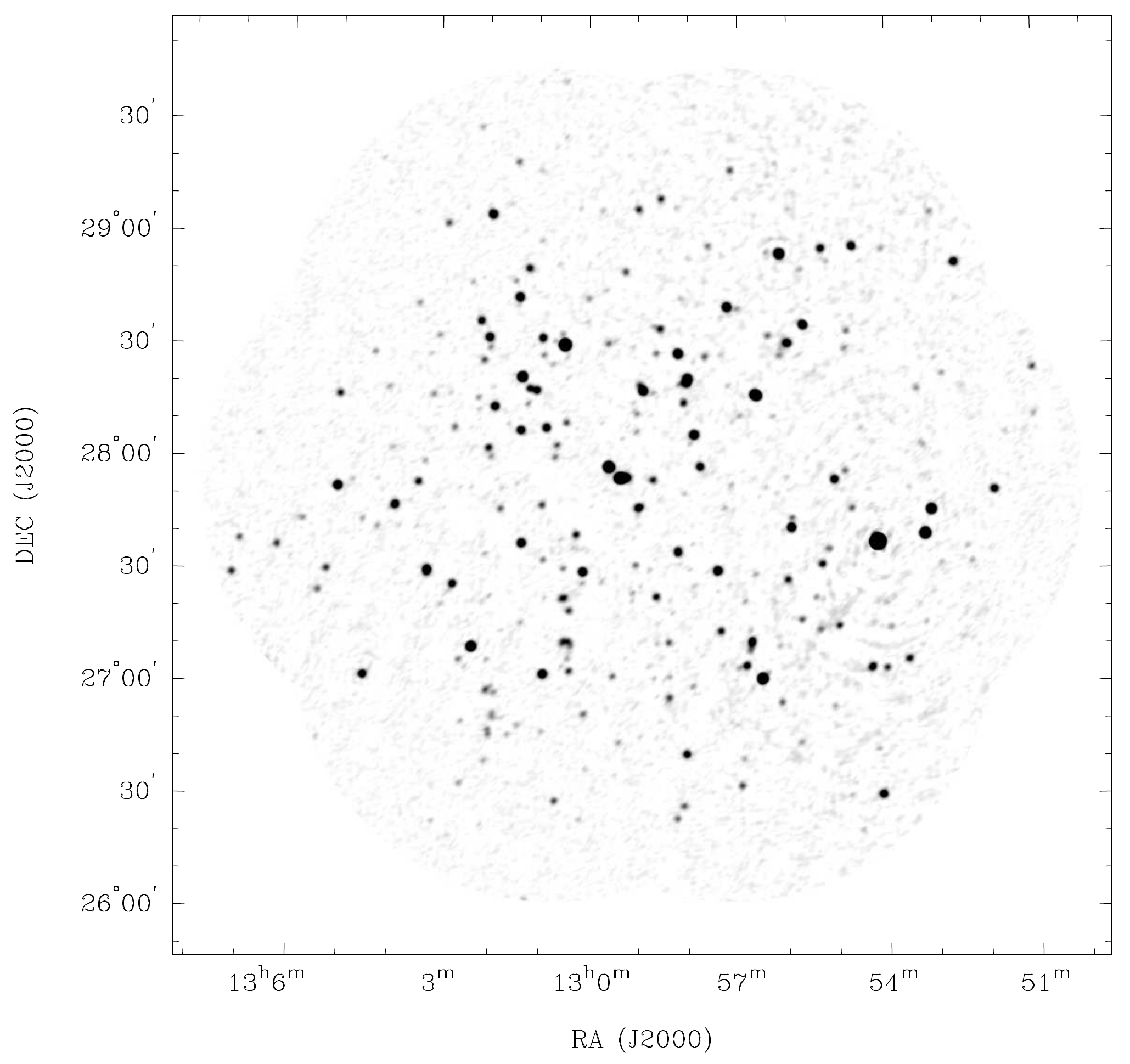}%
\includegraphics[width=0.49\linewidth,draft=false]{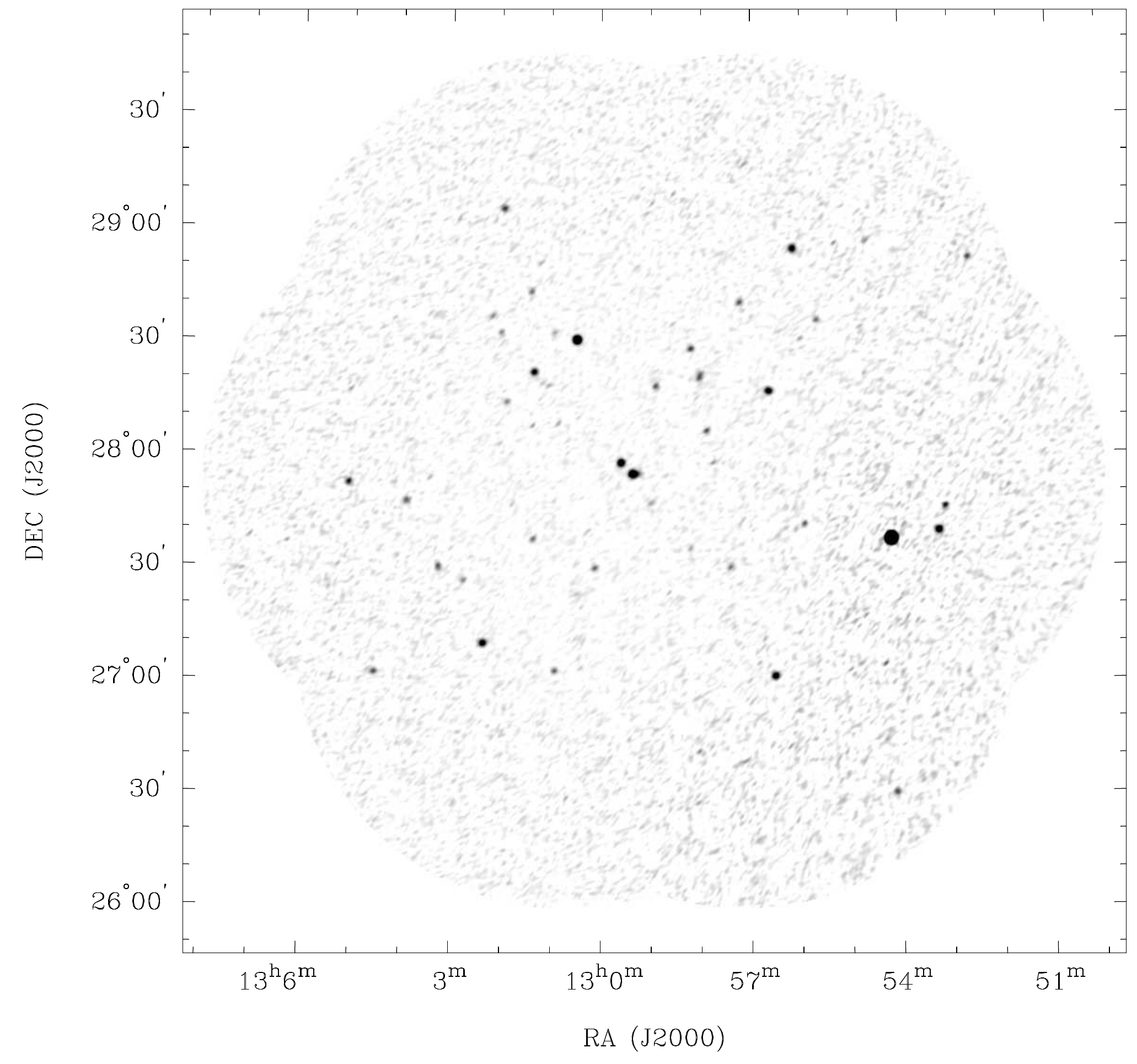}
\caption{\label{fig:comadeep}
{\em Left:}
Coma deep field, made with data at both 3040 and 3140\,MHz.
The greyscale runs from 0 to 10\,\mjypbm.
\label{fig:comatyp}
{\em Right:}
Typical epoch (90\%\ completeness 15.2\,mJy) of the Coma field at 3040\,MHz.
The greyscale runs from 0 to 50\,\mjypbm.
}
\end{figure*}

This reduction process generated over 6000 images --- two (one for each frequency band) for each of the seven mosaic pointings at each of the tens of epochs for each of the four fields. We computed the background rms for each of these images using a skewed Gaussian fit to the pixel histogram, with iterative clipping to remove source pixels. Those of the $\sim 6000$ datasets which resulted in images with rms $\leq 3.5$\,\mjypbm\ (Fig.~\ref{fig:rmshist}) were considered to be good datasets, for the purposes of creating deep images (Figs.~\ref{fig:lockmandeep}, \ref{fig:elaisdeep}, and \ref{fig:comadeep}; except that the NDWFS deep image was constructed using the same single-epoch datasets used in \paperi). One image was made for each pointing for each frequency in each field. Due to small variations in image quality, the number of epochs which satisfied the 3.5\,\mjypbm\ threshold varied by $\sim \pm10$\%\ across the seven pointings, resulting in slight ($\sim 5$\%) variations in limiting sensitivity across the mosaic, which we neglect in subsequent analysis. In Table~\ref{tab:fields}, we report the median number of epochs that went into each pointing.

The resulting seven all-epoch single-pointing images at each frequency were linearly combined into two all-epoch single frequency band ``master mosaics'', using a tapered primary beam correction in LINMOS. We also linearly combined these pairs of mosaic images into a dual-frequency mosaic for each field.

In addition to these three deep images for each of the fields, we made seven-pointing mosaic images of each epoch at each frequency, and linearly combined these into a dual-frequency mosaic for each field. Typical images, here defined as mosaics with the median completeness (Section~\ref{sec:completeness}) for each field, are shown in Figs.~\ref{fig:lockmantyp}, \ref{fig:elaistyp}, and \ref{fig:comatyp}. We also made seven-pointing mosaics of all good data taken in a given month for each field, again inverting data for each pointing separately, and mosaicking in the image domain with a tapered primary beam.

Reprojection to an equal-area coordinate system is necessary for large mosaics, in order that flux densities are measured correctly. However, for the relatively small fields studied here this effect is negligible ($< 0.1$\%\ variation in pixel area in the default SIN projection), and we chose not to reproject in order to avoid possible aliasing as well as the additional processing required.

The single-epoch, single-pointing images had a median rms of 2.42\,mJy (compared to theoretical expectations of $\sim 1.4$\,mJy).
The single-epoch mosaics had a median rms of 2.02\,mJy.
The monthly images had a median rms of 0.43\,mJy.
The deep images had rms of a few hundred \ujy\ (as expected when scaling the rms of the individual images by the square root of the number of epochs), as shown in Table~\ref{tab:fields}. The ELAIS-N1 field reaches a sensitivity close to our estimate (\paperii) of the rms confusion limit, $\sim 150$\,\ujy.

\begin{figure*}[htp]
\centering
\includegraphics[width=0.45\linewidth,draft=false]{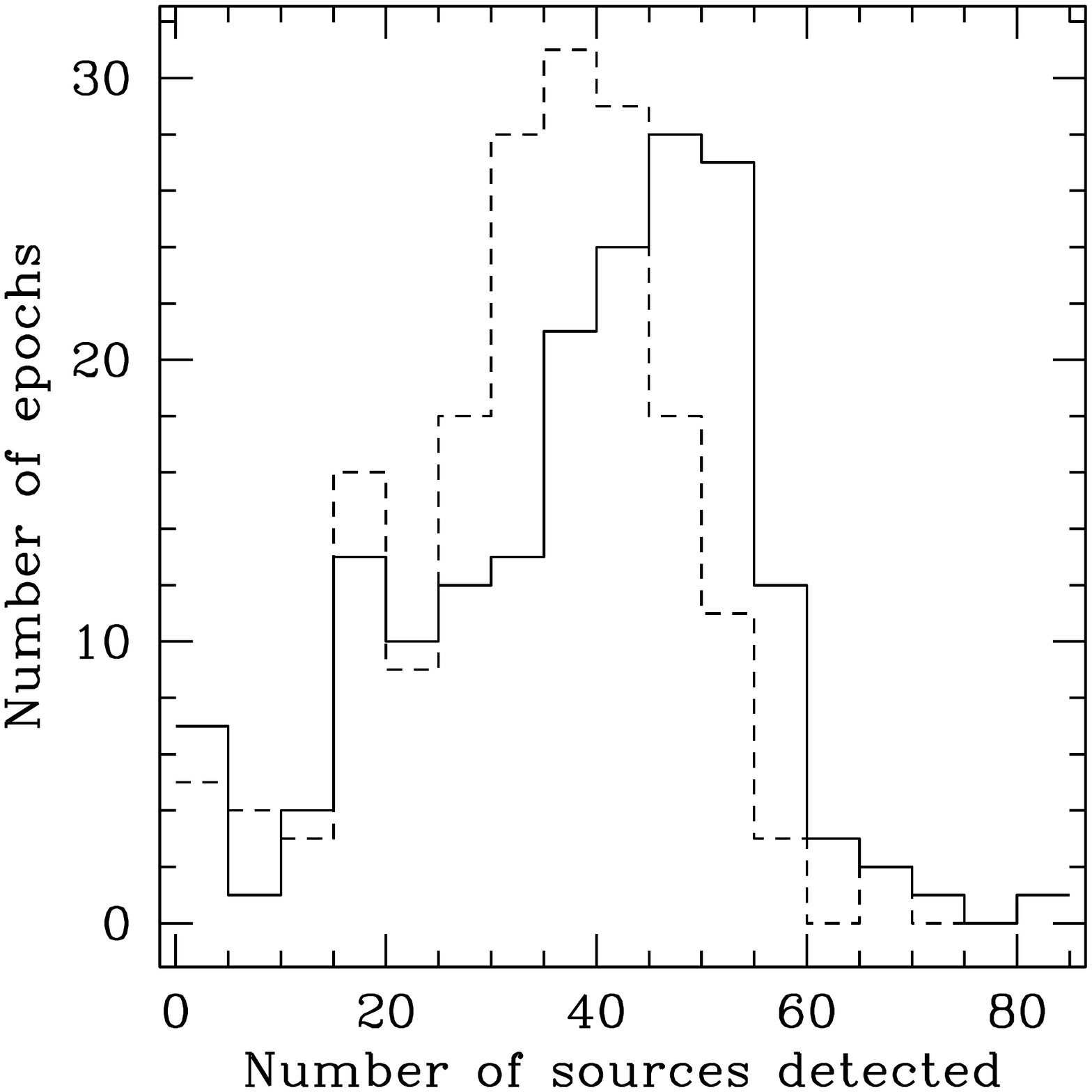}\hspace{0.03\linewidth}%
\includegraphics[width=0.45\linewidth,draft=false]{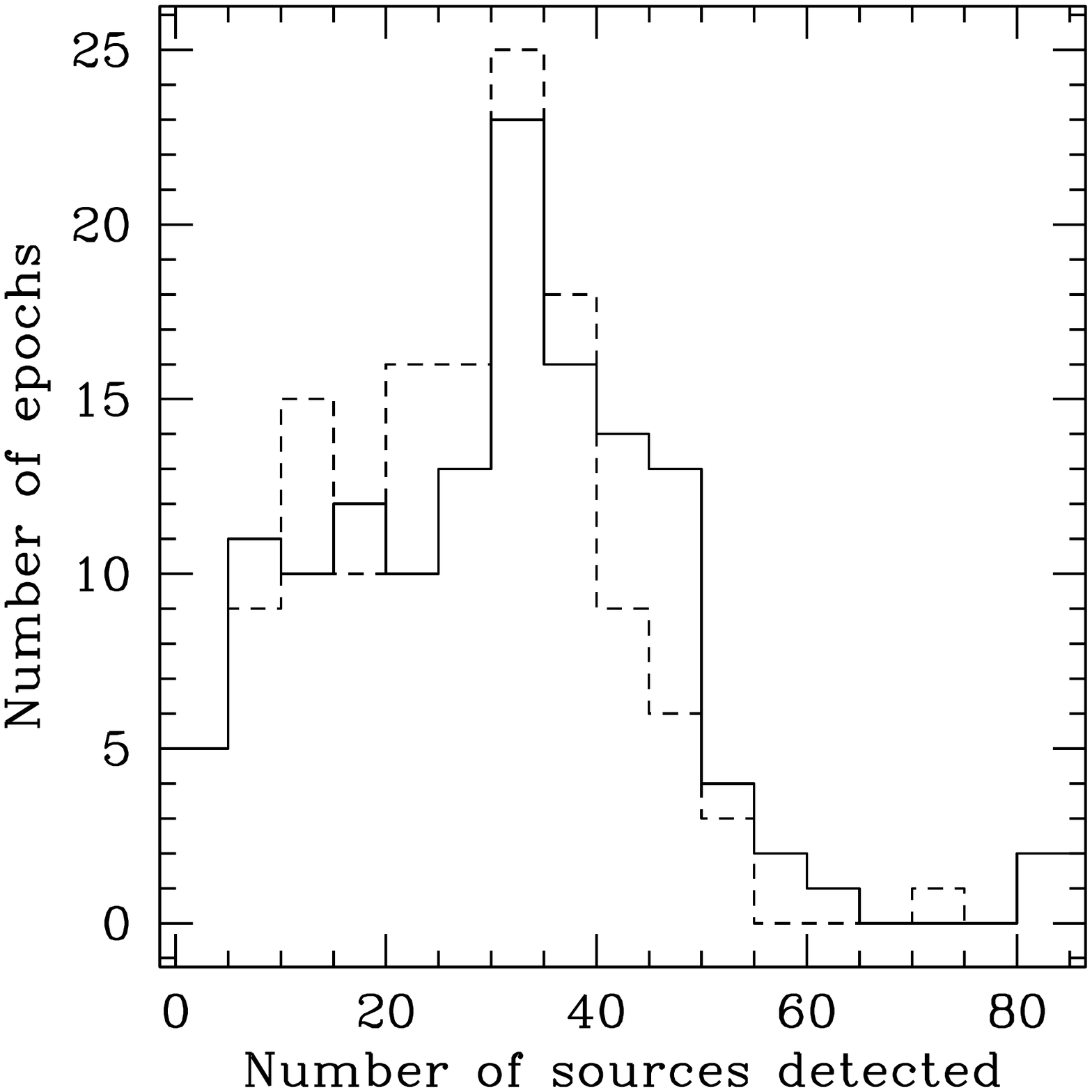}
\includegraphics[width=0.45\linewidth,draft=false]{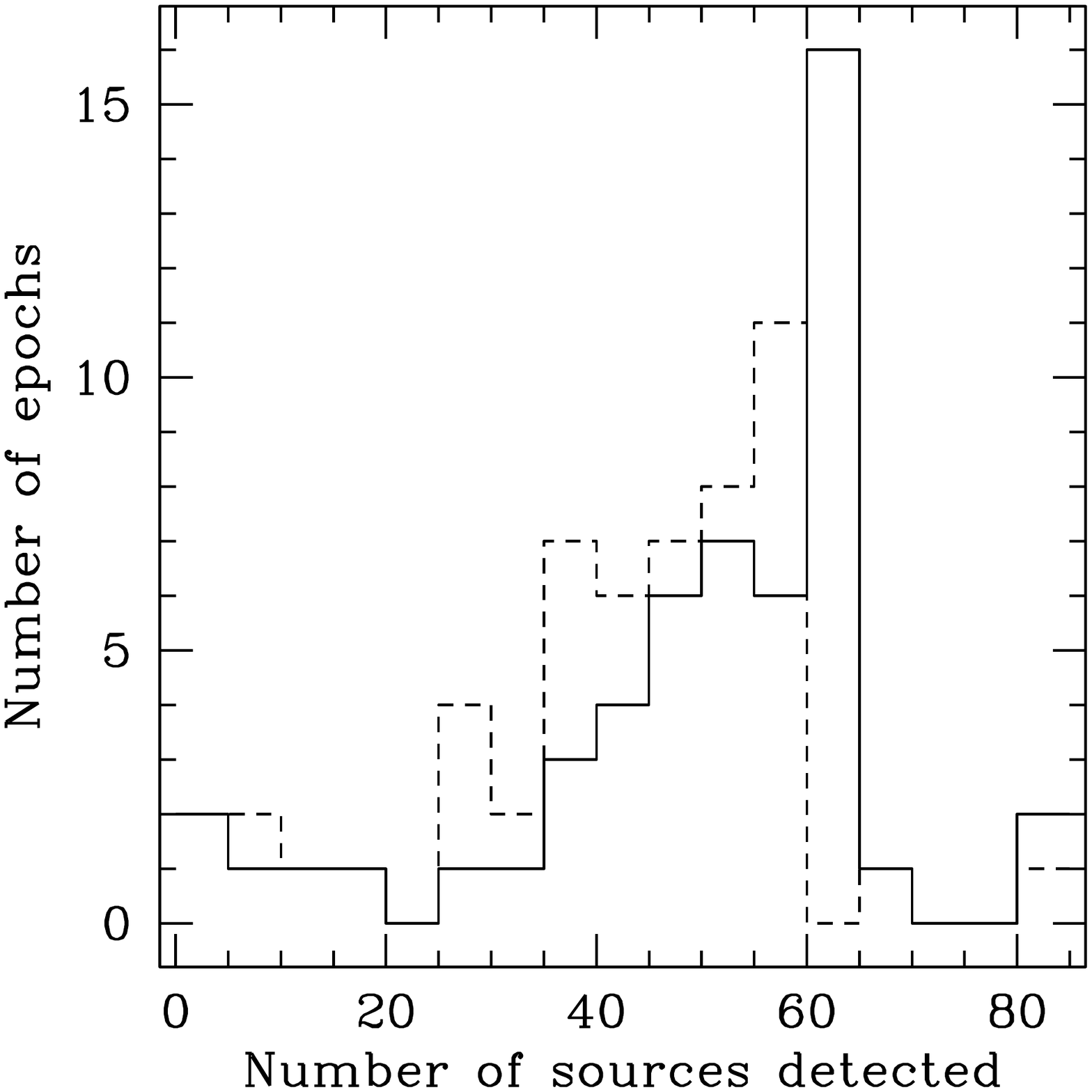}\hspace{0.03\linewidth}%
\includegraphics[width=0.45\linewidth,draft=false]{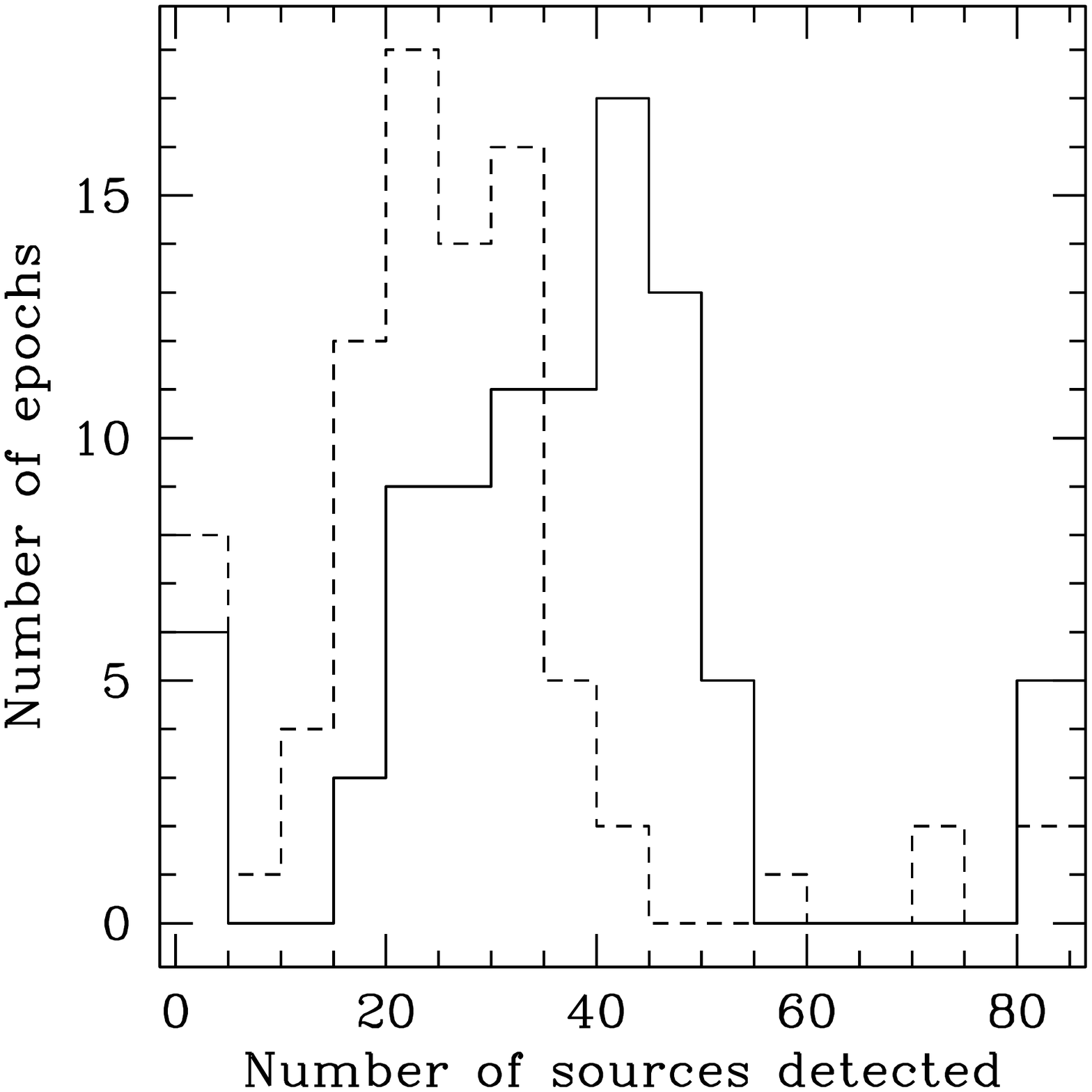}
\caption{\label{fig:sfindhist}
Number of sources detected by SFIND in images of all individual single-epoch mosaics at 3040 (solid line) and 3140\,MHz (dashed line) for ELAIS-N1, Lockman, Coma, and NDWFS.
}
\end{figure*}

Image quality can be assessed in a variety of ways. We inspected all of the images by hand to ensure that selecting particular methods and thresholds did not inadvertently result in the inclusion of bad data (or the exclusion of significant amounts of good data). For the next generation of large surveys, however, inspecting more than a small fraction of the images will become impractical. Fortunately, good images have certain characteristics: they have noise that is reasonably low, they contain a reasonable number of real sources (Fig.~\ref{fig:sfindhist}), and the sources have reasonable flux densities. 

As discussed above, we generated deep field images based on the noise measured in images made from data for individual pointings, keeping only images with rms noise $< 3.5$\,\mjypbm. As shown in Fig.~\ref{fig:rmshist}, and in Table~\ref{tab:fields}, around 85\%\ of the ELAIS and Coma single-epoch images satisfy this criterion, and around 65\%\ of the Lockman images. Some epochs have rms higher than this cutoff, but can nevertheless be used to search for transients and variable sources, whereas a handful of epochs are completely unusable.

\begin{figure*}[htp]
\centering
\includegraphics[width=0.45\linewidth,draft=false]{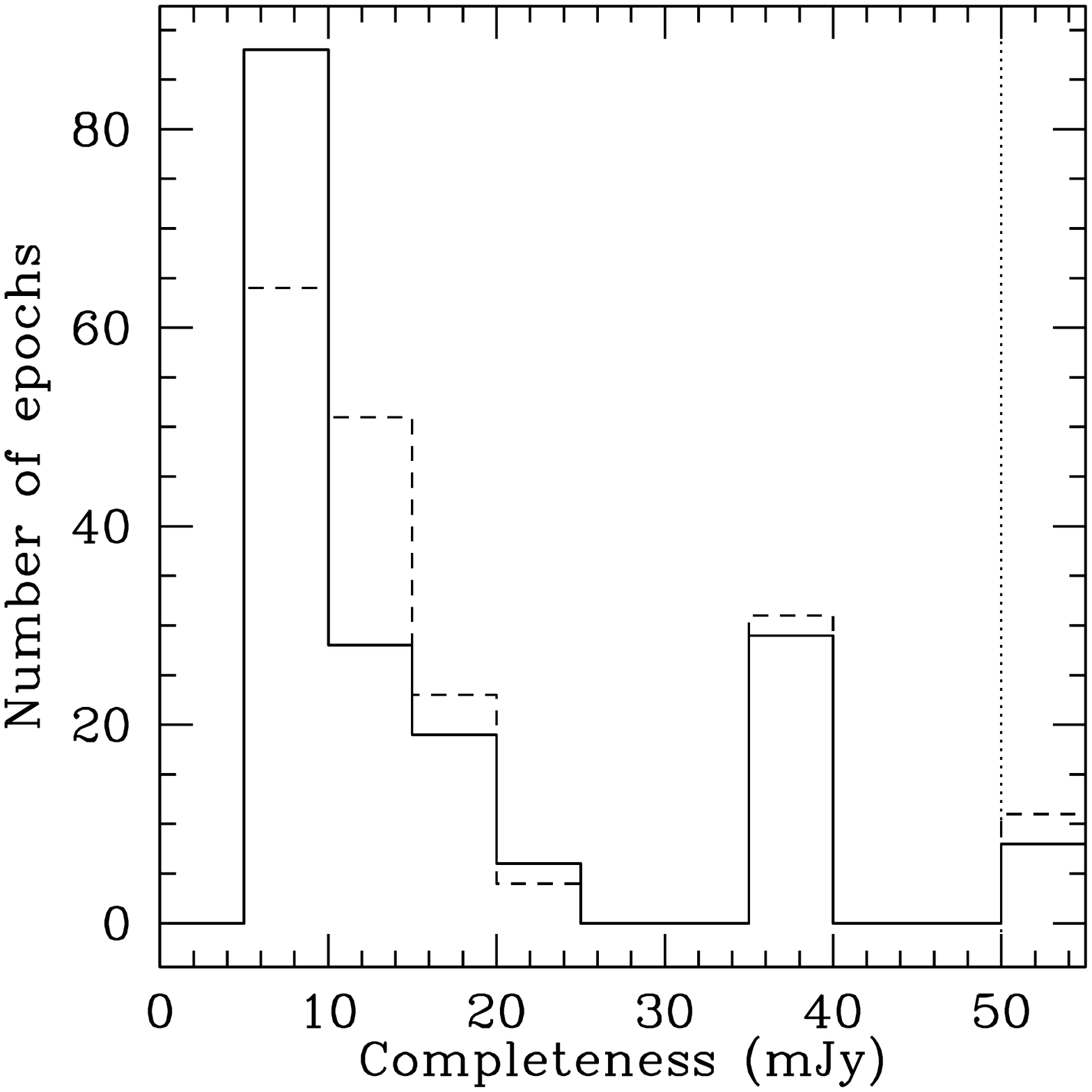}\hspace{0.03\linewidth}%
\includegraphics[width=0.45\linewidth,draft=false]{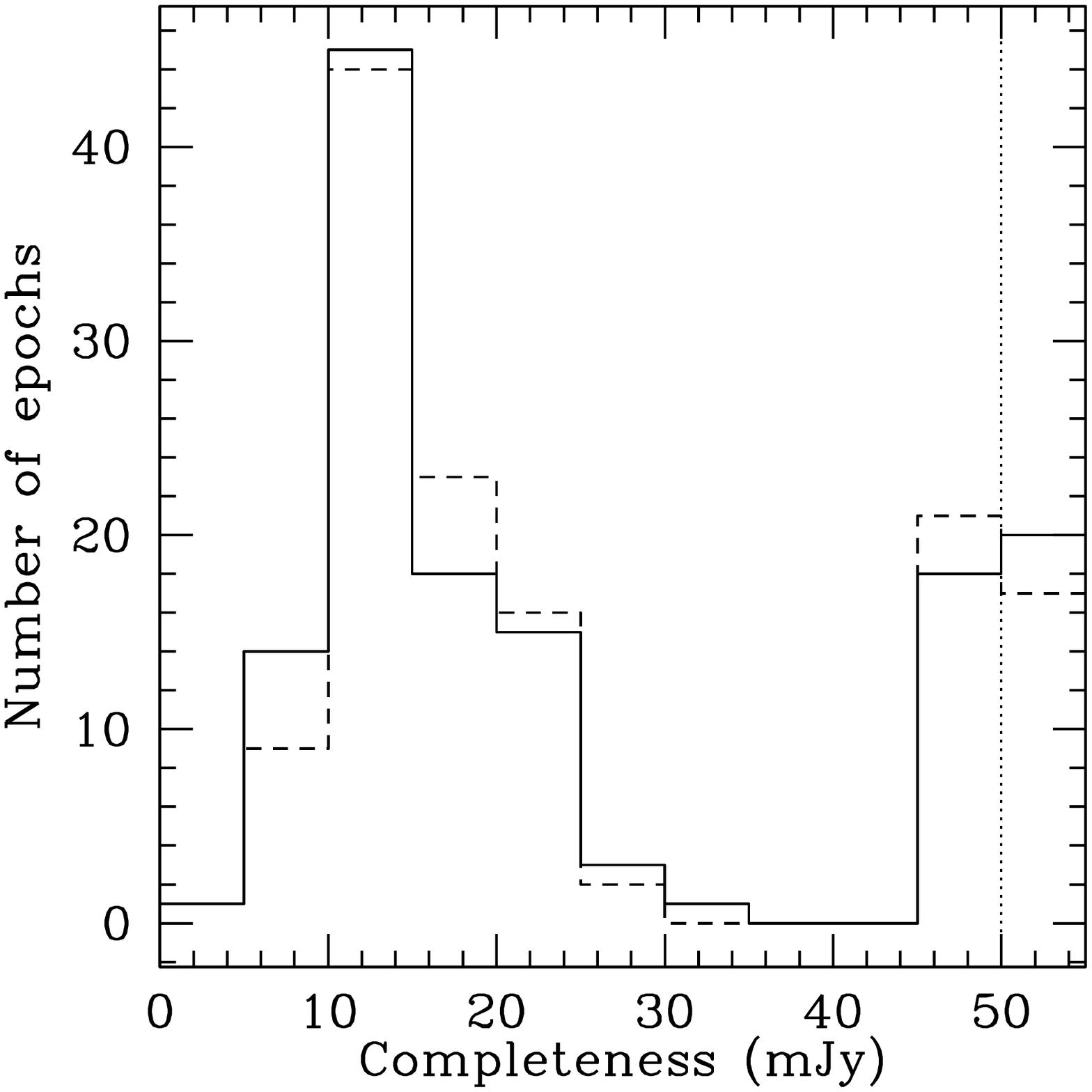}
\includegraphics[width=0.45\linewidth,draft=false]{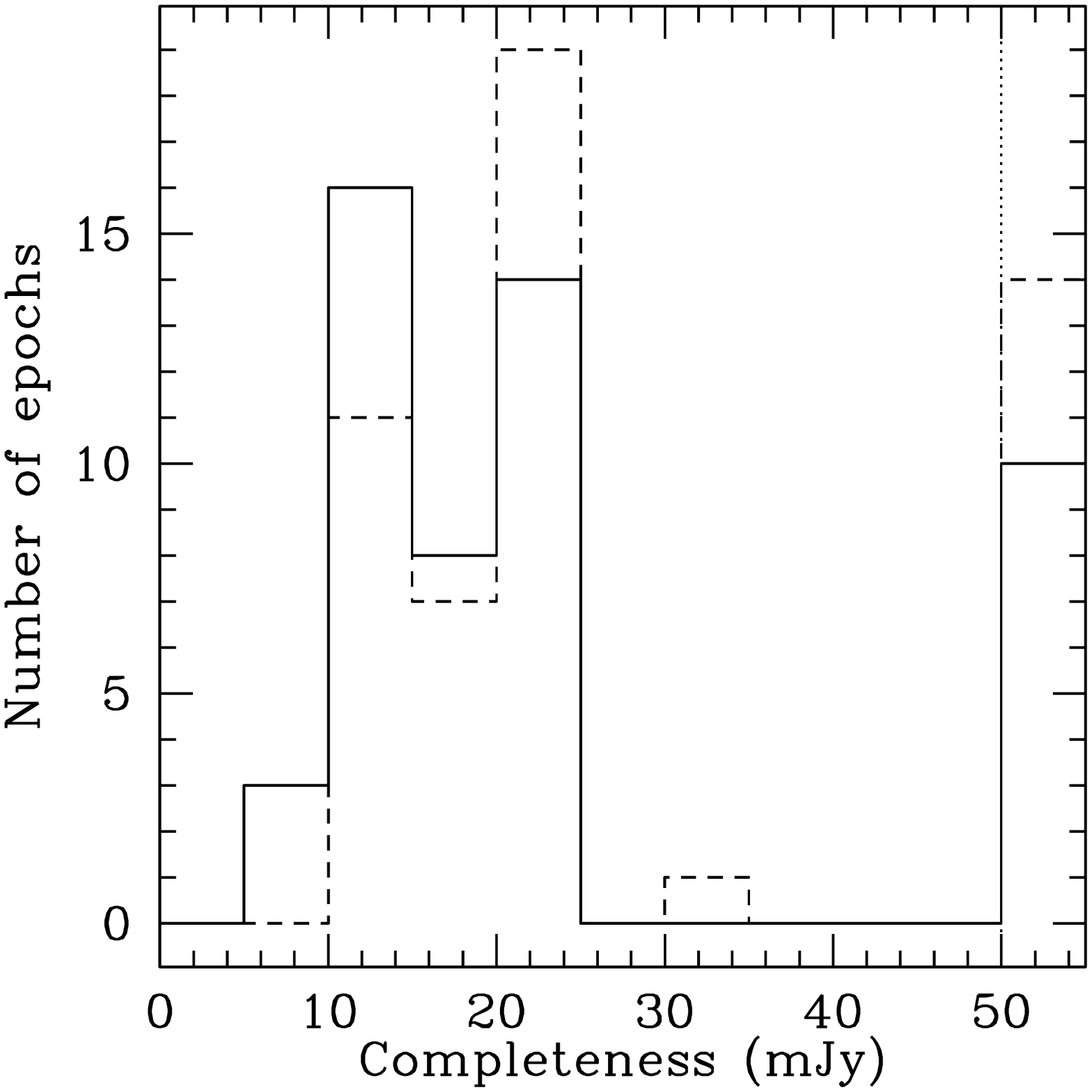}\hspace{0.03\linewidth}%
\includegraphics[width=0.45\linewidth,draft=false]{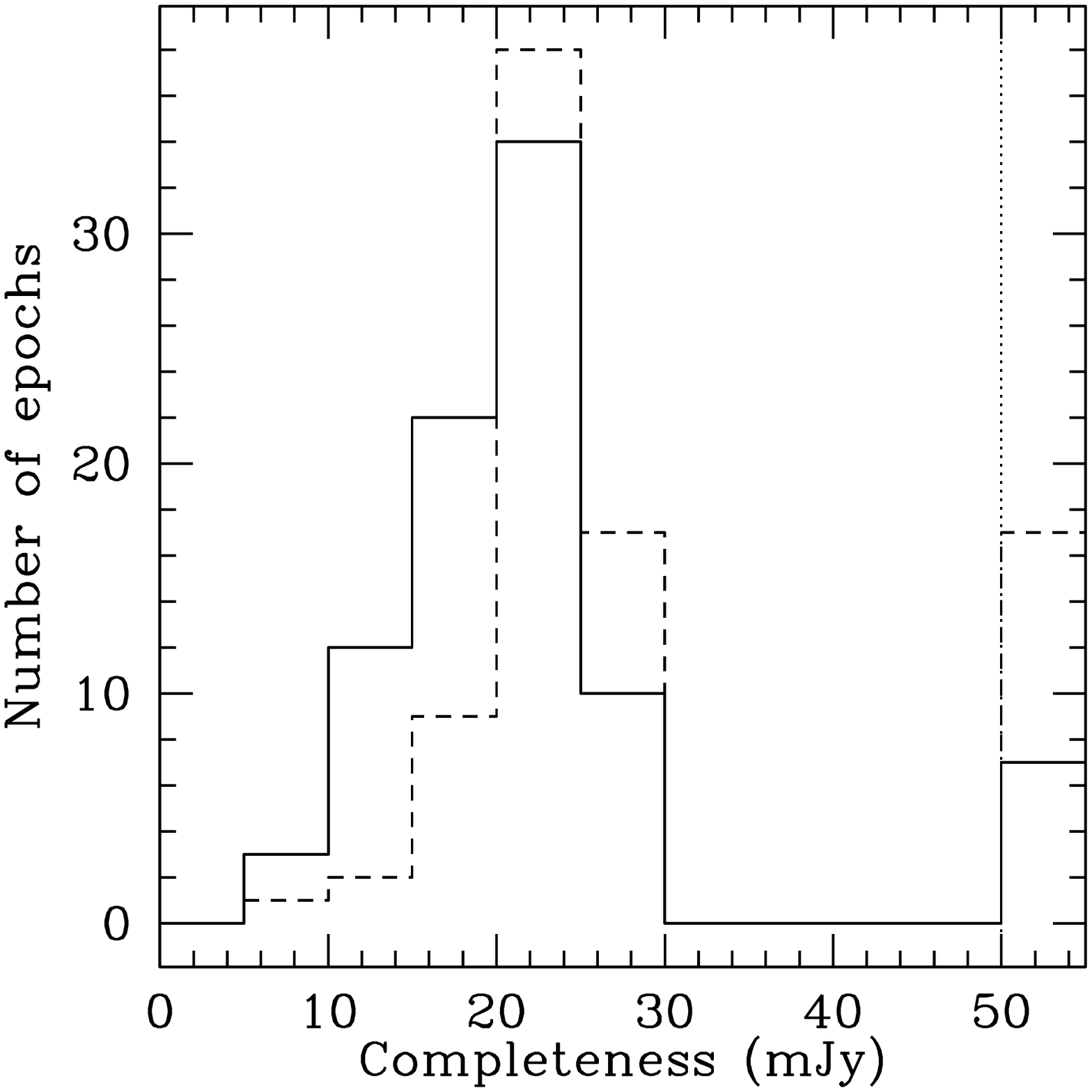}
\caption{\label{fig:comphist}
90\%\ completeness limits for all individual single-epoch mosaics at 3040 (solid line) and 3140\,MHz (dashed line) for ELAIS-N1, Lockman, Coma, and NDWFS. The dotted lines show the 50\,mJy cut used to select epochs for the transient search. 
}
\end{figure*}

To determine transient source counts or limits we wish to determine the effective area covered by the survey as a function of completeness limit (Section~\ref{sec:completeness}). Images which appear cosmetically very poor (some of which also contain spurious sources in the catalogs as a result of image defects) will also tend to have poor completeness limits. For our transient search, we discard epochs where 90\%\ completeness is not reached by 50\,mJy (Fig.~\ref{fig:comphist}).  The number of epochs surviving these cuts for each field are shown in the ``Transients'' column in Table~\ref{tab:fields}.

To measure variability, we require relatively good completeness for all images, as well as reliably measured flux densities. For our variability analysis, we adopt a completeness threshold of 50\,mJy, and require post imaging calibration gain corrections (Section~\ref{sec:picgain}) within $\pm 0.1$ of the median value for that field. The number of epochs surviving these cuts for each field is shown in the ``Variability'' column in Table~\ref{tab:fields}.

\subsection{Catalog Creation and Matching}

Source detection, matching, catalog creation, source databases and matching, and lightcurve and postage stamp creation were performed by our custom software known as ``SLOW'', a suite of Python and Perl scripts calling MIRIAD routines and building and querying sqlite3 databases.
 
SLOW first runs MIRIAD SFIND on the master mosaic and on the single-epoch mosaics. We used SFIND's false detection rate (FDR) algorithm \citep{sfindfdr} with a box size of 50 pixels, and the default 2\%\ acceptable percentage of false pixels. This corresponds to a detection threshold of around $4.2\sigma$. The ``psfsize'' option was used, which restricts the minimum fitted size of sources to the size of the restoring beam; without this option, some sources in the single-epoch images tend to be fit by ellipses with minor axis $< 100\arcsec$, causing their flux densities to be underestimated relative to those from the master mosaic. Sources poorly fit by SFIND (indicated by asterisks in the SFIND output), typically comprising $\sim 3$\%\ of the total sources detected in each epoch, were rejected from the catalog. We computed total flux density uncertainties by adding the reported background rms and fit uncertainties from SFIND in quadrature.

We corrected the attenuation at the edge of the mosaics by creating a gain map for each epoch in LINMOS, and correcting the measured flux densities and uncertainties in the catalogs for the appropriate gains (since SFIND's built-in primary beam correction does not work for linear mosaics). Sources in regions with gain corrections greater than a factor 5 were discarded from the catalog\label{sec:biggain}. This helps to minimize the problem of noise variations close to image edges with large gain corrections being interpreted as real sources.

Sources were added to the database one epoch at a time, and paired with existing sources if there was a match within 150\arcsec. The full sqlite3 databases are available from the first author on request.

\subsection{Post-imaging calibration}

\begin{figure}[htp]
\centering
\includegraphics[width=\linewidth,draft=false]{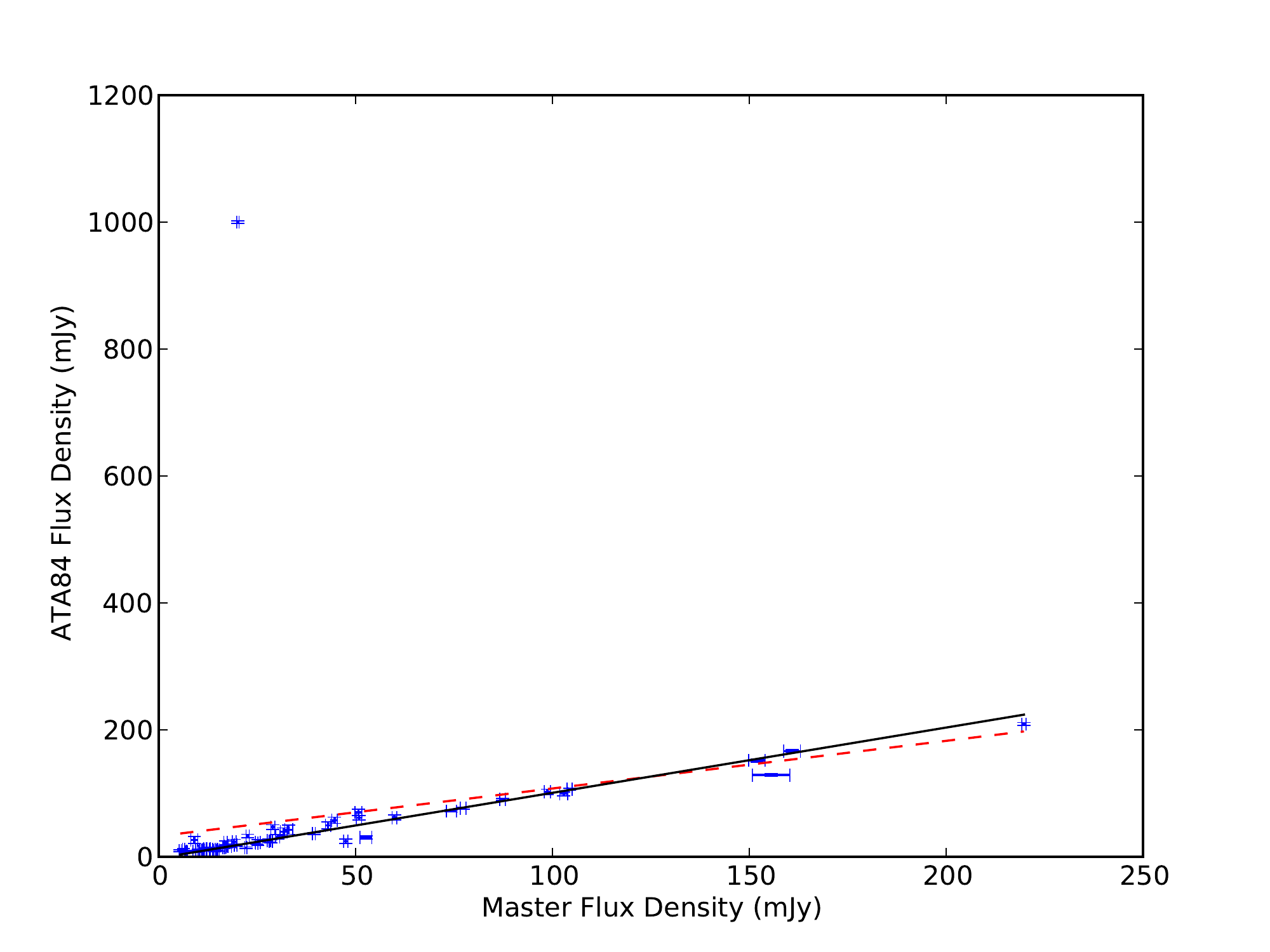}
\caption{\label{fig:ransac}
Flux densities of sources detected in both a typical epoch (epoch 84) and also in the master catalog of PiGSS ELAIS-N1 3040\,MHz. One of the sources, detected with 19.4\,mJy in the master catalog, has had its flux density in the single epoch catalog manipulated to make it into an outlier with 1000\,mJy, in order to test the robustness of our fitting procedure. The dashed line shows the results of a weighted linear least squares fit: a slope of 0.75, and an intercept of 33\,mJy, which were influenced by this single outlier. The solid line shows the RANSAC fit: slope 1.03, and intercept -2\,mJy, which are more representative of the inliers in the dataset. When determining fits for the entire dataset, we use the RANSAC method with fits constrained to pass through zero.
}
\end{figure}

We performed a modified version of the post-imaging calibration (PIC) of \citet{bannister:11} on our catalogs. Sources in the individual epoch catalogs were matched to those from the master catalog, and a linear least squares fit to the primary beam corrected flux densities was performed. An ordinary least squares fit (with weights given by the image uncertainties) gives good results for the majority of epochs, but we were concerned that real variations in source flux densities could cause some epochs to have poor fits. While this should not be a problem for transient detection (since transients will not have a match in the master catalog), sources with extreme flux density variations could cause poor fits. To test this, we made a copy of the SLOW database which we then manipulated for testing purposes. For a source detected in most of the individual epochs, and also detected in the master mosaic with flux density 15\,mJy, we increased its cataloged flux value at each individual epoch to 1000\,mJy, and then ran the fitting procedure again. Figure~\ref{fig:ransac} shows data from a typical epoch, where this artificially modified source (designed to simulate an extreme variable or a source with an erroneous flux density measurement in real data) causes the least squares fit to fail. Performing a RANSAC fit \citep{ransac}, however, results in fits to the data which are very robust to outliers, and so we adopt this technique for all of our fits. No sources in our data show variations in flux density as extreme as our simulated example, so our fits should describe the data very well.

If we fit for both the slope and intercept, we find that the absolute value of the median intercepts are $\lesssim 2$\,mJy for the Lockman and ELAIS-N1 fields at both frequencies, which suggests that CLEAN bias is not significant. For the Coma field, median intercepts are $\sim -6$\,mJy. Coma exhibits extended structure and a bright source, which make imaging and source detection more challenging and probably contribute to the worse fits and possible CLEAN bias here. For subsequent analysis, we follow \citet{bannister:11}, and constrain the fits to pass through zero. 

\begin{figure*}[htp]
\centering
\includegraphics[width=0.47\linewidth,draft=false]{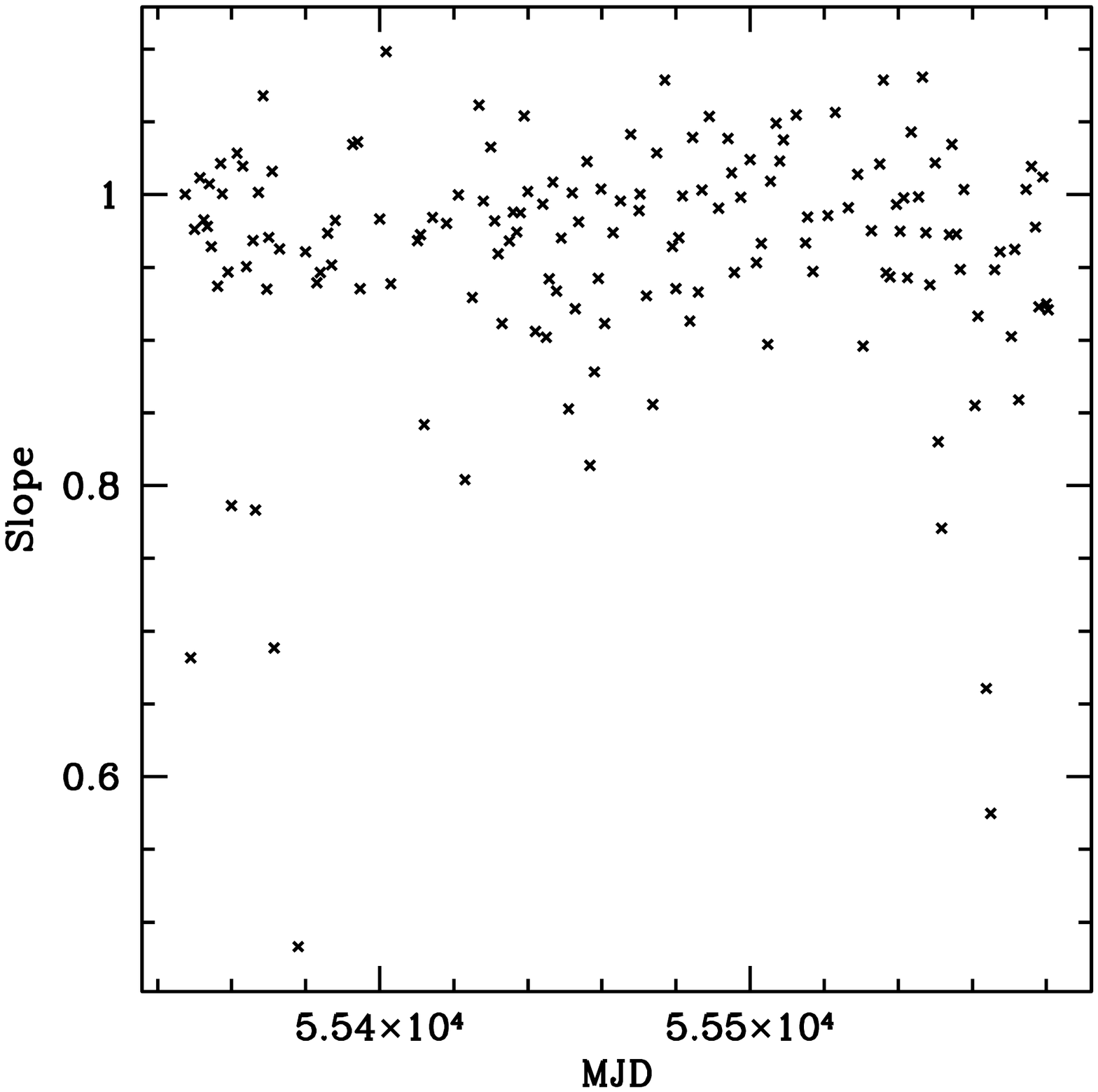}\hspace{0.05\linewidth}%
\includegraphics[width=0.47\linewidth,draft=false]{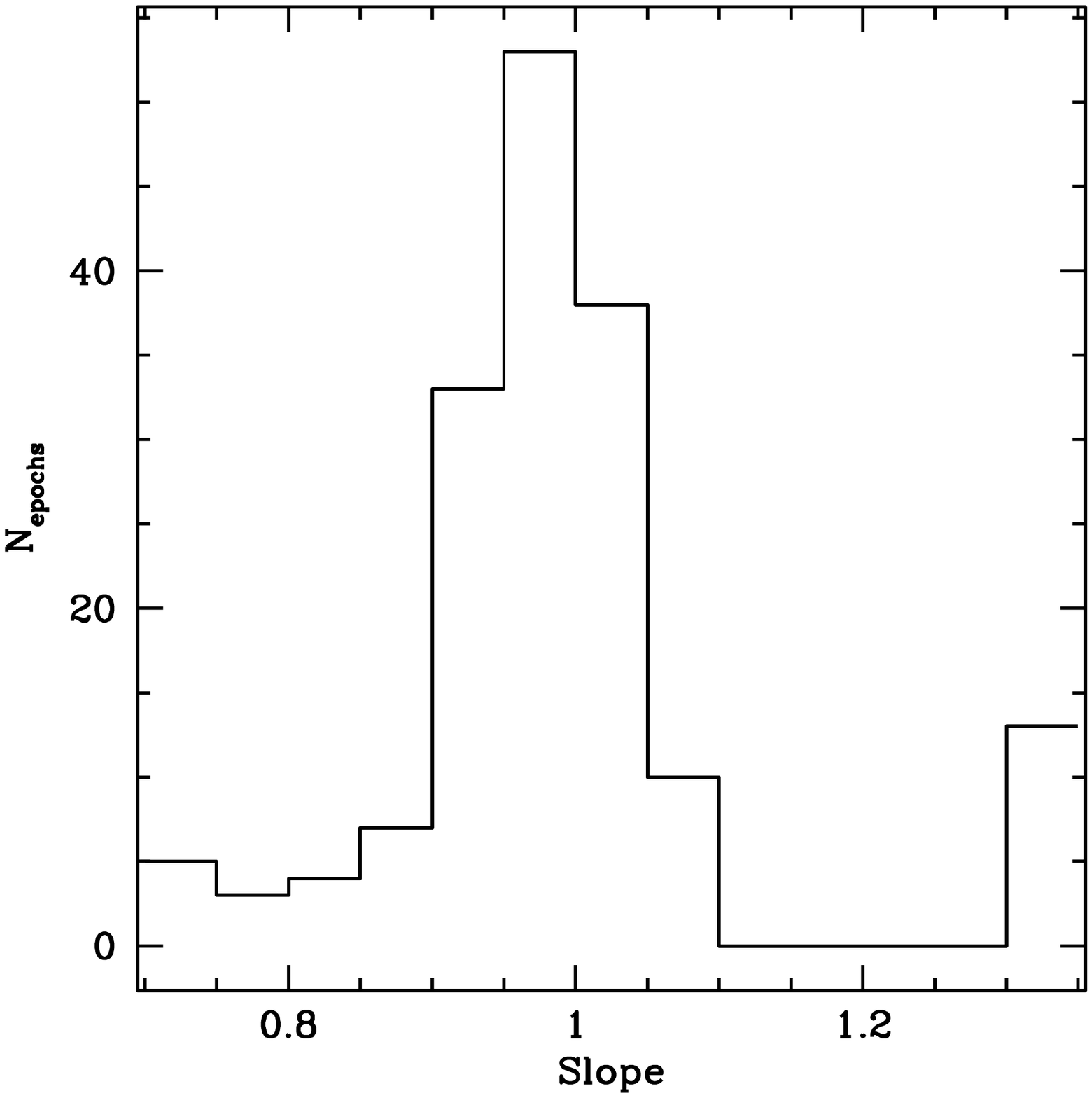}
\caption{\label{fig:elaisslopes}
Left: Slopes of RANSAC fits to the ELAIS 3140\,MHz single epoch data when matched to the ELAIS 3140\,MHz master catalog. A slope, or post-imaging calibration (PIC) gain of 1 indicates epochs where flux densities in the individual epochs were unbiased relative to the deep image. Large excursions represent epochs with poor fits, typically due to calibration problems or other issues with the array. Periods where array performance was not optimal (for example, the excursions around MJD 2455560, \ie, late December 2010) can be seen. Checking the ATA email archive, we find that it was snowing at the observatory at that time, which was probably the reason for the poor calibration in this particular case.
Right: Histogram of the PIC gains from the left panel. Epochs with good data tend to have PIC gains between about 0.9 and 1.1. 
}
\end{figure*}

We use the slope of the least squares fit to estimate a PIC gain correction\label{sec:picgain} for the single epoch mosaics relative to the master mosaic. 
We obtain median PIC gains of 0.94 for Lockman, 0.98 for ELAIS-N1, 0.96 for NDWFS, and 0.89 for Coma.\label{sec:medianpic} PIC gains smaller than one mean that single-epoch flux densities somewhat underestimate deep image flux densities.
The standard deviation (measured for each field and frequency) of PIC gains is  $\sim 0.1$ (\eg, Fig.~\ref{fig:elaisslopes}), suggesting that the overall calibration is good to about this level. We can also use these gain variations to estimate a cutoff for the convincing detection of variability in lightcurves; the epoch-to-epoch variations in gain suggest that a sensible threshold (for well-detected sources) is $3\sigma \sim 30\%$, although some sources do have discrepant measurements that are larger than this, even in epochs where the PIC gain suggests an overall calibration that is reasonable. In a small fraction ($\sim 7$\%) of epochs with a large amount of scatter between the single-epoch and deep-image flux densities, the RANSAC fit fails to converge to within our chosen parameters (at least 70\%\ of sources selected as inliers after 1000 iterations) and these epochs are rejected in our variability analysis (Section~\ref{sec:variability}).

Since the PIC gain corrections are relatively small (albeit larger in Coma due to the extended structure and bright sources), and uncertain themselves to at least a few percent due to the small number of sources present in the fits, the intrinsic variability of individual sources, and the variation in thresholds from epoch to epoch, we do not apply corrections to fluxes measured in individual epochs. The PIC gains do, however, provide a good means by which to identify bad epochs. For variability studies, PIC gain stability is relatively important, lest PIC gain variations be mistaken for intrinsic source variability (Section~\ref{sec:variability}). For transient searches, larger discrepancies in PIC gains may be tolerated in order to increase the effective area of the transient search at the expense of increased systematic errors in determining transient flux densities. 

\subsection{Removal of Spurious Sources}\label{sec:dualimageimp}

As discussed in Section~\ref{sec:dualimageint}, dividing up the data before imaging can result in a decrease in the number of spurious sources (particularly those due to image defects that do not obey Gaussian statistics). This helps in the detection of transients, but also helps to generate purer single-epoch catalogs. We therefore accept only sources that are independently detected in both frequencies in at least one epoch (with a position matching tolerance of 50\arcsec, corresponding to a false match probability of $< 2$\%). In many cases such sources are detected in additional epochs, although not necessarily always at both frequencies in every epoch, particularly if they are faint or if a single-epoch image is of poor quality.

Our threshold of $\sim 4.2\sigma$ for detection in a single image corresponds to a threshold of $\sim 5.9\sigma$ in the dual-image catalog. We generated catalogs for the deep fields, consisting only of sources detected at both frequencies. The catalogs for the ELAIS-N1 (Table~\ref{tab:elaiscat}), Lockman (Table~\ref{tab:lockmancat}), and Coma (Table~\ref{tab:comacat}) deep fields contain 238, 189, and 186 sources respectively. The new catalog for the NDWFS deep field (Table~\ref{tab:ndwfscat}) contains 195 sources.

\begin{deluxetable*}{lllllllllllllll}
\tablewidth{0pt}
\tabletypesize{\scriptsize}
\tablecaption{\label{tab:elaiscat} ELAIS N1 catalog}
\tablehead {
\colhead{PiGSS ID} &
\colhead{R.\ A.\ } &
\colhead{Decl.\ } &
\colhead{$b_{maj}$} &
\colhead{$b_{min}$} &
\colhead{$\phi$} &
\multicolumn{2}{c}{$S_{deep}$ (mJy)\tablenotemark{a}} &
\multicolumn{3}{c}{$n_{det}$\tablenotemark{b}} &
\multicolumn{2}{c}{$\chisq_{\nu}$\tablenotemark{c}} &
\multicolumn{2}{c}{$\sigma_S / \bar{S}$\tablenotemark{d}} \\
&
(hr) &
(deg) &
(arcsec) &
(arcsec) &
(deg) & 
\colhead{3040} &
\colhead{3140} &
3040&
3140 &
both & 
3040 &
3140 &
3040 &
3140
}
\startdata
J160007p540202 & 16.00209 & 54.03392 & 100.0 & 100.0 &   0.0 & $     9.31 \pm  0.90$ & $     8.38 \pm  0.94$ &    1 &    0 &      0 & \ldots & \ldots & \ldots & \ldots \\
J160013p550244 & 16.00393 & 55.04559 & 100.0 & 100.0 &   0.0 & $     5.89 \pm  0.89$ & $     7.25 \pm  0.89$ &    0 &    0 &      0 & \ldots & \ldots & \ldots & \ldots \\
J160027p543634 & 16.00775 & 54.60948 & 100.0 & 100.0 &   0.0 & $     6.10 \pm  0.67$ & $     4.90 \pm  0.57$ &    0 &    1 &      0 & \ldots & \ldots & \ldots & \ldots \\
J160034p542253 & 16.00949 & 54.38160 & 181.1 & 137.7 &  69.6 & $    22.66 \pm  0.98$ & $    24.43 \pm  0.78$ &    2 &    1 &      0 & \ldots & \ldots & \ldots & \ldots \\
J160037p542734 & 16.01066 & 54.45914 & 100.0 & 100.0 &   0.0 & $     3.28 \pm  0.47$ & $     4.06 \pm  0.51$ &    0 &    0 &      0 & \ldots & \ldots & \ldots & \ldots \\
J160043p551327 & 16.01203 & 55.22435 & 330.1 & 101.8 &  79.3 & $    26.54 \pm  1.49$ & $    14.89 \pm  1.76$ &    0 &    1 &      0 & \ldots & \ldots & \ldots & \ldots \\
J160121p545319 & 16.02259 & 54.88886 & 100.0 & 100.0 &   0.0 & $     3.19 \pm  0.45$ & $     2.50 \pm  0.63$ &    0 &    0 &      0 & \ldots & \ldots & \ldots & \ldots \\
J160126p543931 & 16.02405 & 54.65880 & 100.0 & 100.0 &   0.0 & $     4.18 \pm  0.35$ & $     4.00 \pm  0.44$ &    0 &    1 &      0 & \ldots & \ldots & \ldots & \ldots \\
J160127p535640 & 16.02427 & 53.94520 & 115.0 & 106.4 &  69.1 & $    24.11 \pm  0.93$ & $    25.15 \pm  0.97$ &   60 &   29 &     10 &   0.20 &   3.15 &   0.11 &   0.36 \\
J160139p545516 & 16.02777 & 54.92278 & 116.9 & 109.8 & 177.8 & $    10.28 \pm  0.54$ & $    11.09 \pm  0.66$ &    2 &    0 &      0 & \ldots & \ldots & \ldots & \ldots \\
\enddata
\tablecomments{Only a portion of this table is shown here to demonstrate its form and content. A machine-readable version of the full table, which shows all sources detected at both frequencies in the deep image, is available.}
\tablenotetext{a}{Flux densities as measured from the deep images made from all good data at each of the two frequencies, 3040 and 3140\,MHz. Quoted uncertainties are the sum in quadrature of the uncertainties on the fit and on the background measurement as reported by SFIND.}
\tablenotetext{b}{Number of single-epoch images with completeness better than 50\,mJy in which this source was detected, at each frequency (3040 and 3140\,MHz). Number of well-behaved epochs (PIC gain within $\pm 0.1$ of the median value; Section~\ref{sec:picgain}) with completeness better than 50\,mJy in which the source was simultaneously detected in each of the single-frequency images.}
\tablenotetext{c}{Reduced chi-squared, for the hypothesis of no variability, computed using only well-behaved epochs where the source is detected at both frequencies in 5 or more epochs.}
\tablenotetext{d}{Standard deviation divided by the mean of the single-epoch flux densities, computed using only well-behaved epochs where the source is detected at both frequencies in 5 or more epochs.}
\end{deluxetable*}

\begin{deluxetable*}{lllllllllllllll}
\tablewidth{0pt}
\tabletypesize{\scriptsize}
\tablecaption{\label{tab:lockmancat} Lockman catalog}
\tablehead {
\colhead{PiGSS ID} &
\colhead{R.\ A.\ } &
\colhead{Decl.\ } &
\colhead{$b_{maj}$} &
\colhead{$b_{min}$} &
\colhead{$\phi$} &
\multicolumn{2}{c}{$S_{deep}$ (mJy)\tablenotemark{a}} &
\multicolumn{3}{c}{$n_{det}$\tablenotemark{b}} &
\multicolumn{2}{c}{$\chisq_{\nu}$\tablenotemark{c}} &
\multicolumn{2}{c}{$\sigma_S / \bar{S}$\tablenotemark{d}} \\
&
(hr) &
(deg) &
(arcsec) &
(arcsec) &
(deg) & 
\colhead{3040} &
\colhead{3140} &
3040&
3140 &
both & 
3040 &
3140 &
3040 &
3140
}
\startdata
J103814p583014 & 10.63751 & 58.50404 & 104.0 & 100.7 &  94.0 & $    25.55 \pm  1.20$ & $    27.53 \pm  1.00$ &   25 &   13 &      6 &   0.80 &   1.20 &   0.15 &   0.18 \\
J103834p580259 & 10.64299 & 58.04998 & 100.0 & 100.0 &   0.0 & $    31.36 \pm  1.24$ & $    31.76 \pm  1.10$ &   40 &   35 &     17 &   0.92 &   1.20 &   0.17 &   0.17 \\
J103856p575246 & 10.64891 & 57.87946 & 141.9 & 108.9 & 145.9 & $   147.13 \pm  2.78$ & $   139.52 \pm  2.33$ &  110 &  107 &     63 &   8.33 &   9.22 &   0.21 &   0.24 \\
J103900p590543 & 10.65016 & 59.09553 & 100.0 & 100.0 &   0.0 & $    26.45 \pm  2.13$ & $    22.12 \pm  2.04$ &    1 &    1 &      1 & \ldots & \ldots & \ldots & \ldots \\
J103911p581435 & 10.65339 & 58.24363 & 100.0 & 100.0 &   0.0 & $     9.36 \pm  0.69$ & $     9.50 \pm  0.77$ &    2 &    3 &      1 & \ldots & \ldots & \ldots & \ldots \\
J103913p591424 & 10.65398 & 59.24062 & 100.0 & 100.0 &   0.0 & $    15.97 \pm  1.72$ & $    16.06 \pm  2.28$ &    0 &    0 &      0 & \ldots & \ldots & \ldots & \ldots \\
J103922p582909 & 10.65619 & 58.48596 & 100.0 & 100.0 &   0.0 & $    32.63 \pm  0.81$ & $    31.86 \pm  0.72$ &   84 &   69 &     44 &   1.85 &   1.84 &   0.19 &   0.21 \\
J103942p580614 & 10.66180 & 58.10429 & 104.2 & 101.0 &  69.1 & $   106.05 \pm  1.45$ & $   108.84 \pm  1.47$ &  110 &  110 &     65 &   2.85 &   1.85 &   0.09 &   0.08 \\
J103953p583214 & 10.66488 & 58.53742 & 100.0 & 100.0 &   0.0 & $     2.31 \pm  0.52$ & $     2.07 \pm  0.54$ &    0 &    0 &      0 & \ldots & \ldots & \ldots & \ldots \\
J103958p582350 & 10.66636 & 58.39744 & 100.0 & 100.0 &   0.0 & $     5.18 \pm  0.53$ & $     4.03 \pm  0.61$ &    0 &    0 &      0 & \ldots & \ldots & \ldots & \ldots \\
\enddata
\tablecomments{Only a portion of this table is shown here to demonstrate its form and content. A machine-readable version of the full table, which shows all sources detected at both frequencies in the deep image, is available.}
\tablenotetext{a}{Flux densities as measured from the deep images made from all good data at each of the two frequencies, 3040 and 3140\,MHz. Quoted uncertainties are the sum in quadrature of the uncertainties on the fit and on the background measurement as reported by SFIND.}
\tablenotetext{b}{Number of single-epoch images with completeness better than 50\,mJy in which this source was detected, at each frequency (3040 and 3140\,MHz). Number of well-behaved epochs (PIC gain within $\pm 0.1$ of the median value; Section~\ref{sec:picgain}) with completeness better than 50\,mJy in which the source was simultaneously detected in each of the single-frequency images.}
\tablenotetext{c}{Reduced chi-squared, for the hypothesis of no variability, computed using only well-behaved epochs where the source is detected at both frequencies in 5 or more epochs.}
\tablenotetext{d}{Standard deviation divided by the mean of the single-epoch flux densities, computed using only well-behaved epochs where the source is detected at both frequencies in 5 or more epochs.}
\end{deluxetable*}

\begin{deluxetable*}{lllllllllllllll}
\tablewidth{0pt}
\tabletypesize{\scriptsize}
\tablecaption{\label{tab:comacat} Coma catalog}
\tablehead {
\colhead{PiGSS ID} &
\colhead{R.\ A.\ } &
\colhead{Decl.\ } &
\colhead{$b_{maj}$} &
\colhead{$b_{min}$} &
\colhead{$\phi$} &
\multicolumn{2}{c}{$S_{deep}$ (mJy)\tablenotemark{a}} &
\multicolumn{3}{c}{$n_{det}$\tablenotemark{b}} &
\multicolumn{2}{c}{$\chisq_{\nu}$\tablenotemark{c}} &
\multicolumn{2}{c}{$\sigma_S / \bar{S}$\tablenotemark{d}} \\
&
(hr) &
(deg) &
(arcsec) &
(arcsec) &
(deg) & 
\colhead{3040} &
\colhead{3140} &
3040&
3140 &
both & 
3040 &
3140 &
3040 &
3140
}
\startdata

J125137p280644 & 12.86017 & 28.11004 & 119.9 & 102.4 & 169.6 & $     6.47 \pm  1.84$ & $    14.74 \pm  2.07$ &    0 &    0 &      0 & \ldots & \ldots & \ldots & \ldots \\
J125150p275119 & 12.86404 & 27.85549 & 100.0 & 100.0 &   0.0 & $    61.31 \pm  1.85$ & $    59.30 \pm  2.05$ &   10 &    7 &      2 & \ldots & \ldots & \ldots & \ldots \\
J125218p273219 & 12.87162 & 27.53900 & 100.0 & 100.0 &   0.0 & $     9.59 \pm  1.36$ & $     9.63 \pm  1.63$ &    0 &    0 &      0 & \ldots & \ldots & \ldots & \ldots \\
J125224p273613 & 12.87360 & 27.60361 & 100.0 & 100.0 &   0.0 & $    12.27 \pm  1.89$ & $     6.55 \pm  1.96$ &    0 &    0 &      0 & \ldots & \ldots & \ldots & \ldots \\
J125252p282224 & 12.88133 & 28.37352 & 100.0 & 100.0 &   0.0 & $    10.13 \pm  1.39$ & $     9.05 \pm  1.57$ &    1 &    1 &      0 & \ldots & \ldots & \ldots & \ldots \\
J125254p273141 & 12.88200 & 27.53374 & 100.0 & 100.0 &   0.0 & $     5.67 \pm  1.14$ & $     5.78 \pm  1.16$ &    0 &    0 &      0 & \ldots & \ldots & \ldots & \ldots \\
J125306p274607 & 12.88522 & 27.76888 & 105.7 & 101.2 & 179.6 & $    97.19 \pm  1.60$ & $    95.19 \pm  1.26$ &   37 &   36 &     23 &   2.21 &   2.30 &   0.15 &   0.18 \\
J125314p273944 & 12.88734 & 27.66226 & 105.4 & 103.4 & 171.0 & $   164.66 \pm  2.43$ & $   150.22 \pm  1.42$ &   38 &   37 &     23 &  12.88 &   1.64 &   0.18 &   0.09 \\
J125316p280328 & 12.88842 & 28.06191 & 145.1 & 106.9 &  28.0 & $     5.63 \pm  0.85$ & $     7.98 \pm  0.78$ &    0 &    0 &      0 & \ldots & \ldots & \ldots & \ldots \\
J125324p281827 & 12.89001 & 28.30763 & 114.3 & 101.6 &  45.9 & $     9.60 \pm  0.95$ & $     7.42 \pm  1.18$ &    0 &    0 &      0 & \ldots & \ldots & \ldots & \ldots \\
\enddata
\tablecomments{Only a portion of this table is shown here to demonstrate its form and content. A machine-readable version of the full table, which shows all sources detected at both frequencies in the deep image, is available.}
\tablenotetext{a}{Flux densities as measured from the deep images made from all good data at each of the two frequencies, 3040 and 3140\,MHz. Quoted uncertainties are the sum in quadrature of the uncertainties on the fit and on the background measurement as reported by SFIND.}
\tablenotetext{b}{Number of single-epoch images with completeness better than 50\,mJy in which this source was detected, at each frequency (3040 and 3140\,MHz). Number of well-behaved epochs (PIC gain within $\pm 0.1$ of the median value; Section~\ref{sec:picgain}) with completeness better than 50\,mJy in which the source was simultaneously detected in each of the single-frequency images.}
\tablenotetext{c}{Reduced chi-squared, for the hypothesis of no variability, computed using only well-behaved epochs where the source is detected at both frequencies in 5 or more epochs.}
\tablenotetext{d}{Standard deviation divided by the mean of the single-epoch flux densities, computed using only well-behaved epochs where the source is detected at both frequencies in 5 or more epochs.}
\end{deluxetable*}

\begin{deluxetable*}{lllllllllllllll}
\tablewidth{0pt}
\tabletypesize{\scriptsize}
\tablecaption{\label{tab:ndwfscat} NDWFS catalog}
\tablehead {
\colhead{PiGSS ID} &
\colhead{R.\ A.\ } &
\colhead{Decl.\ } &
\colhead{$b_{maj}$} &
\colhead{$b_{min}$} &
\colhead{$\phi$} &
\multicolumn{2}{c}{$S_{deep}$ (mJy)\tablenotemark{a}} &
\multicolumn{3}{c}{$n_{det}$\tablenotemark{b}} &
\multicolumn{2}{c}{$\chisq_{\nu}$\tablenotemark{c}} &
\multicolumn{2}{c}{$\sigma_S / \bar{S}$\tablenotemark{d}} \\
&
(hr) &
(deg) &
(arcsec) &
(arcsec) &
(deg) & 
\colhead{3040} &
\colhead{3140} &
3040&
3140 &
both & 
3040 &
3140 &
3040 &
3140
}
\startdata
J142444p341819 & 14.41246 & 34.30544 & 105.5 & 100.6 & 123.3 & $    39.70 \pm  1.71$ & $    35.04 \pm  2.41$ &   15 &    4 &      3 & \ldots & \ldots & \ldots & \ldots \\
J142448p341000 & 14.41334 & 34.16667 & 100.0 & 100.0 &   0.0 & $     7.55 \pm  1.42$ & $     5.33 \pm  1.50$ &    0 &    0 &      0 & \ldots & \ldots & \ldots & \ldots \\
J142515p345256 & 14.42094 & 34.88238 & 112.0 & 102.8 &  76.1 & $    37.53 \pm  2.43$ & $    26.92 \pm  2.75$ &    2 &    0 &      0 & \ldots & \ldots & \ldots & \ldots \\
J142516p341558 & 14.42156 & 34.26741 & 136.0 & 100.5 &  26.5 & $     9.33 \pm  0.94$ & $    11.47 \pm  1.40$ &    1 &    0 &      0 & \ldots & \ldots & \ldots & \ldots \\
J142522p340938 & 14.42299 & 34.16064 & 100.0 & 100.0 &   0.0 & $     7.31 \pm  0.91$ & $     5.66 \pm  1.26$ &    0 &    1 &      0 & \ldots & \ldots & \ldots & \ldots \\
J142540p345835 & 14.42789 & 34.97645 & 124.0 & 111.8 &  66.3 & $   104.21 \pm  3.21$ & $    84.29 \pm  2.90$ &   48 &   11 &      9 &   1.29 &   1.00 &   0.17 &   0.18 \\
J142543p335535 & 14.42868 & 33.92650 & 100.0 & 100.0 &   0.0 & $   124.18 \pm  2.80$ & $   121.76 \pm  2.90$ &   67 &   67 &     59 &   2.40 &   2.25 &   0.10 &   0.13 \\
J142607p340429 & 14.43540 & 34.07408 & 100.0 & 100.0 &   0.0 & $    25.98 \pm  0.85$ & $    26.72 \pm  0.83$ &   37 &   25 &     20 &   1.56 &   1.49 &   0.19 &   0.18 \\
J142609p333941 & 14.43592 & 33.66152 & 100.0 & 100.0 &   0.0 & $    11.52 \pm  1.50$ & $    10.28 \pm  1.74$ &    1 &    1 &      0 & \ldots & \ldots & \ldots & \ldots \\
J142620p335126 & 14.43910 & 33.85727 & 100.0 & 100.0 &   0.0 & $     5.57 \pm  0.99$ & $     4.57 \pm  1.17$ &    0 &    0 &      0 & \ldots & \ldots & \ldots & \ldots \\
\enddata
\tablecomments{Only a portion of this table is shown here to demonstrate its form and content. A machine-readable version of the full table, which shows all sources detected at both frequencies in the deep image, is available.}
\tablenotetext{a}{Flux densities as measured from the deep images made from all good data at each of the two frequencies, 3040 and 3140\,MHz. Quoted uncertainties are the sum in quadrature of the uncertainties on the fit and on the background measurement as reported by SFIND.}
\tablenotetext{b}{Number of single-epoch images with completeness better than 50\,mJy in which this source was detected, at each frequency (3040 and 3140\,MHz). Number of well-behaved epochs (PIC gain within $\pm 0.1$ of the median value; Section~\ref{sec:picgain}) with completeness better than 50\,mJy in which the source was simultaneously detected in each of the single-frequency images.}
\tablenotetext{c}{Reduced chi-squared, for the hypothesis of no variability, computed using only well-behaved epochs where the source is detected at both frequencies in 5 or more epochs.}
\tablenotetext{d}{Standard deviation divided by the mean of the single-epoch flux densities, computed using only well-behaved epochs where the source is detected at both frequencies in 5 or more epochs.}
\end{deluxetable*}

\begin{figure*}[htp]
\centering
\includegraphics[width=0.5\linewidth,draft=false]{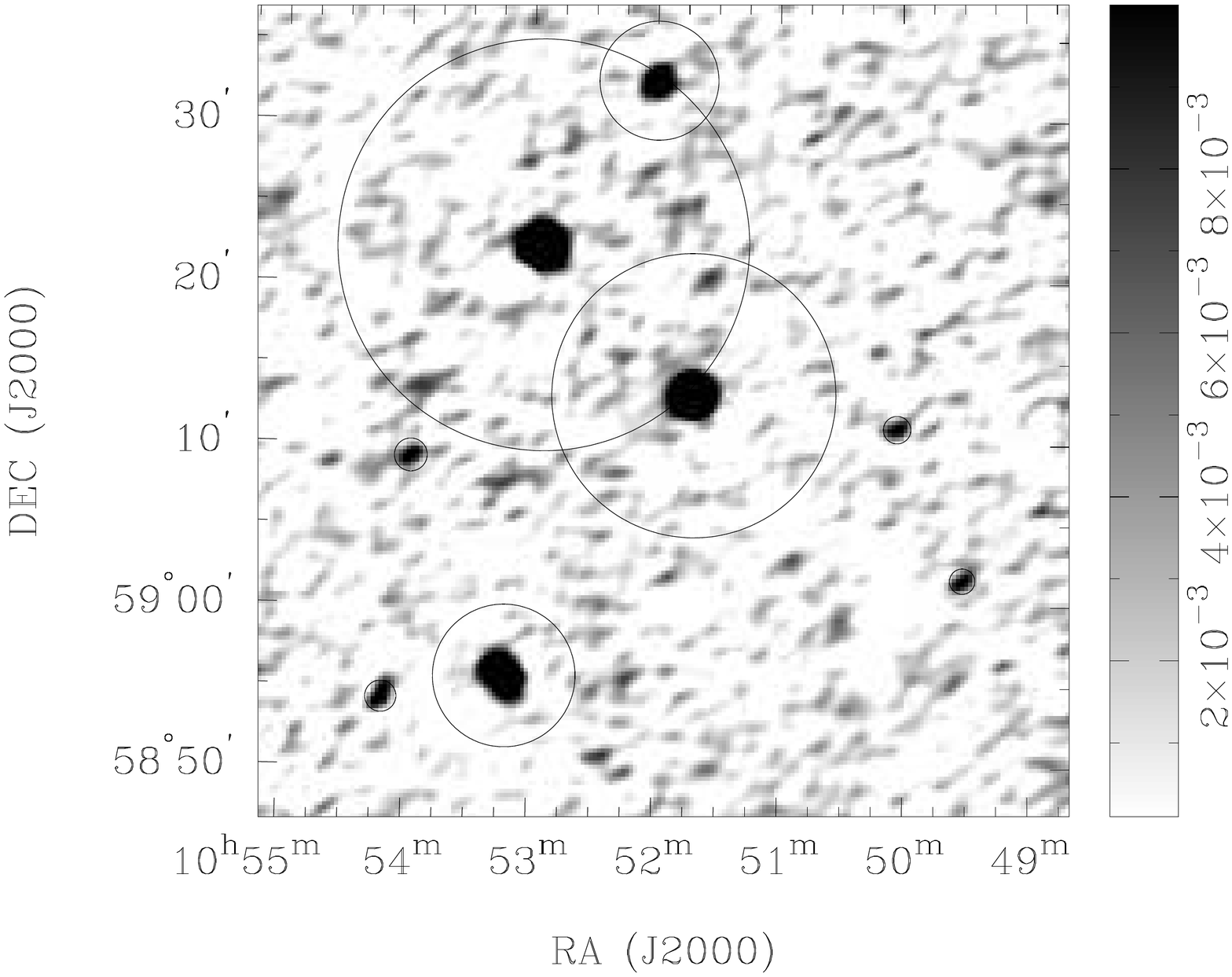}%
\includegraphics[width=0.5\linewidth,draft=false]{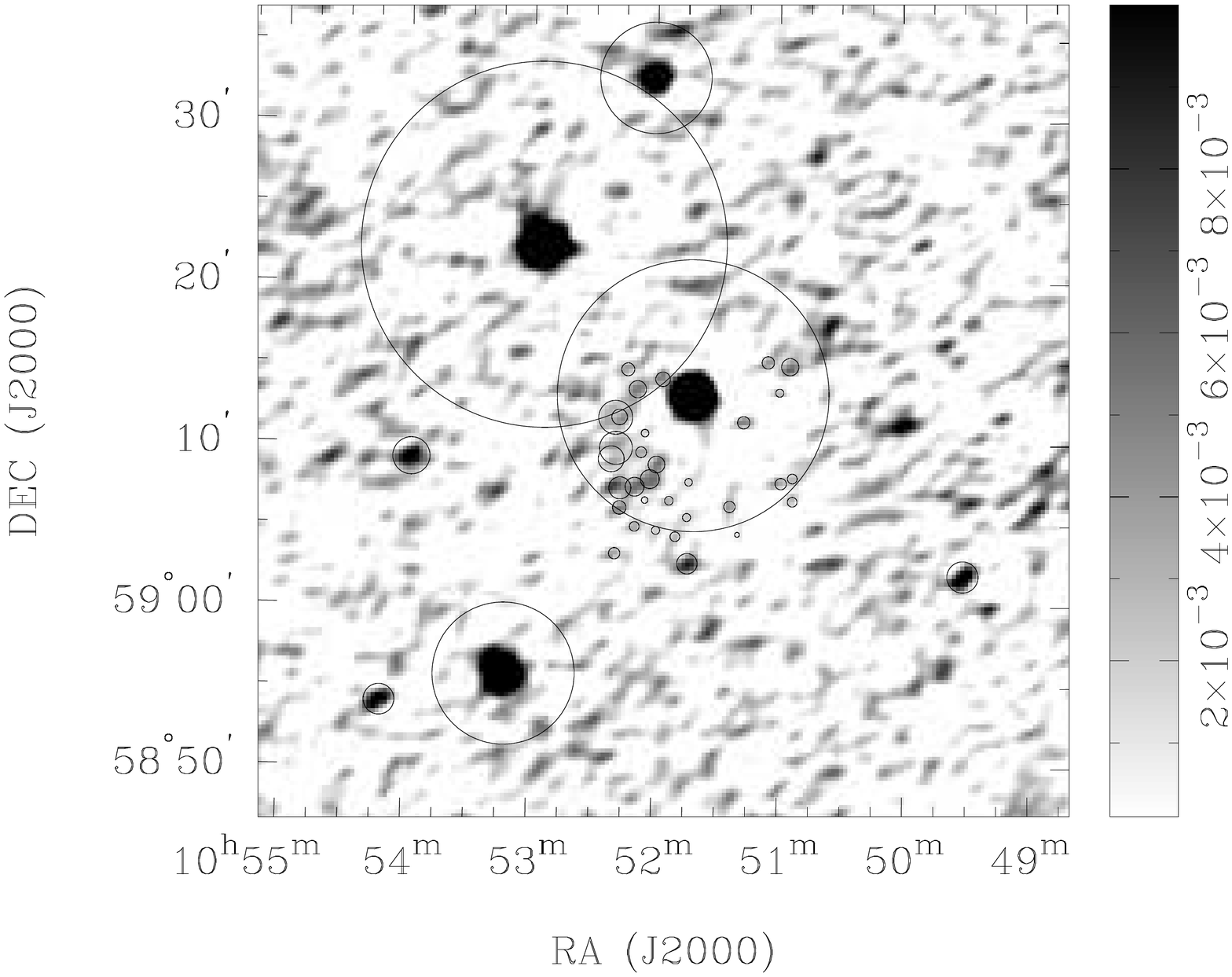}%
\caption{\label{fig:sfindfail}
Left: Part of the single-epoch 3040\,MHz image for PiGSS Lockman 2010 April 7, with source positions from SFIND marked as circles of radius scaled with measured flux density. The image is 90\%\ complete to 11\,mJy.
Right: Same region of the 3140\,MHz image from the same epoch, also 90\%\ complete to 11\,mJy. A cluster of faint spurious sources is visible where the SFIND algorithm failed, and misidentified noise peaks as real sources. By keeping only sources that are present at both 3040 and 3140\,MHz, we can exclude spurious sources in cases like this, and make our catalogs much cleaner. Examples like this one are rare in our data in the sense that they make up only a small fraction of detected sources, but if not identified and excluded, they dominate lists of transient candidates.
Requiring two-frequency detections also excludes some faint real sources, such as the one at \hmt{10}{50}, \dmt{59}{11} in the left panel (also visible but not detected above the chosen threshold in the right panel), but in essence this is equivalent to increasing the overall threshold for detection.
Greyscales are 0 to 10\,\mjypbm.
}
\end{figure*}

Examples of defects successfully removed from our data by the two-frequency requirement are spurious detections near bright sources, amplified noise close to the edges of images, and other failures of the source finder, including those due to underestimates of the background rms (\eg, Fig.~\ref{fig:sfindfail}). Although this method will not remove all spurious sources, examination of our data shows that it is successful enough to greatly reduce the number of false positive sources detected, which is extremely helpful in the search for transients.

\begin{figure*}[htp]
\centering
\includegraphics[height=0.45\linewidth,draft=false,angle=-90]{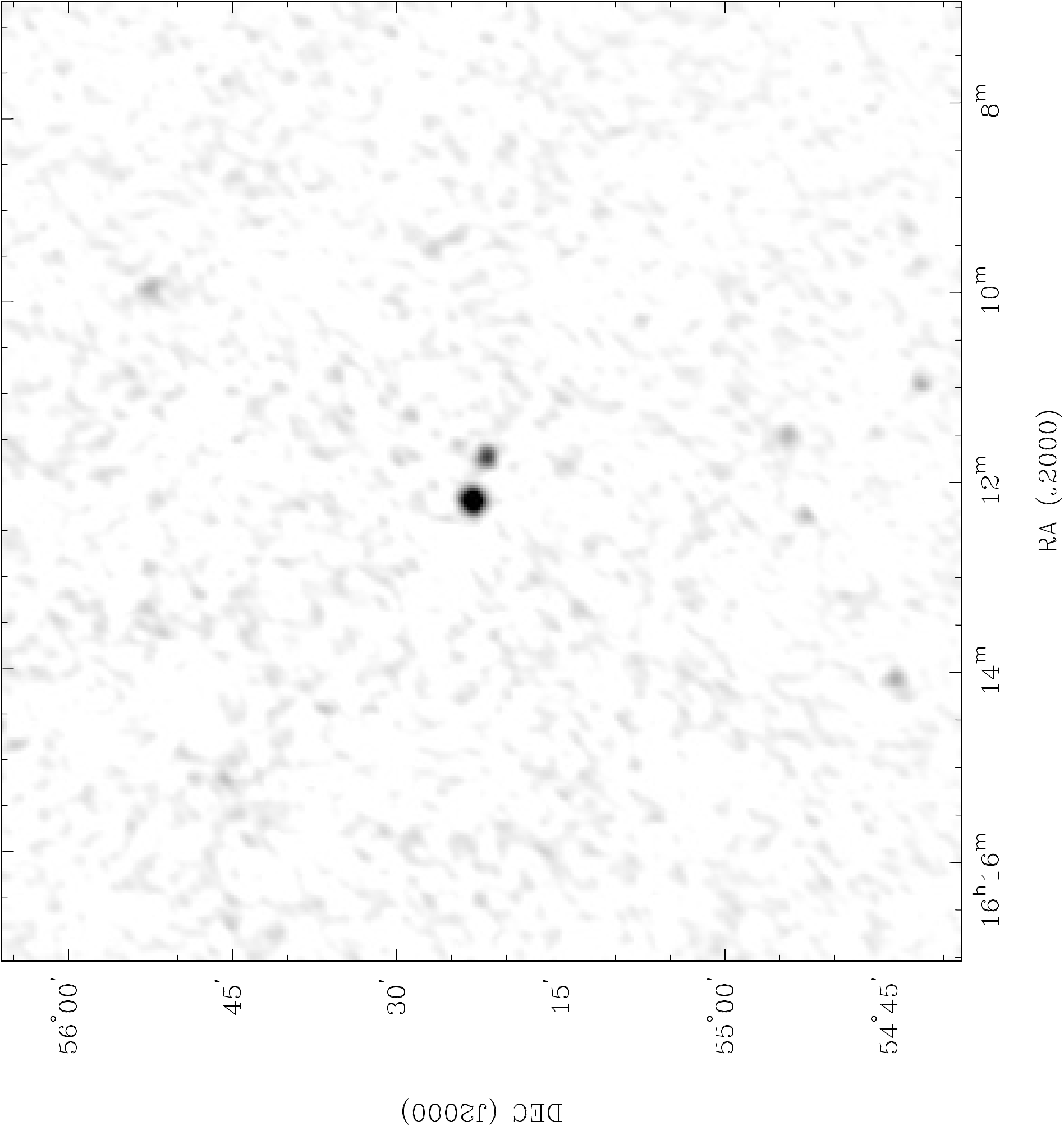}\hspace{0.08\linewidth}%
\includegraphics[height=0.45\linewidth,draft=false,angle=-90]{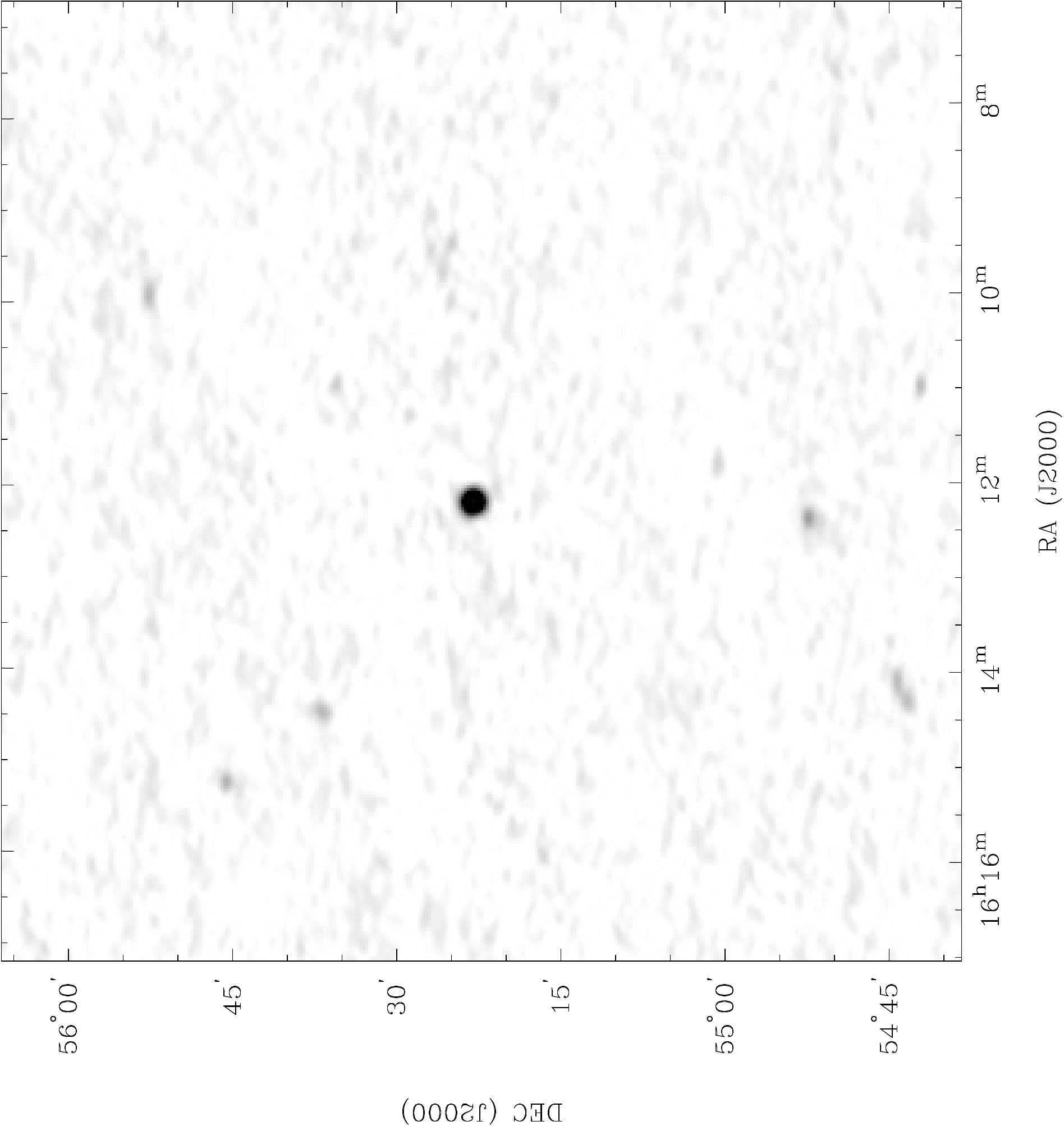}%
\caption{\label{fig:badpoint}
Subregion of the single-epoch 3040\,MHz image for PiGSS ELAIS-N1 2010 October 5 (left). Comparison with the same frequency on epoch 2010 October 7 (right) shows a bright transient candidate (which is not present in any other epoch of ELAIS-N1, but is also present in the 3140\,MHz image on 2010 October 5). More detailed examination shows that this is due to the inclusion of a small amount of data where the telescope was not pointed correctly. Greyscales are 0 to 50\,\mjypbm.
}
\end{figure*}

This method does not result in the removal of all spurious transient candidates, however. An example of a strong transient candidate seen in our data is shown in Fig.~\ref{fig:badpoint} --- the candidate is present in images at both 3040 and 3140\,MHz. Further examination, however, showed that this apparent transient is an image of a bright source at a nearby position, and that the data were erroneously included in this image due to a  telescope pointing error. Similar kinds of pernicious errors were also noted in VLA data by \citet{ofek:10}, \citet{thyagarajan:11}, and \citet{frail:12}, so even transient candidates that survive all other cuts must be considered with appropriate skepticism until they are examined in detail. 

\subsection{Completeness}\label{sec:completeness}

As in \paperii, we can assess the completeness and reliability of our data by plotting the source count histograms of the individual epochs and comparing to those from the master catalogs. The small size of the field compared to ATATS results in a larger fraction of sources detected in regions where the sensitivity falls off due to the primary beam. We therefore compute completeness  considering only sources from central regions of the mosaic with gain equal to one.

\begin{figure*}[htp]
\centering
\includegraphics[width=0.55\linewidth,draft=false]{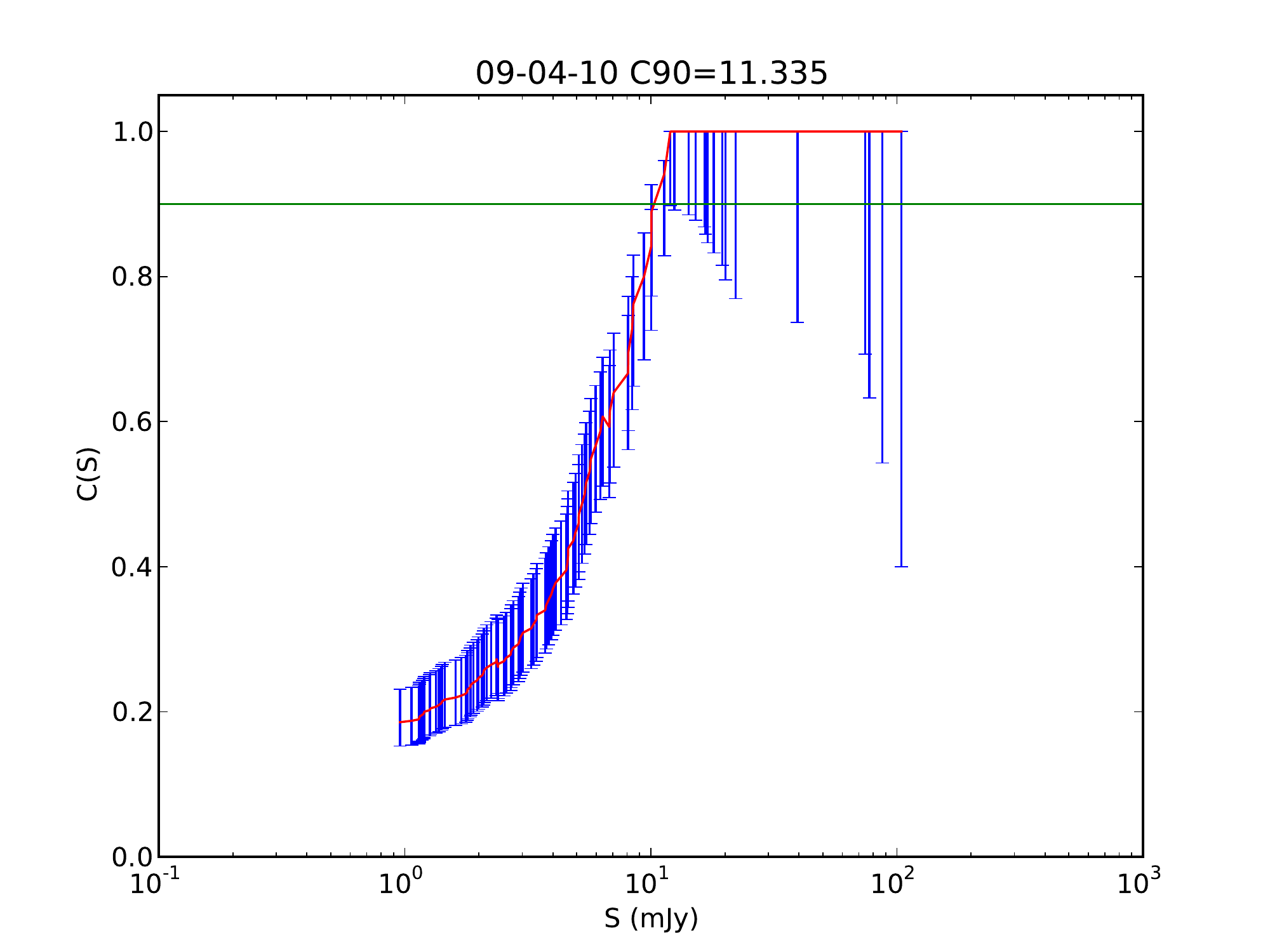}
\includegraphics[width=0.445\linewidth,draft=false]{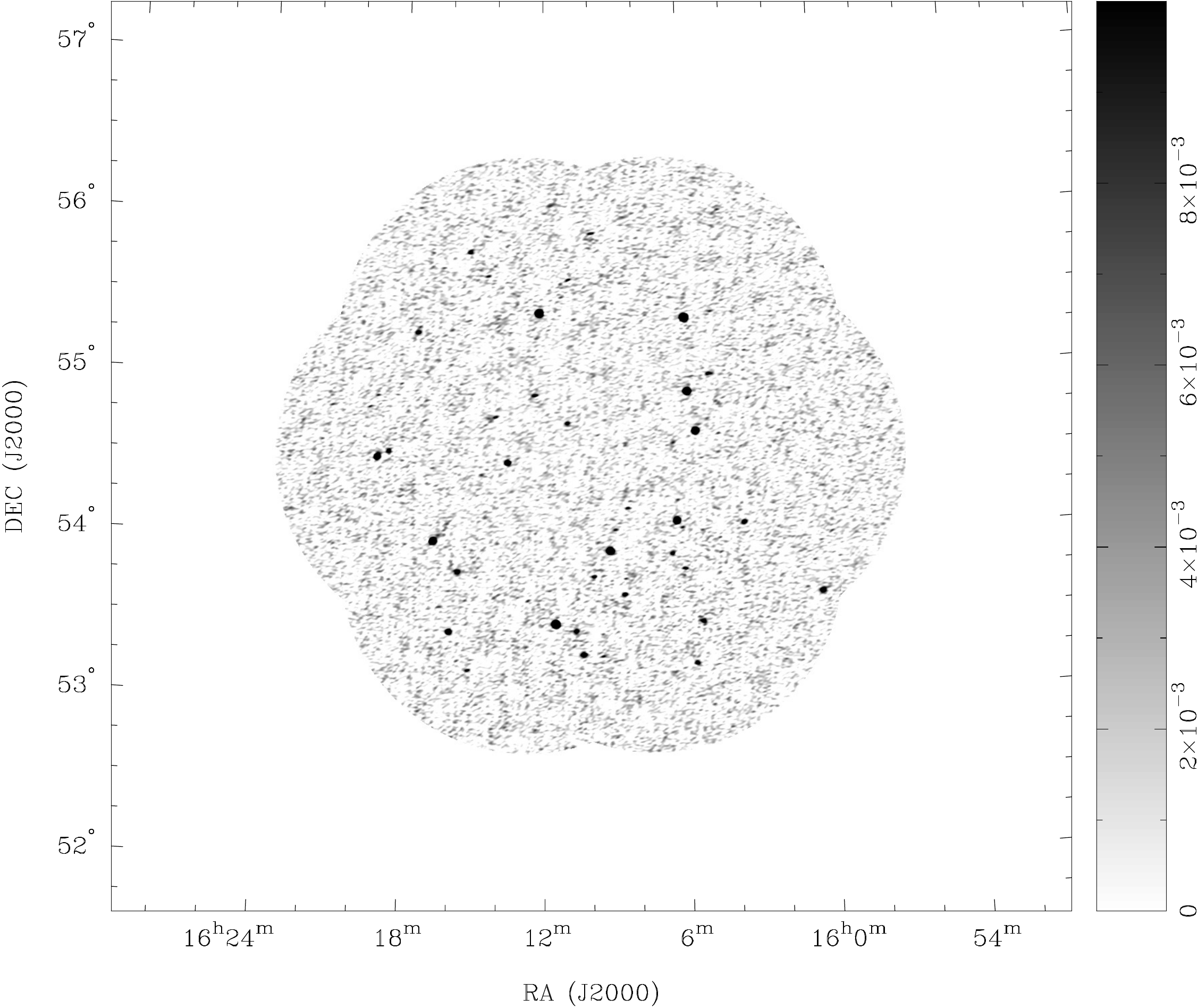}%
\caption{\label{fig:goodcomplete}
Left: Completeness, $C(S) = 1 - (N_{\rm u}(S) / N_{\rm a}(S))$, \ie, the fraction of sources in the master catalog brighter than flux density $S$ (in a region with gain equal to one) with a match in the single epoch catalog for ELAIS-N1 3140\,MHz epoch 2010 September 4.
The horizontal line shows 90\%\ completeness, and the error bars were calculated following \citet{cameron:11}. The image quality in this field is typical (see Fig.~\ref{fig:rmshist}), and 90\%\ completeness is reached at 12\,mJy. 
Right: Image of the field. The greyscale runs from 0 to 10\,\mjypbm.
}
\end{figure*}

\begin{figure*}[htp]
\centering
\includegraphics[width=0.55\linewidth,draft=false]{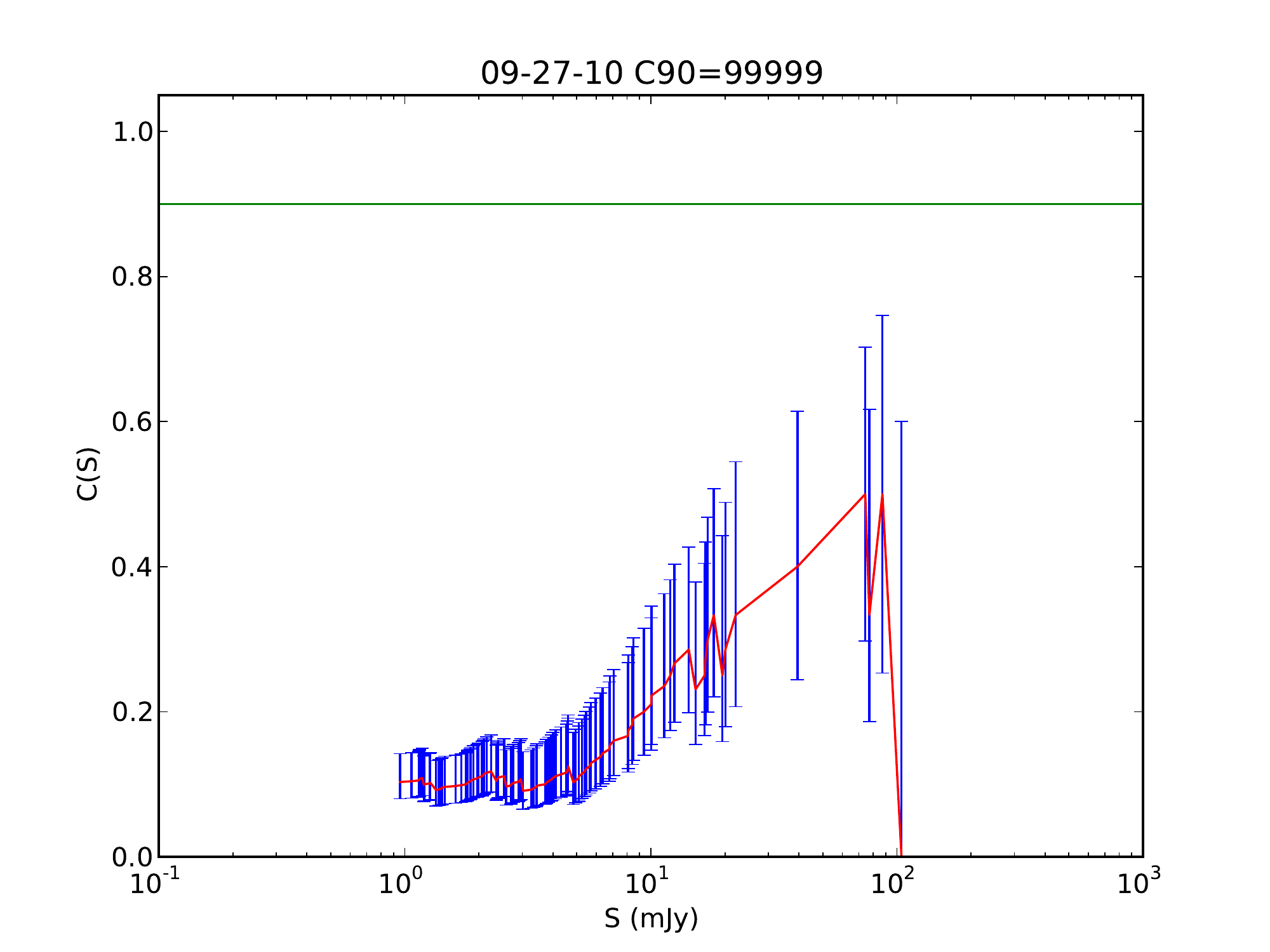}
\includegraphics[width=0.445\linewidth,draft=false]{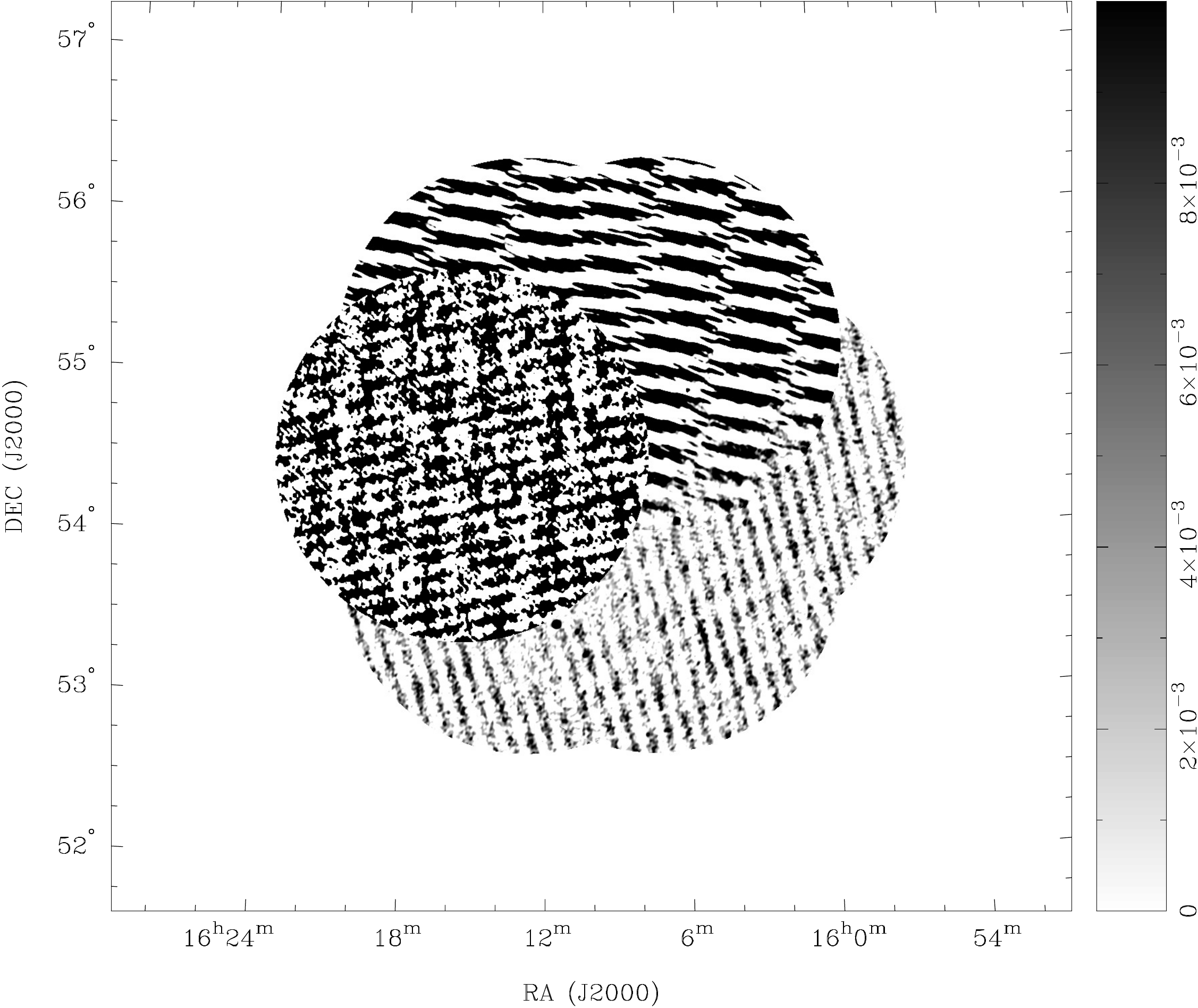}%
\caption{\label{fig:badcomplete}
Left: Completeness, $C(S) = 1 - (N_{\rm u}(S) / N_{\rm a}(S))$, \ie, the fraction of sources in the master catalog brighter than flux density $S$ (in a region with gain equal to one) with a match in the single epoch catalog for ELAIS-N1 3140\,MHz epoch 2010 September 27.
The horizontal line shows 90\%\ completeness, and the error bars were calculated following \citet{cameron:11}. The image quality in this field is unusually poor (see Fig.~\ref{fig:rmshist}), and 90\%\ completeness is never reached. As a result this field was rejected from our analysis.
Right: Image of the field, where the defects causing poor completeness are obvious. This epoch would be easily rejected upon manual inspection, but in large surveys where it is not possible to examine each image, completeness proves to be a useful method for rejecting bad images such as this one. It would be possible to implement an iterative approach to rejecting such data when constructing the deep image, but in practice the 3.5\,\mjypbm\ rms cut proves adequate for this purpose.
The greyscale runs from 0 to 10\,\mjypbm.
}
\end{figure*}

For each epoch, we define completeness as

\begin{equation}
C(S) = 1 - \frac{N_{\rm u}(S)}{N_{\rm a}(S)}
\end{equation}

where $N_{\rm u}(S)$ is the number of sources in the master catalog (from regions with primary beam gain equal to one) with flux density brighter than $S$, and no match in the single-epoch catalog within 150\arcsec; and $N_{\rm a}(S)$ is the total number of sources in the master catalog with flux density brighter than $S$. This statistic does not rely on an accurate measure of the flux densities in individual epochs; it simply determines the fraction of sources in the master catalog that are matched in a particular epoch, as a function of master epoch flux density. For a given epoch, completeness tends to increase as flux density increases (Fig.~\ref{fig:goodcomplete}). For most epochs, completeness crosses a threshold of 90\%\ at some point, which we determine to be the completeness value for that epoch. For a handful of particularly bad images, completeness never crosses the 90\%\ threshold (\eg, Fig.~\ref{fig:badcomplete}).

\section{Transients}\label{sec:transients}

\subsection{Comparing individual epochs to the deep image}

We searched for sources detected in one or more epochs (at both frequencies in at least one of these epochs) without a counterpart at either frequency in the deep image. As previously noted, the requirement of a detection in both frequencies dramatically lowers the false detection rate while preserving true transients. Nevertheless, any candidates must be examined carefully. 

\subsubsection{Lockman}

In the Lockman field, a single transient candidate was detected, with flux density $\sim 200$\,mJy, in nine (non-consecutive) epochs, but not in the catalog made from the deep image. Examination of postage stamps shows that the source is in fact clearly visible in all single epoch images, and the deep image, but was close to the edge of the mosaic, in a region where the mosaic gain correction is a factor $\sim 5$. Small variations in the centroid position (and hence the gain) resulted in the source not meeting our cutoff of a factor 5 gain correction in many epochs (despite being detected by the source finder in 89 of them), but just meeting it in the nine where it was detected. We conclude that there are no transients of one or more epochs in Lockman without a counterpart in the deep image.

We also searched for sources with counterparts in the deep image that brightened in any epoch to 10 times or more the measured flux density in the deep image. As we argue in Section~\ref{sec:fac10}, such sources ought to be considered along with transients when calculating statistics. No such sources were found in the Lockman field.

\subsubsection{ELAIS-N1}

In the ELAIS-N1 field, 16 transient candidates were detected. Six were in a single epoch that showed imaging defects that were coincident in the 3040 and 3140\,MHz data, and so are not astronomical.
Five were in another epoch where the source finder failed near a bright source at both frequencies, and spurious sources coincided. 
Two of the candidates were in regions where the primary beam gain correction was a factor $\sim 5$. The remaining three candidates occurred due to inclusion of data from mispointed observations (\eg, Fig.~\ref{fig:badpoint}). We conclude that there are no transients of one or more epochs in ELAIS-N1 without a counterpart in the deep image. 

Two sources in ELAIS-N1 appeared to brighten to 10 times or more their flux density in the deep image.
In one case a 30\,mJy source in the 3040\,MHz deep image was matched to a nearby faint (4\,mJy) source in the 3140\,MHz deep image. The fainter source was not detected in the individual epoch 3140\,MHz images, but the brighter source was (sometimes at $> 40$\,mJy, since it was close to the edge of the mosaic where gain variations also play a part). This lead to the source being flagged as a candidate highly variable object, although in reality it is not.

The second candidate was in fact two sources which were resolved in some epochs but not in others, leading to apparent large variability. We conclude that there are no sources in ELAIS-N1 which truly vary by a factor of 10 or more.

\subsubsection{Coma}

In the Coma field, the transient search identified 19 candidates without a match in the deep image. Of these, 16 were present in the same single epoch. Many of the single-epoch images exhibited defects to varying degrees near the bright radio source Coma\,A, but this epoch was unusually bad, and 16 of the defects coincided at both frequencies and so were identified as transient candidates. Another two of the 19 candidates also appear near to Coma\,A in other epochs. On visual inspection, they appear to be sidelobes (they appear in clusters of other apparent sidelobes, although these candidates are the only two that coincide in position at both frequencies). The remaining candidate is close to the edge of the mosaic where the primary beam gain correction is a factor $\sim 5$. Once again, none of these candidates survives further examination.

Three sources in Coma appeared to brighten by more than a factor 10 compared to their deep field flux densities. Two of these candidates were close pairs that were resolved at some epochs but not in others; summing up the fluxes of the components, very little variability is in fact seen. The third candidate (where a 10\,mJy source appeared to brighten to $> 100$\,mJy) occurred in the same bad epoch (with 16 transient candidates) mentioned above, and is clearly an imaging defect related to Coma\,A. Once again, there are no highly variable sources in our Coma data.

\subsubsection{NDWFS}

In the reprocessed NDWFS field (with corrected astrometry and a slightly different reduction pipeline including the choice of CLEAN iterations), surveying a smaller number of epochs than in \paperii\ (due to the completeness threshold we apply here) and the slightly different method of transient identification we use here (including a larger match radius), we find a single 11\,mJy transient candidate. Further investigation shows that this source was detected at 3\,mJy in the 3040\,MHz deep image, and undetected, although just apparent on visual inspection, in the 3140\,MHz deep image, and so it is not truly a transient, but may have varied by a factor $\sim 4$.

One source in NDWFS appeared to brighten to 10 times or more its flux density in the deep image. In this case a bright (300\,mJy) source in the 3040\,MHz deep image was matched to a nearby faint (11\,mJy) source in the 3140\,MHz deep image. The fainter source was not detected in the individual epoch 3140\,MHz images, but the 300\,mJy source was, making it appear to be highly variable. Our chosen radius for source association is comparable to the synthesized beam size, so such mismatches are not common in our data. However, cases like this could be better discriminated against by fitting Gaussians in every single-epoch image with positions constrained to match the sources detected in the deep image. We intend to implement such a feature in our pipeline for future analysis.

Once again, we find no highly variable sources in this field when comparing single epochs to the deep image. This result is in agreement with the lack of transients seen when using a $\gtrsim 6\sigma$ threshold on the NDWFS analysis presented in \paperii.

\subsection{Comparing monthly images to the deep image}

We searched for sources detected in one or more of the monthly images (at both frequencies in at least one of these epochs) without a counterpart at both frequencies in the deep image.

In ELAIS-N1, seven candidates were found. Six were positionally coincident with the single epoch transients discussed above, which as previously noted, are not true astronomical transients. The seventh was a marginal detection at both frequencies in 2010 December, and at one frequency in 2010 October. Inspection of postage stamps shows that it was, in fact, also faintly visible in the other monthly images and the deep image. There are therefore no true transients in the ELAIS-N1 monthly data.

\begin{figure*}[htp]
\centering
\includegraphics[width=\linewidth,draft=false]{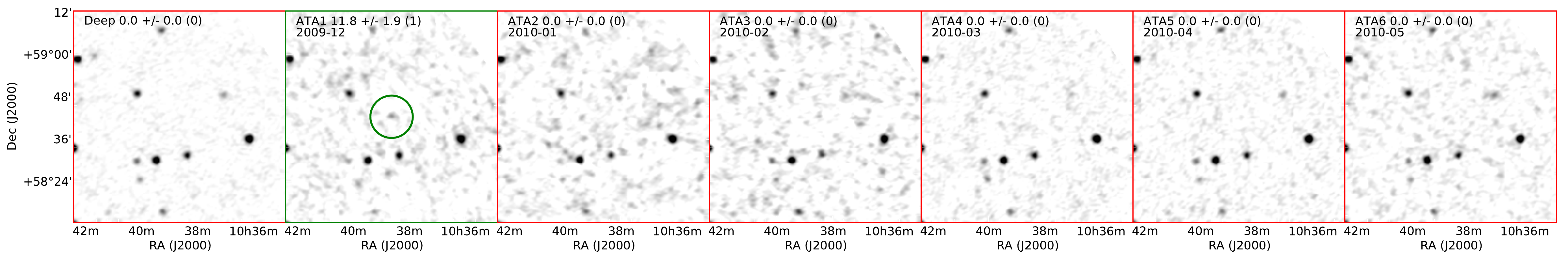}
\includegraphics[width=\linewidth,draft=false]{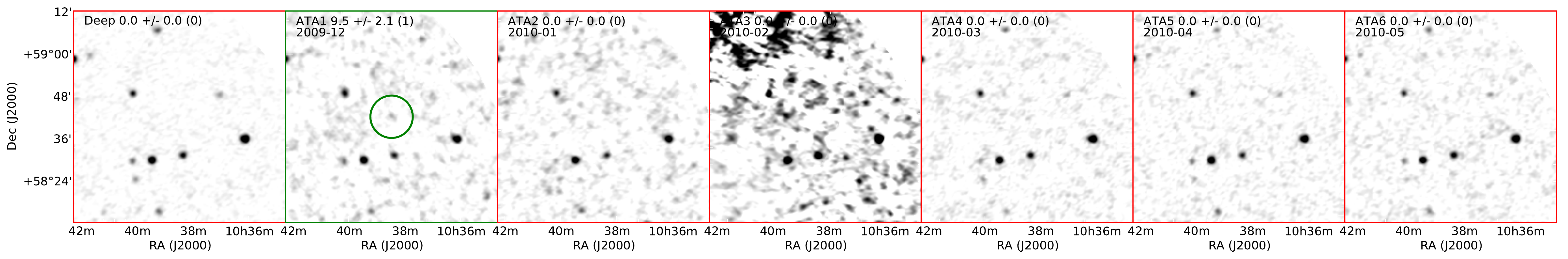}%
\caption{\label{fig:locktrans}
Postage stamp images of Lockman monthly transient candidate PiGSS\,J103818+584118. The deep image is shown in the first panel, and then images from 2009 December, 2010 January, February, March, April, and May. Data are at 3040\,MHz (top row) and 3140\,MHz (bottom row). The greyscale runs from 0 to 10\,\mjypbm. The transient candidate is visible in the center of the 2009 December postage stamp at both frequencies. The upper limit from the deep image ($< 1.6$\,\mjypbm) is not faint enough to determine whether the source brightened by a factor $> 10$, to meet our definition (Section~\ref{sec:factor10}) of a transient.
}
\end{figure*}

In Lockman, two candidates were found. PiGSS\,J103818+584118 is visible only in the 2009 December image, with $S_{3040} = 11.8 \pm 2.1$\,mJy and $S_{3140} = 9.5 \pm 2.1$\,mJy. The probability of such a source being spurious, assuming Gaussian statistics, is $2 \times 10^{-6}$, although as previously noted, image defects and other issues mean that spurious sources do in fact arise more often than pure Gaussian statistics would imply, and in cases where the uncertainties are underestimated, the significance of a detection may be overstated. However, this candidate is detected with a joint significance of $\gtrsim 7\sigma$, which suggests that it is probably astronomical. The source is not extended relative to the restoring beam. The mosaic primary beam gain correction at the source position is a factor 2.8. The source is not obviously detected on visual inspection of postage stamps of individual epoch images from 2009 December, which is to be expected since its measured flux density (uncorrected for the primary beam) is $\sim 4$\,mJy, below the typical single-epoch completeness limit.  Postage stamp images of the monthly data are shown in Fig.~\ref{fig:locktrans}. The closest NVSS source is over 6\arcmin\ away. The source apparently brightened by a factor $\geq 6$ relative to the $2\sigma$ upper limit from the deep image ($1.6$\,\mjypbm), so is not a confirmed transient as we define them here, in the sense of a source that can be shown to have brightened by a factor of at least 10. The upper limit from FIRST is $1.02$\,\mjypbm, however, suggestive of a brightening of a factor $\gtrsim 10$ if the source is assumed to have a flat spectrum. The candidate is 40\arcsec\ away from a Giant Meterwave Radio Telescope (GMRT) source seen at 610\,MHz with a flux density of $1.189\pm 0.137$\,mJy, GMRTLH J103820.1+584153 \citep{gmrtlockman}, but the areal density of GMRT sources is quite high, and we calculate a 30\%\ probability that this association is due to chance. The closest Sloan Digital Sky Survey (SDSS) galaxy is also $\sim 40$\arcsec\ away from our transient candidate.

\begin{figure*}[htp]
\centering
\includegraphics[width=\linewidth,draft=false]{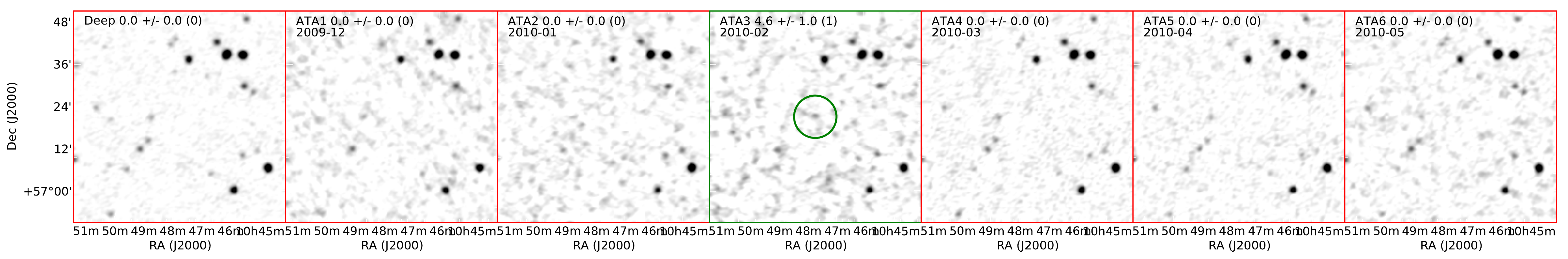}
\includegraphics[width=\linewidth,draft=false]{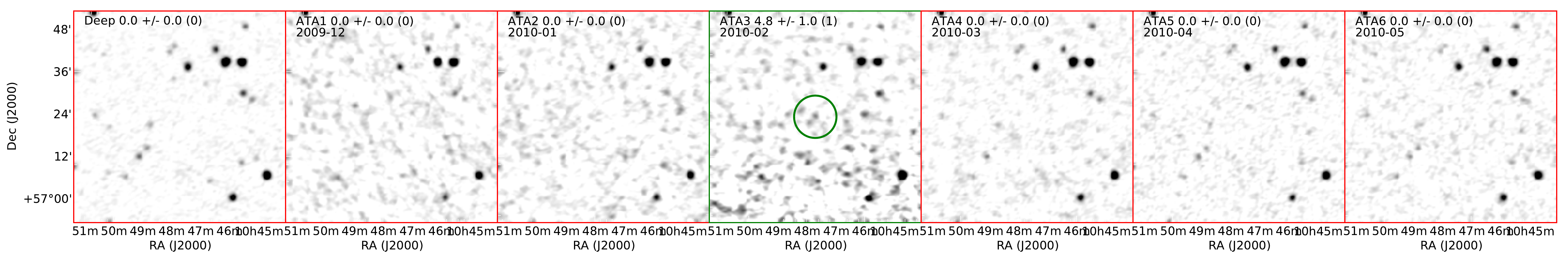}%
\caption{\label{fig:locktrans2}
Postage stamp images of Lockman monthly transient candidate PiGSS\,J104741+572316. The deep image is shown in the first panel, and then images from 2009 December, 2010 January, February, March, April, and May. Data are at 3040\,MHz (top row) and 3140\,MHz (bottom row). The transient candidate is visible in the center of the 2010 February postage stamp at both frequencies. It appears similar to other background fluctuations in the image, and so the significance of the detection is likely overestimated, and it is probably not a real source.}
\end{figure*}

The second candidate in the Lockman monthly data, PiGSS\,J104741+572316, appeared in 2010 February but is quite faint ($S_{3040} = 4.6 \pm 1.1$\,mJy; $S_{3140} = 4.8 \pm 1.1$\,mJy). Visual inspection of the postage stamps (Fig.~\ref{fig:locktrans2}) suggests that it is probably not real. It appears similar to other background fluctuations in the image, particularly at 3140\,MHz where image quality is noticeably worse than for other months.

In Coma, 13 candidates were found. One was positionally coincident with a single-epoch transient discussed above, and so is spurious. Another 9 were found in a single image (2011 January) that showed a rather large number of presumably spurious sources of similar brightness (typically a few mJy). Several had flux ratios in the two bands that would imply unrealistic spectral indices. All, except for the candidate that was coincident with the single-epoch candidate (which we also reject) do not appear to be real candidates after further inspection. Once again the presence of the bright source Coma\,A in the field makes calibration difficult and leads to large numbers of faint sidelobes or other imaging defects in the images, some of which are picked up by our transient pipeline.

In NDWFS, seven candidates were found. All bar one were from regions of the image with primary beam gain $> 2$. On visual inspection, all appeared similar to other background fluctuations in the images.

No sources in the Lockman, ELAIS-N1 or Coma fields were observed to vary by more than a factor of 10 in the monthly images compared to the deep images. One source in the NDWFS field was detected at 282\,mJy in the deep 3040\,MHz image, and matched to a nearby 11\,mJy source in the 3140\,MHz image. The source was detected at $\sim 280$\,mJy at both frequencies in all NDWFS monthly images, but the match to the 11\,mJy source resulted in it being flagged as a candidate highly variable source, although once again it is not truly highly variable.

In summary, we find one transient candidate (although not a confirmed transient that is shown to vary by more than a factor 10) when comparing the monthly images to the deep images.

\subsection{Comparing the deep catalogs to NVSS}

We also compared the catalogs generated from the deep images (\eg, Tables \ref{tab:elaiscat}, \ref{tab:lockmancat}, \ref{tab:comacat}, and \ref{tab:ndwfscat}) to the NVSS catalog. We used a match radius of 45\arcsec, chosen to produce an expectation of less than one false match between the catalogs.

With the corrected astrometry, and our more stringent detection requirements for the final catalog, 99\%\ of PiGSS NDWFS sources have a match in NVSS.

The flux limit of the NVSS is $\sim 2.5$\,mJy, so flat spectrum PiGSS sources fainter than 2.5\,mJy, or PiGSS sources fainter than $\sim 1.4$\,mJy with typical spectral indices, $\alpha \approx -0.7$ ($S_\nu \propto \nu^\alpha$), will not have NVSS counterparts. We find three PiGSS sources $> 2.5$\,mJy without NVSS counterparts in the Lockman field, five in ELAIS-N1, five in Coma, and two in NDWFS. Only a total of five of these sources are brighter than 5\,mJy in PiGSS; the brightest candidate is 11\,mJy. We examined postage stamps for all of the candidates, and all except two have faintly visible counterparts in the NVSS images, which were too faint to be detected in the NVSS catalogs. Of those two, one is clearly a sidelobe of Coma\,A.

\begin{figure*}[htp]
\centering
\includegraphics[width=0.34\linewidth,draft=false]{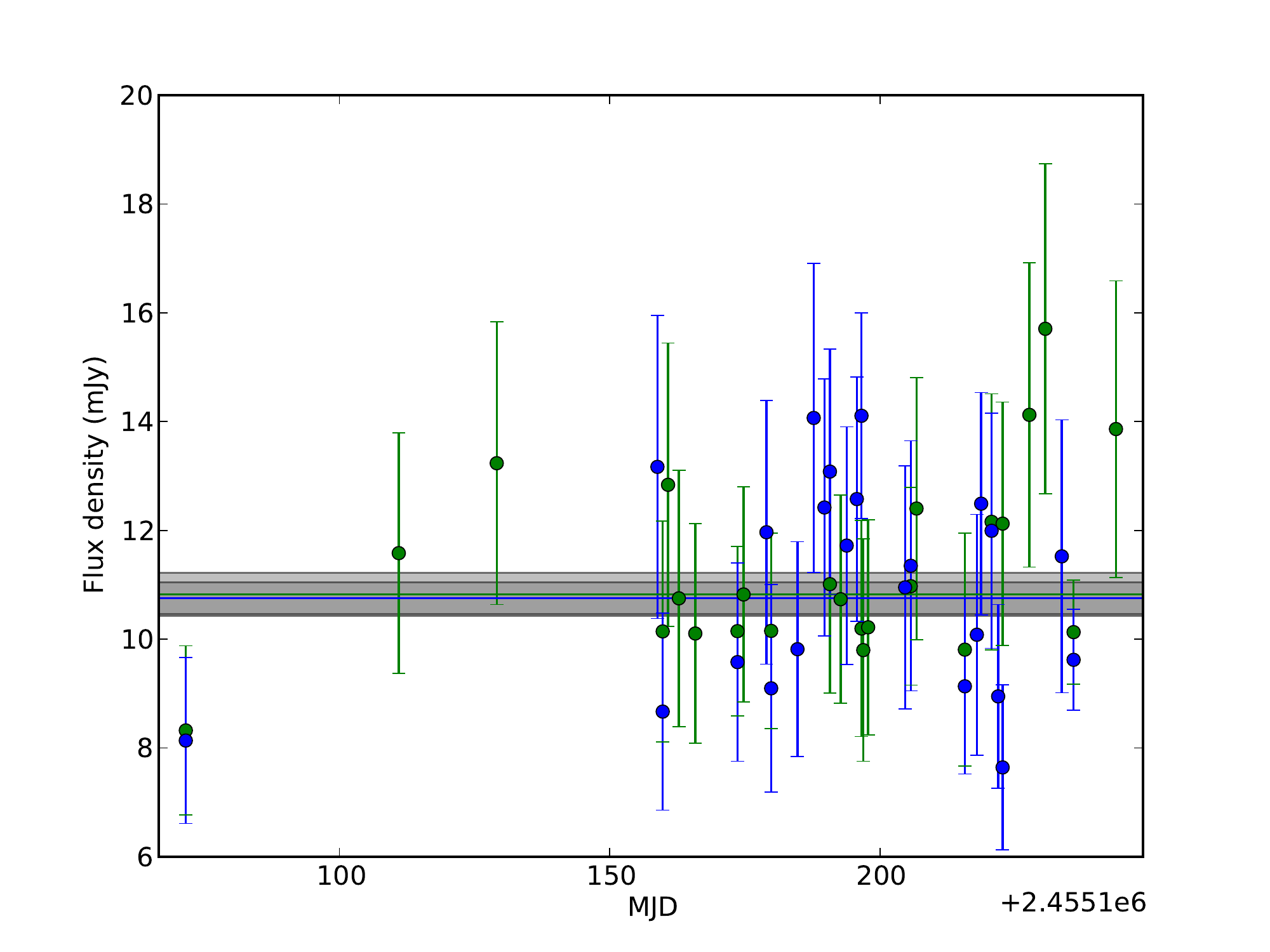}%
\includegraphics[width=0.32\linewidth,draft=false]{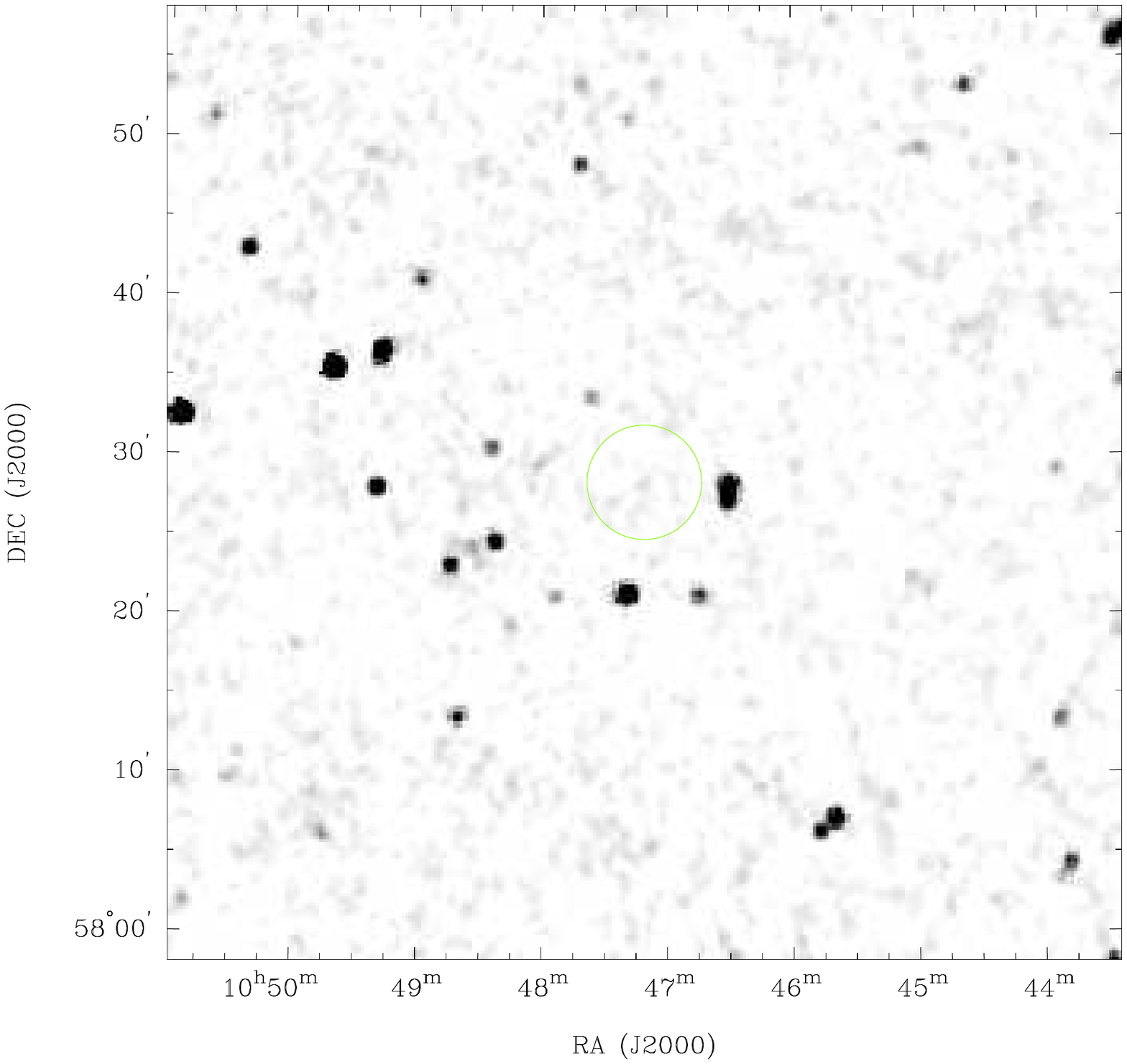}%
\includegraphics[width=0.32\linewidth,draft=false]{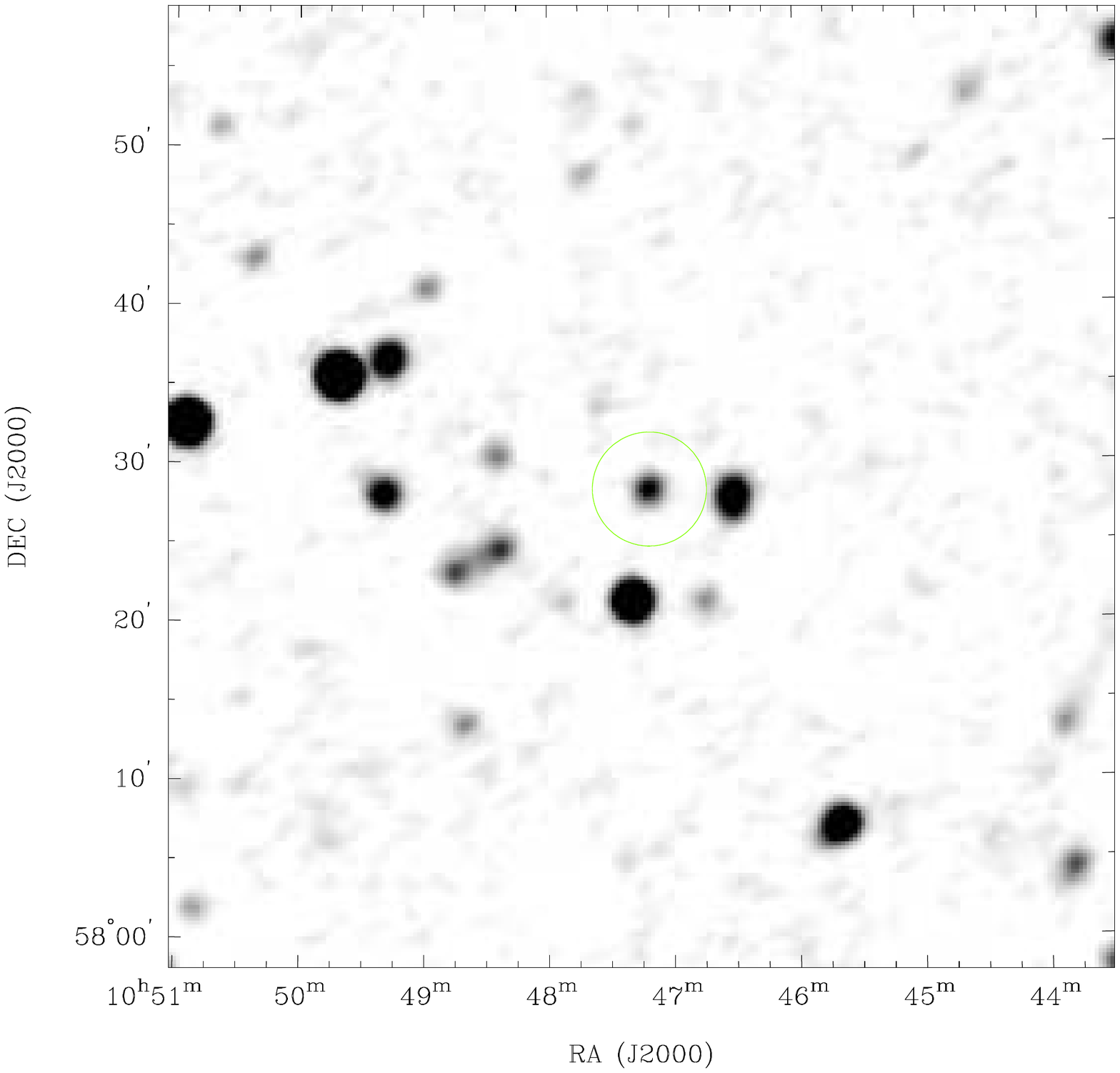}
\caption{\label{fig:j104711}
Left: Lightcurve for PiGSS\,J104711+582817, which appears to be a variable $z = 0.55$ AGN. 3040\,MHz flux densities are shown in green, and 3140\,MHz in blue. The deep field flux densities are shown as horizontal lines, with grey boxes representing the measured uncertainties on the deep field flux densities.
Center: NVSS postage stamp. The green circle shows the position of the ATA source with no match in the NVSS catalog. The greyscale runs from 0 to 10\,\mjypbm.
Right: ATA 3040\,MHz deep field postage stamp. The green circle shows the position of the ATA source with no match in the NVSS catalog. The greyscale runs from 0 to 10\,\mjypbm.
}
\end{figure*}

The remaining source, PiGSS\,J104711+582817, is seen with a flux density of $10.8 \pm 0.5$\,mJy in both the 3040 and 3140\,MHz deep images of the Lockman field. It is also detected in over 20 single epoch images with consistent flux (Fig.~\ref{fig:j104711}), and visible on inspection in almost all epochs. The daily flux density measurements show no clear trend with time. There is no counterpart in the NVSS catalog (which is complete to 2.5\,mJy), and nothing on visual inspection in the NVSS images. We measured a $2\sigma$ upper limit of 2.5\,\mjypbm\ at the ATA position from the NVSS images, consistent with the lack of detection in the NVSS catalog.
A search of the FIRST catalog shows a counterpart at $3.4 \pm 0.15$\,mJy; although the source is present in FIRST and absent from NVSS, it was not included in the analysis of \citet{levinson:02} because it is fainter than the threshold they use to select transient candidates. The 325\,MHz VLA observations of \citet{owen:09} detected a counterpart at $2.88 \pm 0.16$\,mJy, and the 610\,MHz GMRT observations of \citet{gmrtlockman} detected a $2.28 \pm 0.14$\,mJy counterpart. The radio sources are associated with a $z=0.55$ QSO, SDSS\,J104711.15+582820.7 \citep{richards:09}. The FIRST and NVSS flux densities suggest that the source exhibits variability. This would also explain the bright detection in PiGSS, although this may also be partially due to an inverted spectrum between 1.4 and 3.1\,GHz; since we do not have simultaneous radio detections at multiple widely-spaced frequencies we are unable to determine an instantaneous spectral index for the source.

The degeneracy between variability and spectral index also makes it difficult for us to clearly detect sources that vary by more than a factor 10 between NVSS and PiGSS. If we assume a flat spectrum, however, and simply compare the flux densities from NVSS and PiGSS, we find no sources that varied by a factor 10 or more.

Reversing the sense of the match, and considering only NVSS sources in the region where the PiGSS primary beam gain is unity, we find a total of 18 NVSS sources brighter than 10\,mJy without a cataloged PiGSS counterpart in the four deep fields. Of the 18, 10 are in the Coma field. On inspection, all in fact do have counterparts in the PiGSS images, and the lack of matches is, in all cases, due to different deconvolution of source structures at the two frequencies. Many of the NVSS sources fainter than 10\,mJy also appear to have counterparts in the PiGSS images, although some are truly undetected, due to the fainter flux limit of NVSS compared to PiGSS for sources with normal spectral indices.

\subsection{Rate limits}

Following \citet{bower:07}, the surface density (or ``two-epoch rate'') for a survey with $N_e$ epochs that cover a total area $A$ to sensitivity $S$ is 

\begin{equation}
\Sigma ({>}S)=\frac{N_t}{(N_e-1) A({>}S)}
\label{eqn:snapshot}
\end{equation}

where $N_t$ is the number of transients detected. Where no transient is 
detected, the 95\%\ confidence upper limit is $N_t = 3$ \citep{gehrels:86}. Here, $A({>}S)$ refers to 
the solid angle over which a source of flux density $S$ or greater can 
be detected.  Since faint sources are easier to detect close to the mosaic center than in regions where the primary beam gain falls off, we must compute A as a function of flux density.

We use the 90\%\ completeness values determined in the central regions of the mosaic where gain equals one to determine the detection limit for transients in a particular epoch. We then assume that the detection limit gets higher in inverse proportion to the gain falloff towards the mosaic edges, as the area above a certain primary beam gain increases. To determine transient limits for the survey as a whole, we sum the effective survey areas across epochs, as a function of sensitivity.

\begin{figure*}[htp]
\centering
\includegraphics[width=0.8\linewidth,draft=false]{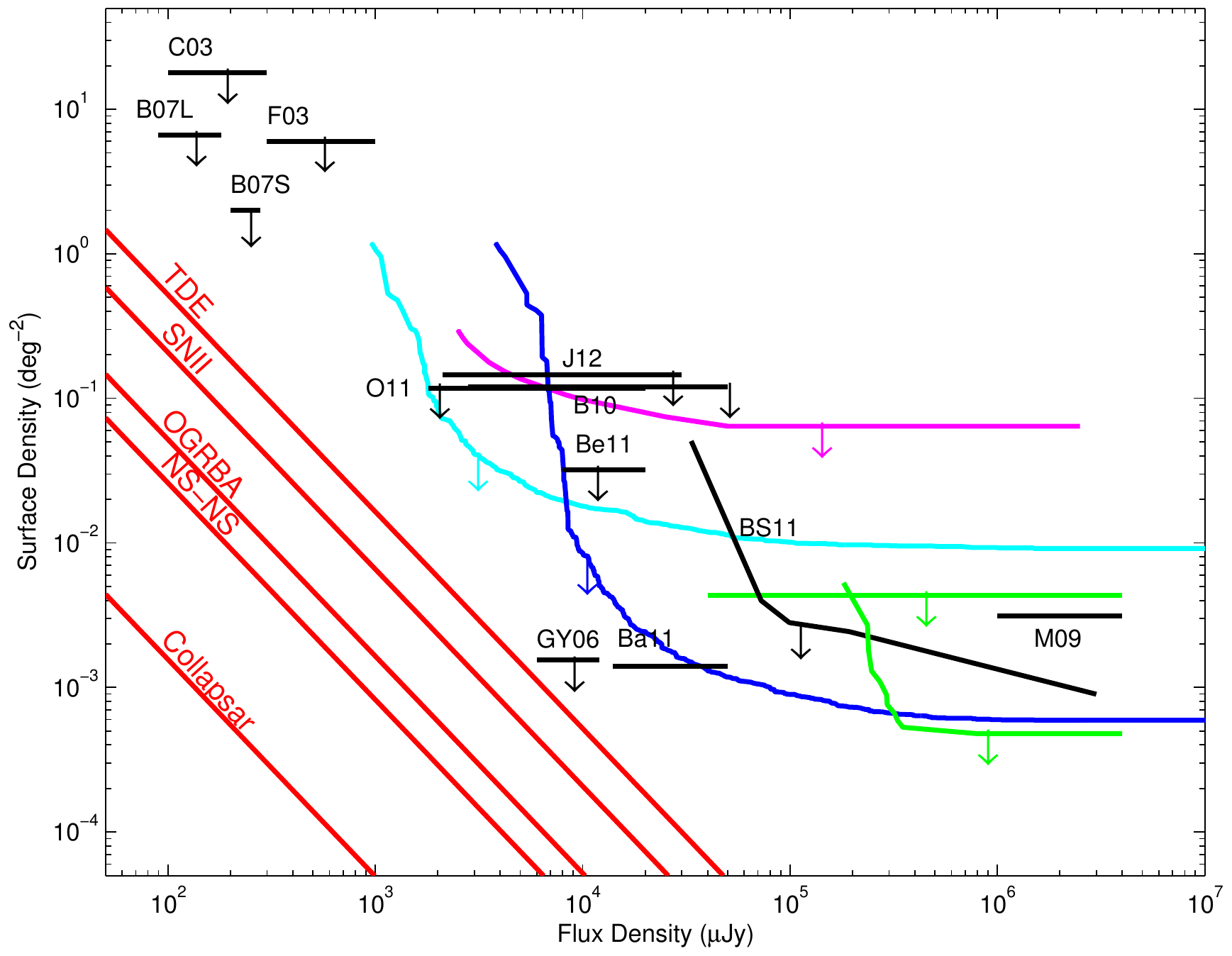}
\caption{\label{fig:rate}
Cumulative two-epoch source surface density for radio transients as a
function of flux density, based on Fig.~9 of \citet{bower:07}, Fig.~20 of \paperi, and updated in the light of \citet{frail:12}. 
Black lines show upper limits or measurements from the literature: from the study of highly variable sources in the Lockman Hole at 1.4\,GHz \citep[][C03]{carilli:03}; from the VLA archival observations of \citeauthor{bower:07}\ with a 1\,yr timescale (labeled B07L) and a 2 month timescale (B07S); from GRB follow-up observations \citep[][F03]{frail:03}; from structure function analysis of NVSS and FIRST \citep[][O11]{ofek:11}, reinterpreted in the light of \citet{frail:12}; from a 325\,MHz VLA Lockman Hole archival transient search \citep[][J12]{jaeger:12}, plotted here as a $2\sigma$ upper limit; from a 5\,GHz Galactic Plane survey \citep[][B10]{becker:10}; from an archival survey of VLA calibrators \citep[][Be11]{bell:trans}; from a VLA archival survey of the 3C\,286 field \citep[][BS11]{3c286}; from comparison of NVSS and FIRST \citep[][GY06]{galyam:06}; from the MOST archive \citep[][Ba11]{bannister:11};  and from drift-scanning surveys \citep[][M09]{matsumura:09}.
Green lines correspond to upper limits from ATATS: the upper straight line from the comparison of the deep field to NVSS \citep{atats}, and the lower curve from analysis of the ATATS single-epoch images by \citet{atatsii}. The magenta curve shows the upper limit on long-duration transients present in our PiGSS catalogs but missing in NVSS. The cyan curve shows the upper limit on PiGSS transients from our monthly images when compared to the deep image. The blue curve shows the upper limit on PiGSS transients from daily images compared to the deep image. All of the PiGSS results are combined limits from all four PiGSS deep fields. 
The red lines show the predicted rates for the various source classes (tidal disruption events, Type II SNe, orphan gamma-ray burst afterglows, neutron star -- neutron star mergers, and collapsar-like events) discussed by \citet{frail:12}. 
}
\end{figure*}

For PiGSS sources missing in NVSS, corresponding to transient candidates on timescales of months to years, we set an upper limit to the transient rate (given that we see no transients brighter than 2.5\,mJy) for long duration transients of 10\,mJy, $\Sigma_{long} \lesssim 0.08$\,deg$^{-2}$. Rate limits for such sources are shown as the magenta curve in Fig.~\ref{fig:rate}.  

For PiGSS transients seen in the monthly data when compared to the deep image, of which we see one possible candidate, although no confirmed transients, we obtain $\Sigma_{monthly} \lesssim 0.02$\,deg$^{-2}$ at 10\,mJy. Rate limits for these sources are shown as the cyan curve in Fig.~\ref{fig:rate}.  

For PiGSS transients seen in the daily data when compared to the deep image, of which we see none, we obtain $\Sigma_{daily} \lesssim 0.009$\,deg$^{-2}$ at 10\,mJy. Rate limits for these sources are shown as the blue curve in Fig.~\ref{fig:rate}. 

\subsubsection{Rates per galaxy in Coma}

The Coma field contains a large overdensity of galaxies, and is therefore more likely to contain supernovae than a random extragalactic field. At the distance of Coma (100\,Mpc, corresponding to the cluster center at $\sim 7000$\,\kms), our median 90\%\ completeness in the single-epoch images, 15.2\,mJy, corresponds to $\sim 4 \times 10^{28}$\,\esh. This is comparable to the peak radio luminosity of the unusually luminous SN\,1998bw \citep{kulkarni:98}.

To estimate the number of galaxies in the Coma cluster surveyed by our observations, we searched the NASA Extragalactic Database for galaxies with redshifts between 5000 and 9000\,\kms, within 60\arcmin\ of the center of our Coma map. This corresponds to approximately the area with mosaic gain unity in the deep image. We found 665 galaxies within this area, which is consistent with the Coma cluster galaxy density determined by \citet{the:86}.

In the 73 days from the beginning to end of our Coma observations, no transients were seen. Taking a $2\sigma$ upper limit of 15 transients per year, we can therefore constrain the rate of SN\,1998bw like events to $\lesssim 0.02$  yr$^{-1}$ per average galaxy in the Coma cluster.

The average supernova rate in local galaxy clusters (including Coma) is approximately 0.01 galaxy$^{-1}$ yr$^{-1}$ \citep{mannucci:08}. The fraction of SNe detectable in the radio is still poorly constrained, and depends on survey sensitivity, but is typically greater than 10\% \citep{Lien:11}. This predicts that the radio transient rate from SNe in Coma ought to be $> 0.001$ galaxy$^{-1}$ yr$^{-1}$. With order of magnitude or more improvements (compared to PiGSS) in sensitivity, area, or length of the observing campaign, seeing ordinary radio supernovae \citep{weiler:98} in such a blind search might be possible.

\section{Variability}\label{sec:variability}

\begin{figure}[htp]
\centering
\includegraphics[width=\linewidth,draft=false]{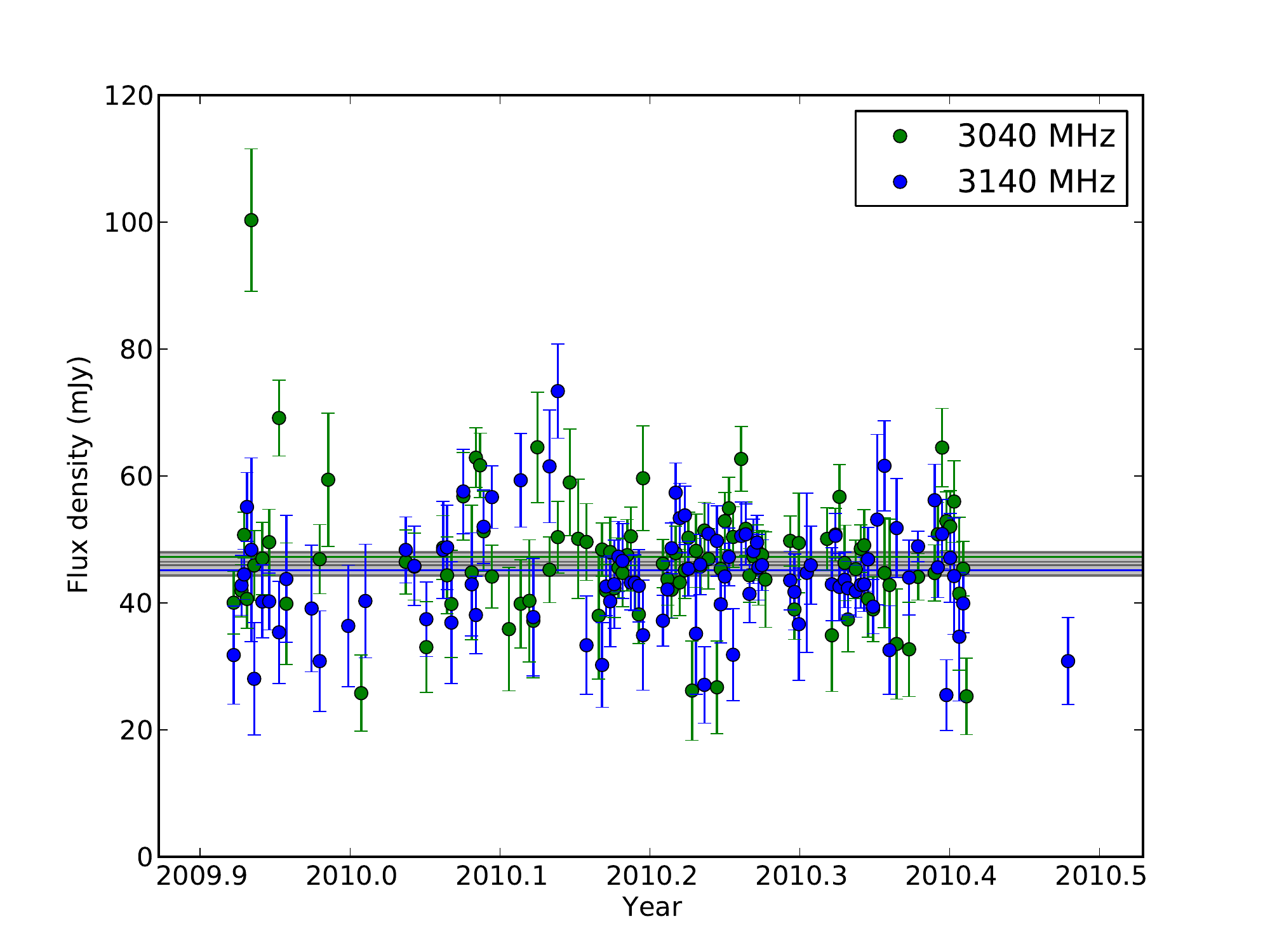}
\caption{\label{fig:bigvarlock}
Lightcurve for PiGSS\,J105159+593240. 3040\,MHz flux densities are shown in green, and 3140\,MHz in blue. The deep field flux densities are shown as horizontal lines, with grey boxes representing the measured uncertainties.
This source appears to brighten in one epoch to more than twice its flux density in the deep Lockman image, although the variation is seen only at 3040\,MHz and not at 3140\,MHz, and is therefore not intrinsic to the source. Some of the variability appears correlated between the two frequencies, whereas some does not. Some of this variability is nevertheless within the expectations from Gaussian uncertainties, although there is also a long tail of outlier points (Fig.~\ref{fig:varrat}). In the example shown here, the $\sim 100$\,mJy outlier is from an epoch where the PIC gain calibration at 3040\,MHz did not converge to a fit (Section~\ref{sec:picgain}), and so the bad epoch is in fact automatically excluded from our variability analysis. 
}
\end{figure}

The PIC gain calibration (Section~\ref{sec:picgain}) helps us exclude apparent variability that is in fact due to bad calibration for particular epochs. For example, only one source in Lockman (Fig.~\ref{fig:bigvarlock}) was detected with a flux density that brightened to more than twice its flux density in the deep image. In this case the brightening was seen in only one of the two frequencies, which immediately arouses suspicion that it is not intrinsic to the source. In fact, the PIC gain fit failed to converge for this epoch and so the epoch was excluded from our variability analysis. The PIC gain cuts remove a number of similar cases where sources that appear highly variable in an individual epoch in fact have poor calibration for all sources in that epoch.

\begin{figure}[htp]
\centering
\includegraphics[width=\linewidth,draft=false]{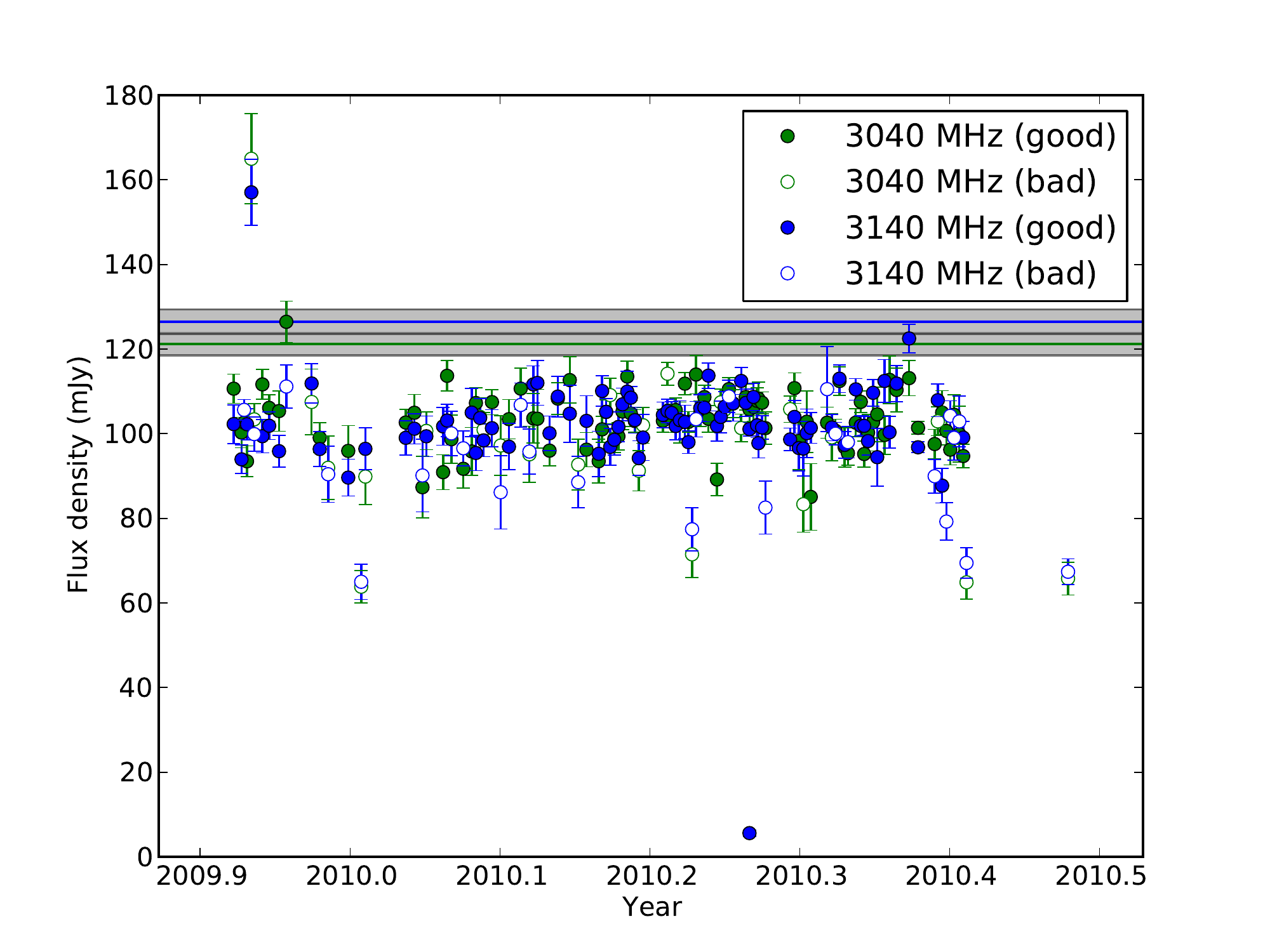}
\caption{\label{fig:bigvarlock2}
Lightcurve for PiGSS\,J105141+591305, a $z = 0.44$ SDSS QSO \citep{schneider:07}.  3040\,MHz flux densities are shown in green, and 3140\,MHz in blue. Those within the acceptable range of PIC gain for Lockman (between 0.835 and 1.035) are shown as solid symbols, and those outside the acceptable range are shown as open symbols. The deep field flux densities are shown as horizontal lines, with grey boxes representing the measured uncertainties.
This source appears to be highly variable when the maximum absolute difference in flux density between any two epochs is calculated, and despite the PIC gain selection rejecting some outlying points (as well as a few which appear, at least from this lightcurve, to be acceptable) the largest outliers at 3040\,MHz are not rejected in this particular case. In many cases, however, selection based on PIC gain enables us to better measure real variability. The deep field flux densities are significantly higher than those in the individual epochs; this appears to be because the source appears marginally resolved in the deep images.
}
\end{figure}

Nonetheless, some erroneous flux densities still contribute to apparent large variability. Fig.~\ref{fig:bigvarlock2} shows the source with the fourth highest variability in the Lockman field (for sources with detections in at least five individual epochs), as measured by the maximum absolute difference in flux density between any two epochs. Although some of the outlying points are rejected due to PIC gains outside our range of acceptable values, remaining outlying flux density measurements are still responsible for this source being identified as highly variable. However, in many cases, the PIC gain selection does improve our ability to accurately measure variability.

For the remainder of our variability analysis, we consider only data from ``good'' epochs with PIC gains within $\pm 0.1$ of the median PIC gain for the field (Section~\ref{sec:medianpic}), that also meet our completeness cut (Table~\ref{tab:fields}). 

\subsection{Fractional modulation}

\begin{figure}[htp]
\centering
\includegraphics[width=\linewidth,draft=false]{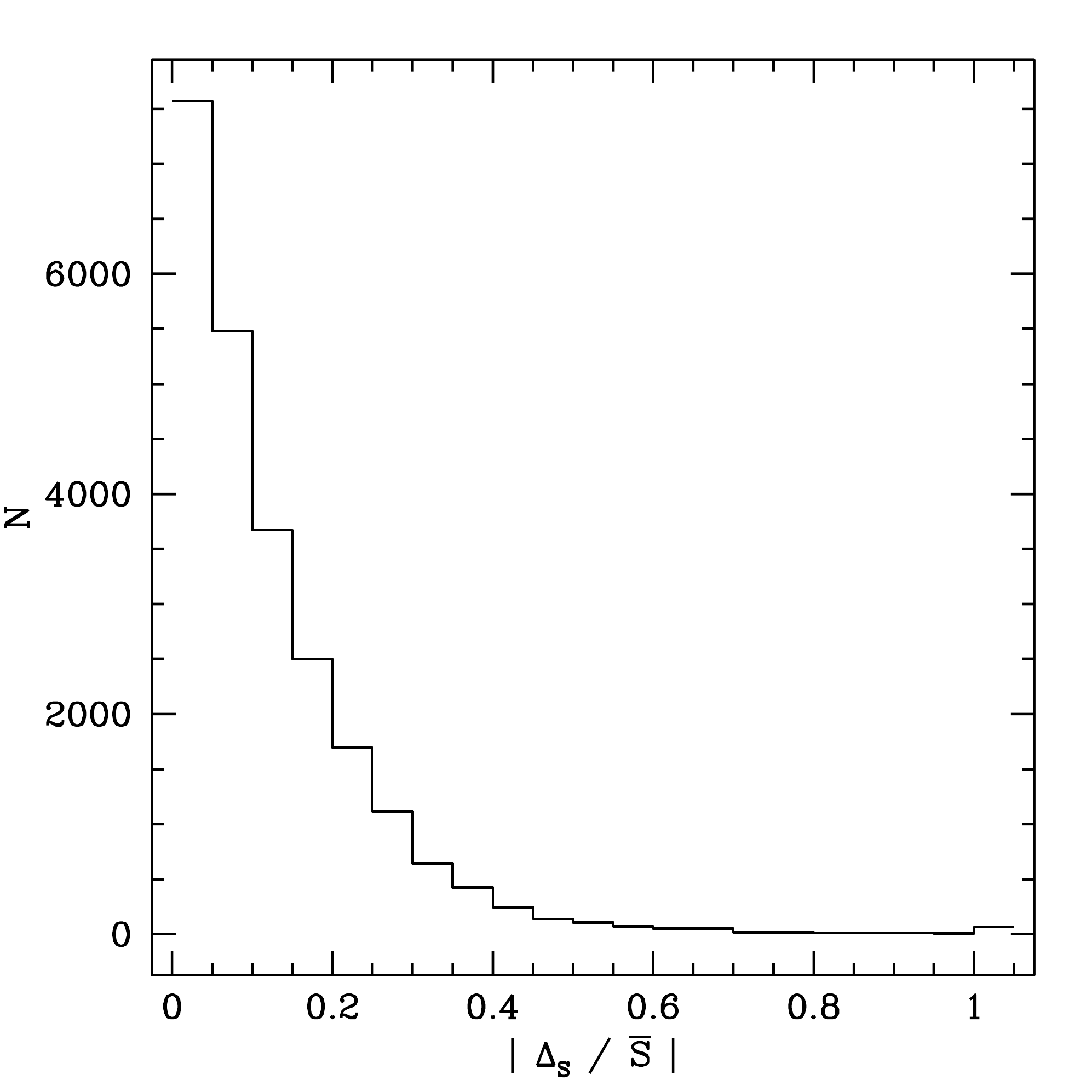}
\caption{\label{fig:fracmod}
Daily fractional modulation, $m_{i}$ (computed for ``good'' epochs; see text) as a function of mean flux density, $\bar{S}$, for sources detected in all four fields.
}
\end{figure}

We compute the \label{sec:fracmod}daily fractional modulation, 
\begin{equation}
m_{i} = \frac{S_i - \bar{S}} {\bar{S}}
\end{equation}
for all sources detected in these good epochs. In Fig.~\ref{fig:fracmod}, we show a histogram of the measured modulation fractions. We see that, as expected, the majority of measurements are consistent with variability less than a fraction of $\sim 0.1$.

\begin{figure}[htp]
\centering
\includegraphics[width=\linewidth,draft=false]{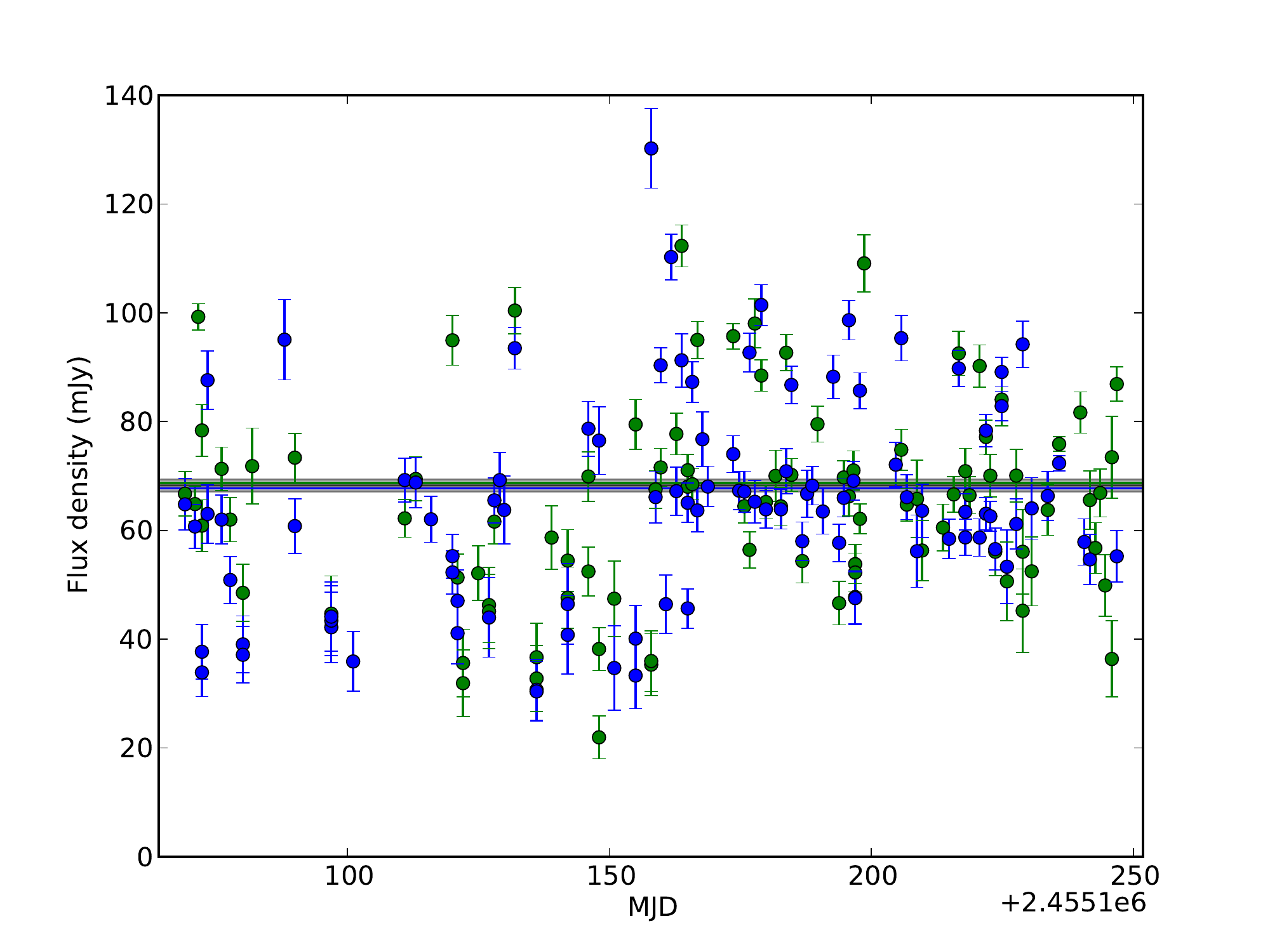}
\includegraphics[width=\linewidth,draft=false]{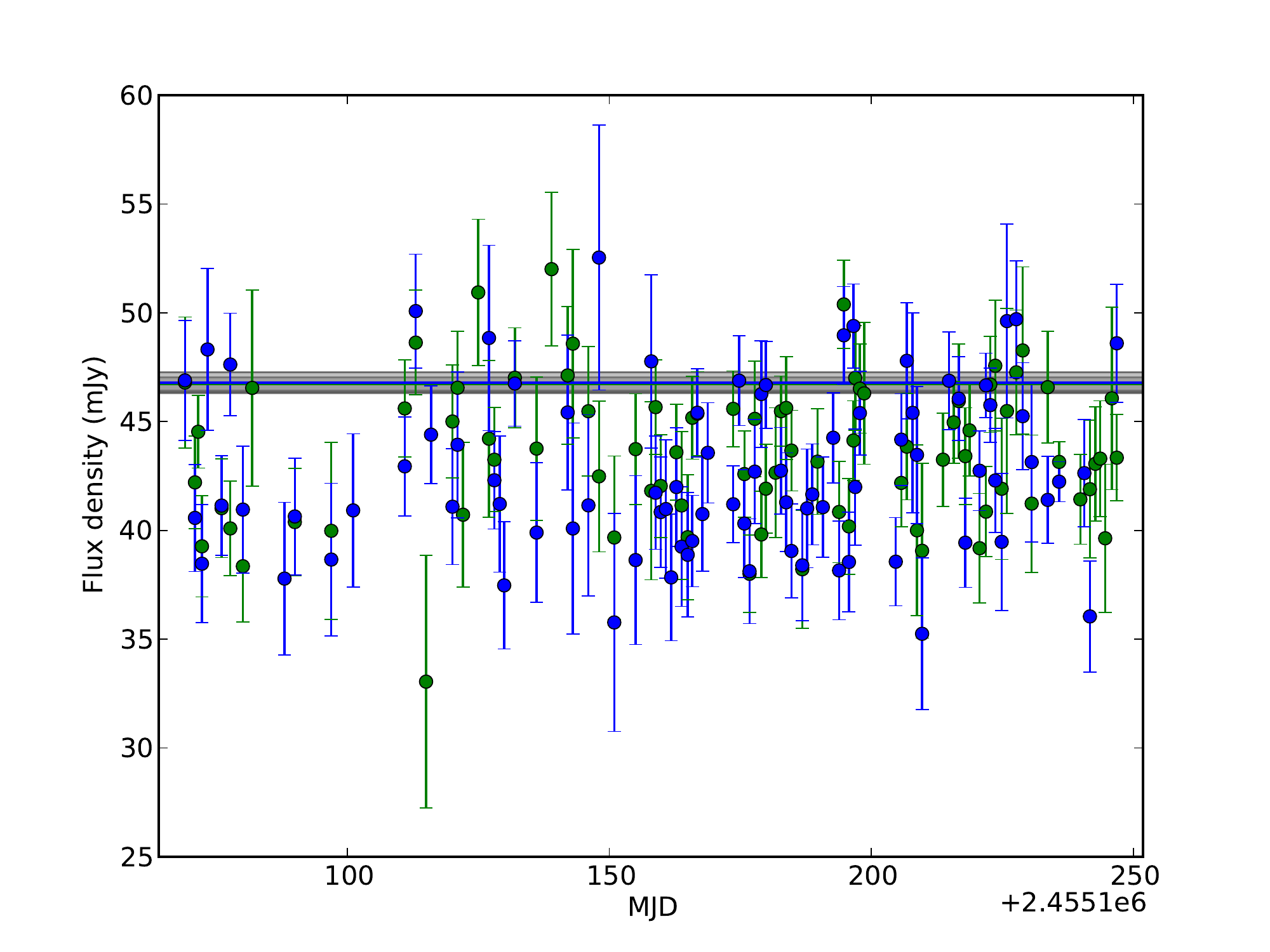}
\caption{\label{fig:maxfrac}
Lightcurve for PiGSS\,J105703+584717 (top), a source with large maximum daily fractional modulation, $m_{max}$, and PiGSS\,J104719+582116 (bottom), a source with small $m_{max}$ associated with a $z = 1.22$ QSO \citep{schneider:07}.
3040\,MHz flux densities are shown in green, and 3140\,MHz in blue. The deep field flux densities are shown as horizontal lines, with grey boxes representing the measured uncertainties.
}
\end{figure}

\begin{figure}[htp]
\centering
\includegraphics[width=\linewidth,draft=false]{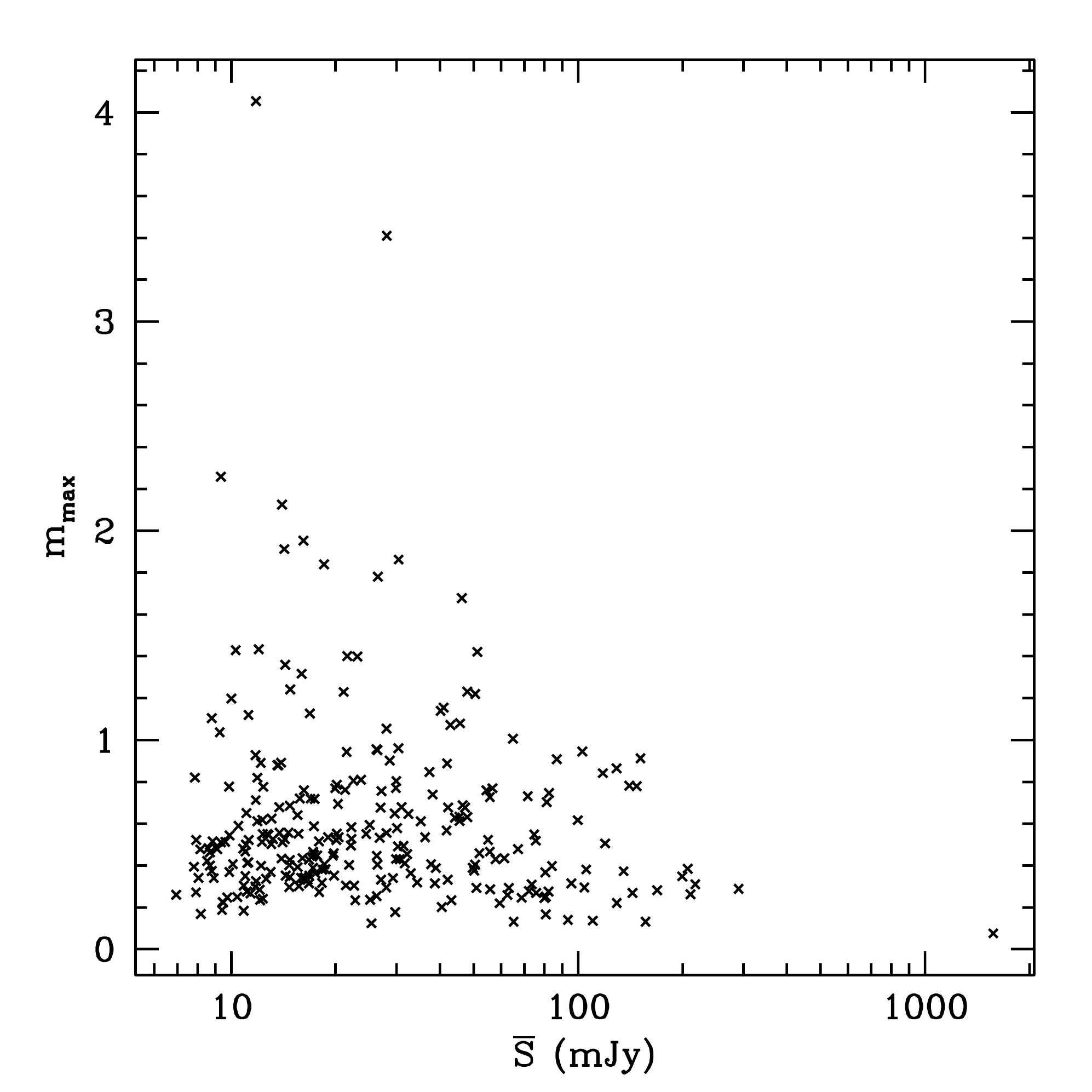}
\caption{\label{fig:maxfracflux}
Maximum daily fractional modulation, $m_{max}$ as a function of mean flux density, $\bar{S}$, for sources with 10 or more detections (in good epochs) in all four fields (using data for both frequencies).
}
\end{figure}

We also compute the \label{sec:maxfracmod}maximum daily fractional modulation in each source, 
\begin{equation}
m_{max} = {\rm MAX} \left\{ \frac{S_i - \bar{S}} {\bar{S}} \right\}
\end{equation}

for sources detected in at least 10 good epochs. In Fig.~\ref{fig:maxfrac} we show examples of sources with large and small $m_{max}$. There appears to still be a tendency for outlier points to be responsible for some of the largest degrees of detected variability, as determined by variability seen at one frequency but not at the other. In Fig.~\ref{fig:maxfracflux} we plot $m_{max}$ as a function of $\bar{S}$. As noted by \citet{ofek:11}, fractional modulation and related variability measures have irregular statistical properties that depend on the number of epochs of observation, and so this statistic is not easily comparable from one survey to another. However, it can still provide a measure of how often variability of some magnitude is seen in an individual survey.

The upper envelope of the points in Fig.~\ref{fig:maxfracflux} shows that variable sources in our dataset do not typically vary by more than a factor $\sim 2$ from epoch to epoch, and in the few cases where more extreme variability appears to be detected, further examination of light curves and postage stamp images shows that it is not intrinsic to the source (for example, because variability is seen at only one of the two frequencies). This suggests that our threshold of a factor 10 (Section~\ref{sec:factor10}) for ``true'' transients (typically explosive or impulsive events), as opposed to the upper envelope of ``normal'' variability, is reasonable. Of course, extreme variables are still interesting in their own right, and may represent sudden changes in accretion rate onto AGNs, for example.

\subsection{Consistency of variability in the two frequency bands}

 Where similar patterns of variability are seen at both frequencies simultaneously, these are more likely to be intrinsic to the source than patterns where the variability is not correlated at the two frequencies. However, variations in overall gain calibration of the telescope, as well as other systematic effects, can also lead to variations in both frequencies simultaneously, so this property alone cannot be used to distinguish intrinsic variability from that due to calibration uncertainties.
 
 To examine whether differences between measured flux densities at the two frequencies are consistent with the measurement uncertainties, we compute the difference between the two flux densities as a multiple of the total uncertainty. We define this difference,
 
 \begin{equation}
 D_i = \frac{S_{i,3040} - S_{i,3140}}{\sqrt{(\Delta S_{i,3040})^2 + (\Delta S_{i,3140})^2}},
 \end{equation}

for  each good epoch, $i$, where a source is detected in both frequencies. The uncertainties at each frequency are the reported uncertainties in the fit and the background for each measurement, added in quadrature.
 
\begin{figure}[htp]
\centering
\includegraphics[width=\linewidth,draft=false]{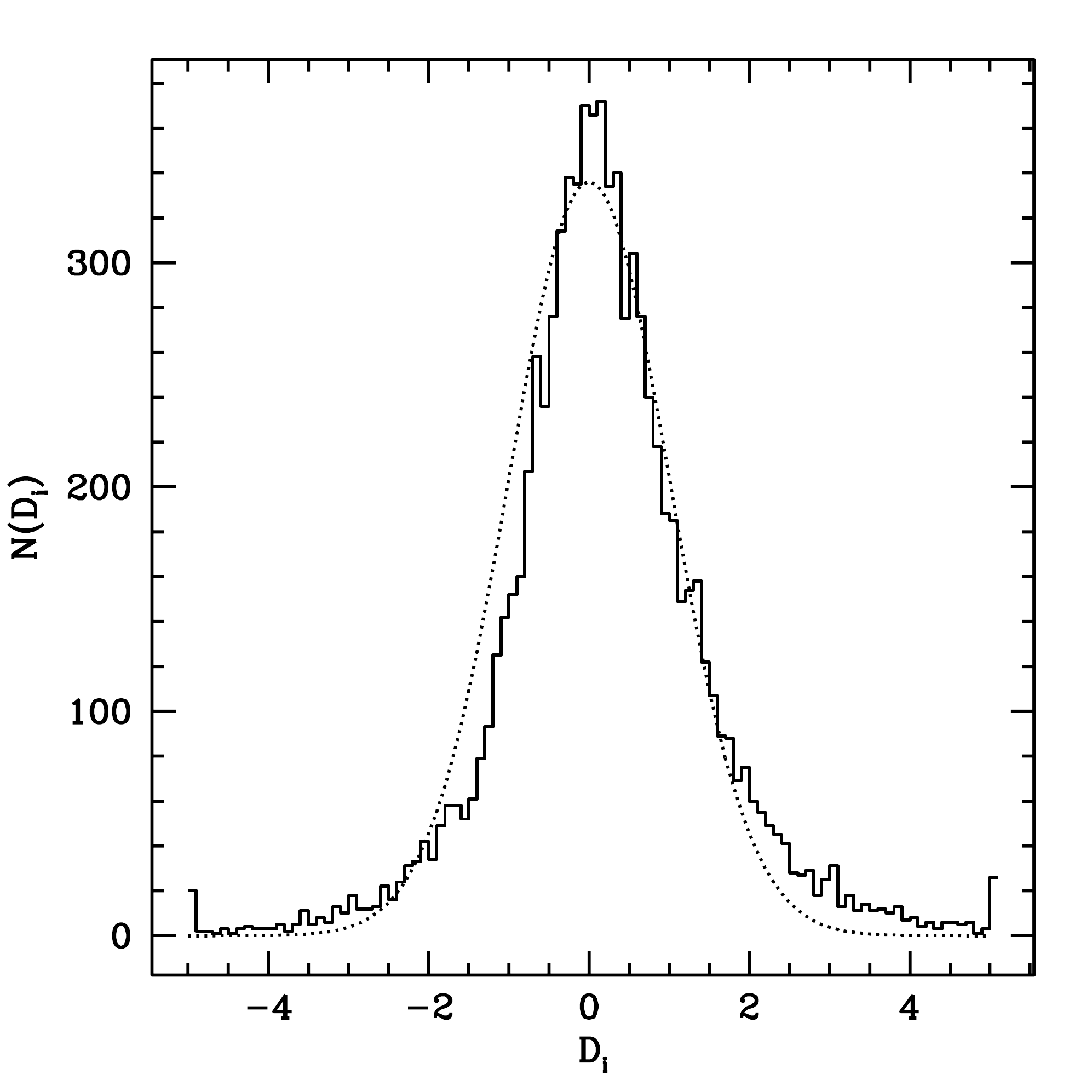}
\caption{\label{fig:varrat}
Difference between flux density measurements at the two frequencies, as a multiple of measured uncertainties on the flux densities (solid histogram). The dotted line shows the standard normal distribution, scaled to match the sample and bin sizes of the solid histogram.
}
\end{figure}

In Fig.~\ref{fig:varrat}, we plot a histogram of $D_i$. If the differences in measured flux densities obey Gaussian statistics, this should match the standard normal distribution. We see that the agreement is quite good, although there is a tail made up of a few points which have larger discrepancies than would be expected from Gaussian statistics, and the data histogram is skewed slightly to the right, implying that sources are measured to be slightly brighter at 3040\,MHz than at 3140\,MHz. In fact, if we divide the 3040\,MHz fluxes by the corresponding 3140\,MHz fluxes for the 8430 dual-frequency detections of sources in the epochs considered in our variability analysis, we find that the sources are 1.9\%\ brighter at 3040\,MHz than at 3140\,MHz. If we assume no systematic errors in overall calibration, this implies a mean spectral index of $\alpha = -0.6$, consistent with expectations.
 
\subsection{Standard deviation}
 
\begin{figure*}[htp]
\centering
\includegraphics[width=0.45\linewidth,draft=false]{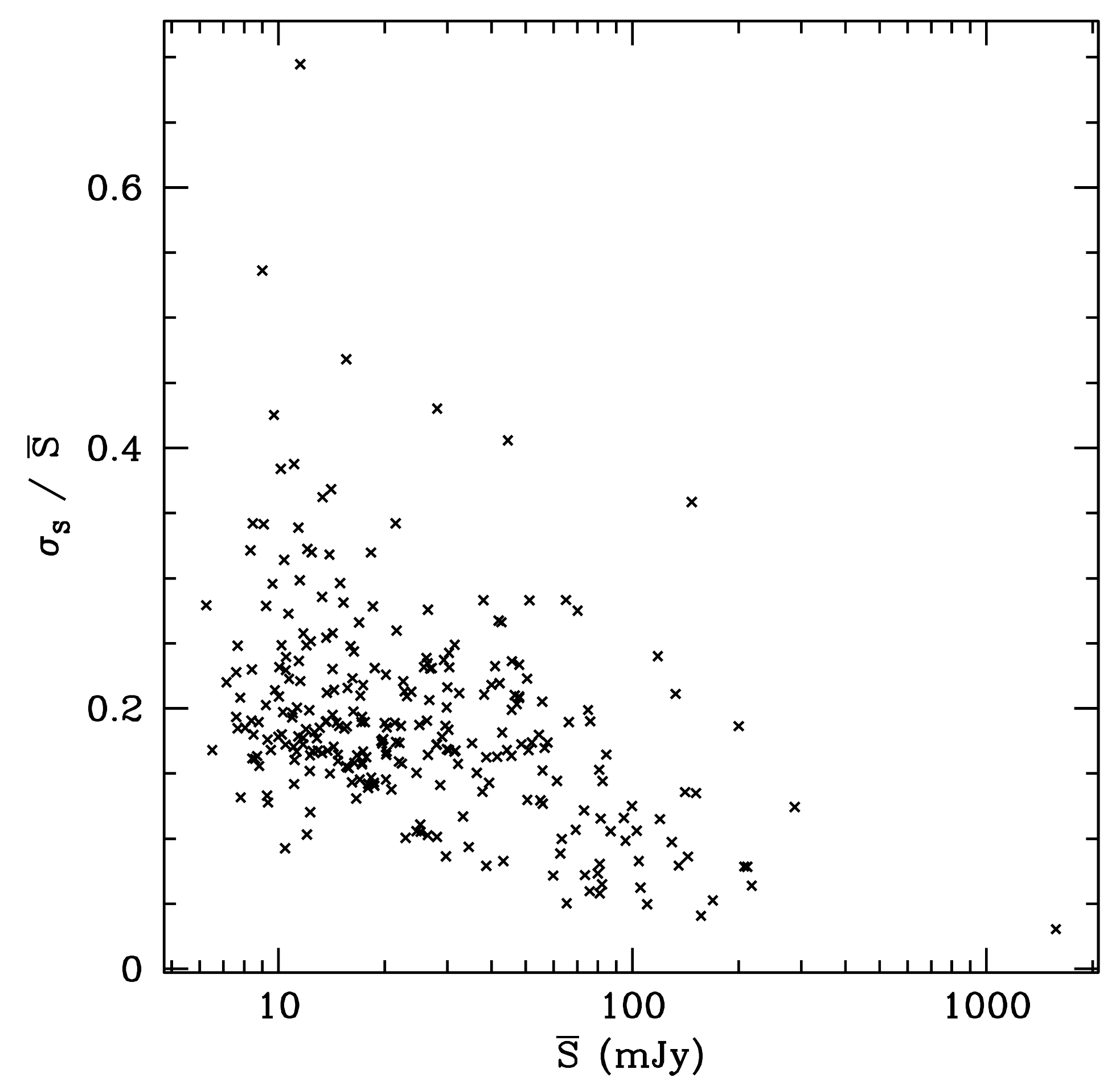}\hspace{0.03\linewidth}%
\includegraphics[width=0.45\linewidth,draft=false]{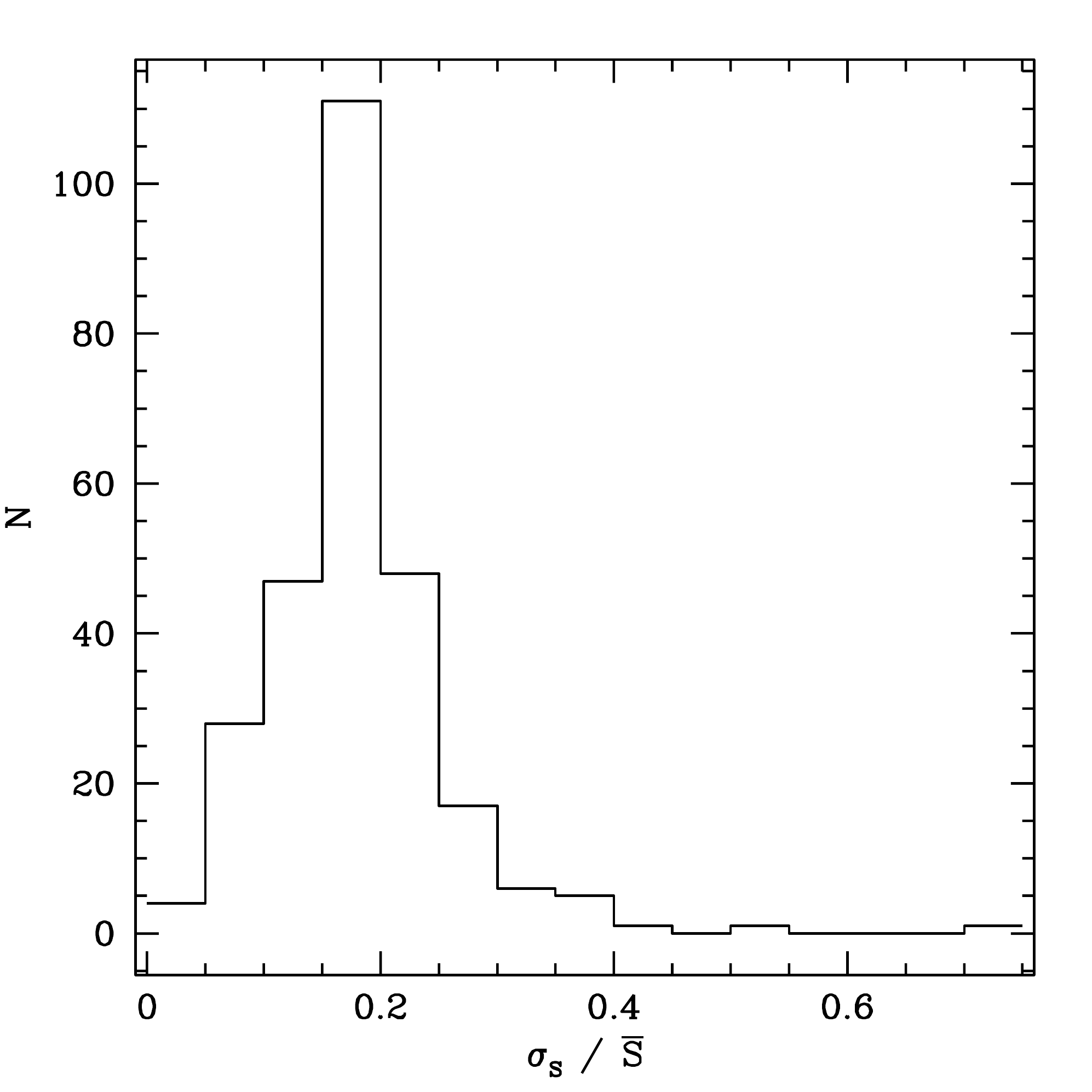}
\caption{\label{fig:stdevmean}
{\em Left:} Standard deviation of the daily flux densities, $\sigma_S$, divided by the mean flux density, $\bar{S}$, for sources in all four fields with 10 or more detections in good epochs (using data from both frequencies), as a function of $\bar{S}$.
{\em Right:} Histogram of $\sigma_S / \bar{S}$.
}
\end{figure*}

We also compute the standard deviation divided by the mean flux density for all sources with 10 or more detections in good epochs, which we plot in Fig.~\ref{fig:stdevmean}. Once again, the sources showing the most extreme variability (for example, the sources with $\sigma_S / \bar{S} = 0.71$ and $0.53$ in this plot) are often due to outliers (in each of these two cases, a single deviant 3040\,MHz point which is not matched by a correspondingly deviant 3140\,MHz point). However, many other sources which would otherwise show up as extreme variables in such plots benefit greatly from the PIC rejection in having their variability accurately measured. For example, the most extreme source using this criterion without PIC rejection would be PiGSS\,J160833p543457 with  $\sigma_S / \bar{S} = 0.96$; after we reject the bad epochs, its  $\sigma_S / \bar{S}$ drops dramatically to 0.16. The median value of  $\sigma_S / \bar{S}$ is $\sim 6$\%\ lower than it would be for the ensemble of all sources if we did not apply the PIC rejection.

\begin{figure*}[htp]
\centering
\includegraphics[width=0.45\linewidth,draft=false]{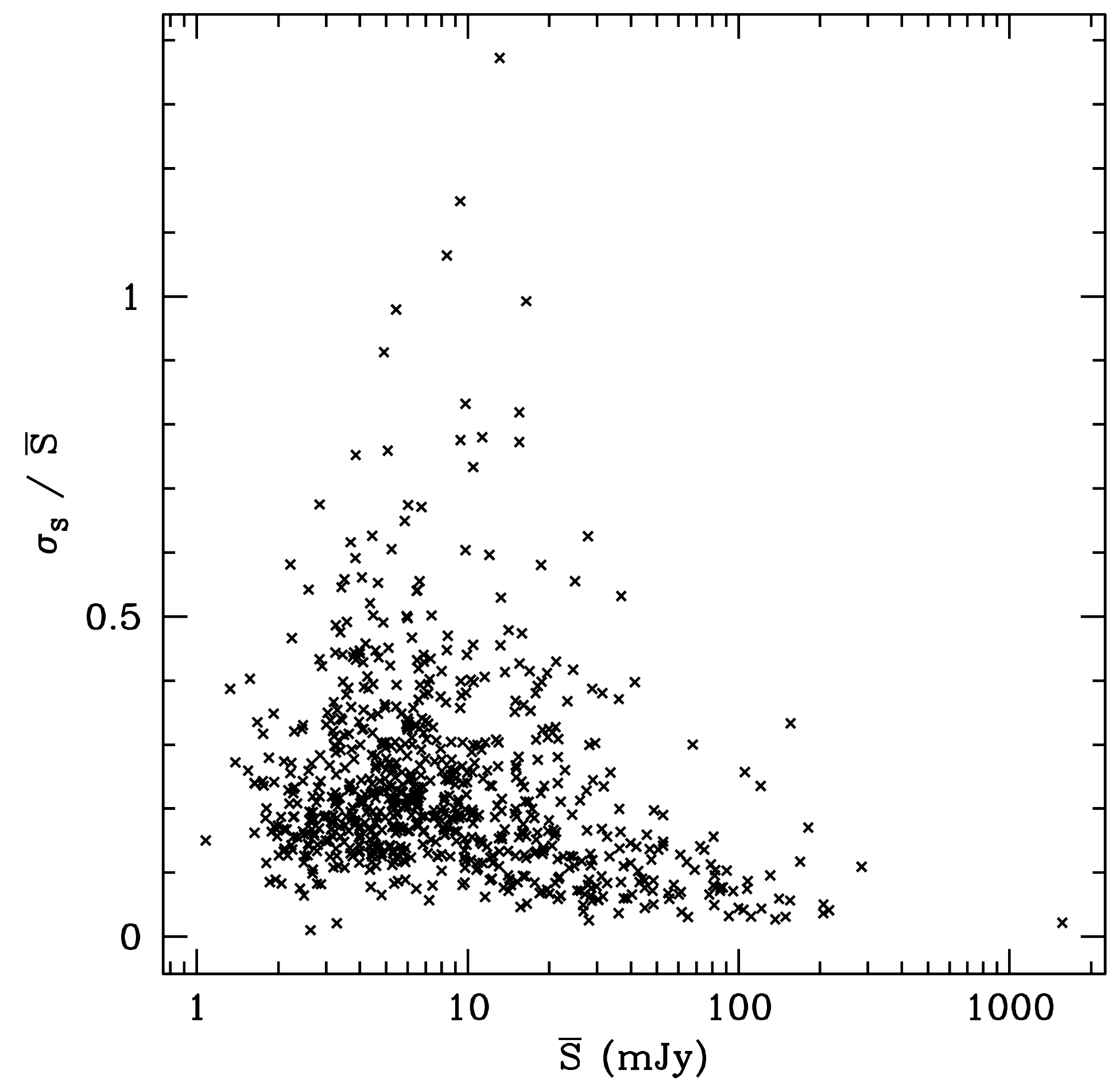}\hspace{0.03\linewidth}%
\includegraphics[width=0.45\linewidth,draft=false]{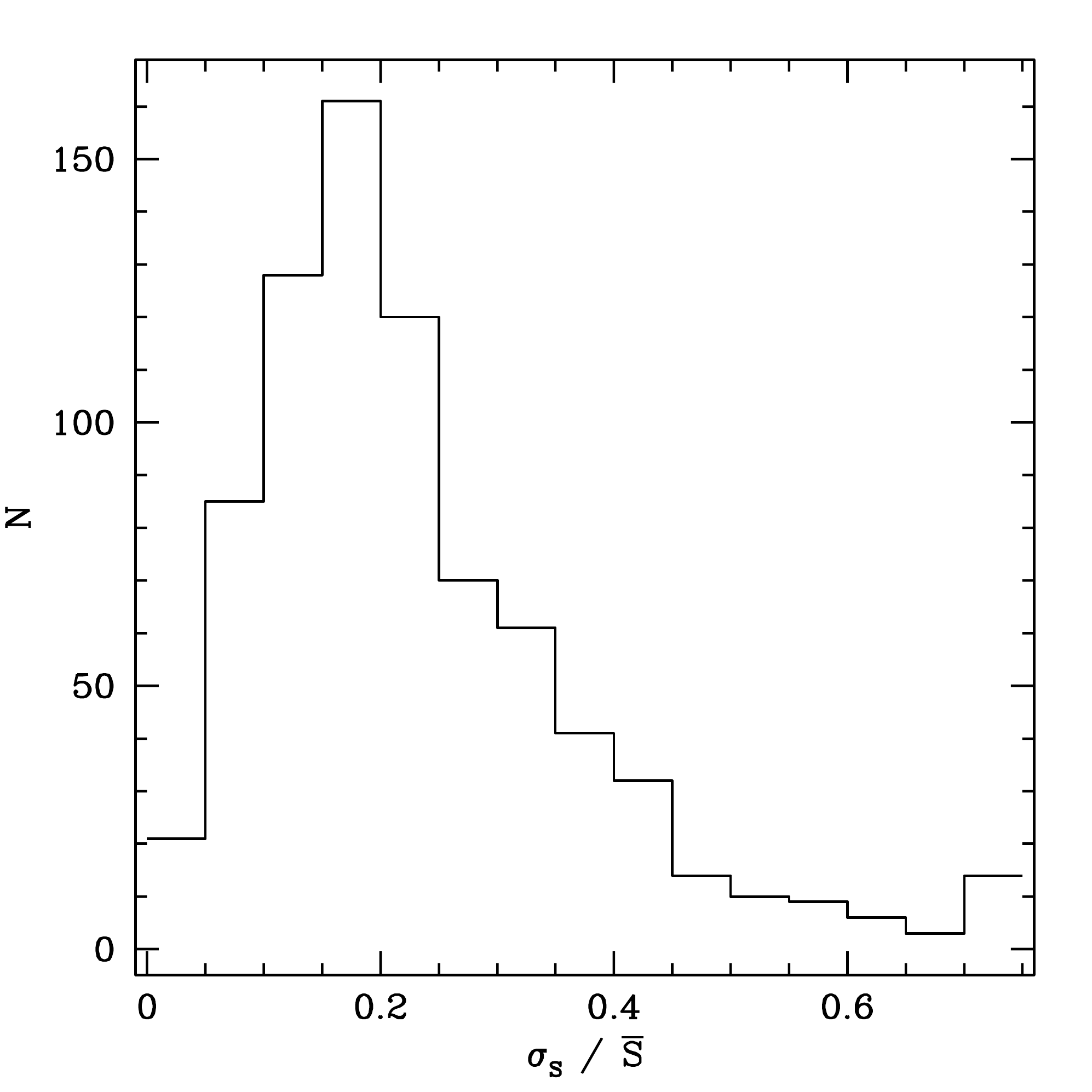}
\caption{\label{fig:stdevmeanmo}
{\em Left:} Standard deviation of the monthly flux densities, $\sigma_S$, divided by the mean flux density, $\bar{S}$, for sources in all four fields with 3 or more detections in the monthly images, as a function of $\bar{S}$ (using data for both frequencies).
{\em Right:} Histogram of $\sigma_S / \bar{S}$.
}
\end{figure*}

In Fig.~\ref{fig:stdevmeanmo}, we show $\sigma_S / \bar{S}$ for the monthly data. Again, many of the outliers show apparent variability in only one of the two frequencies. Some apparently extremely variable sources are resolved into multiple components (or associated with confusing sources) at some epochs. Some appear variable due to the smaller number of detections we require here.

\subsection{Chi-squared}

We also compute a chi-squared statistic, given the hypothesis that the source flux density does not change from epoch to epoch,

\begin{equation}
\chisq = \sum_i \left( \frac{S_i - \bar{S}}{\Delta S_i} \right)^2
\end{equation}

\begin{figure*}[htp]
\centering
\includegraphics[width=0.49\linewidth,draft=false]{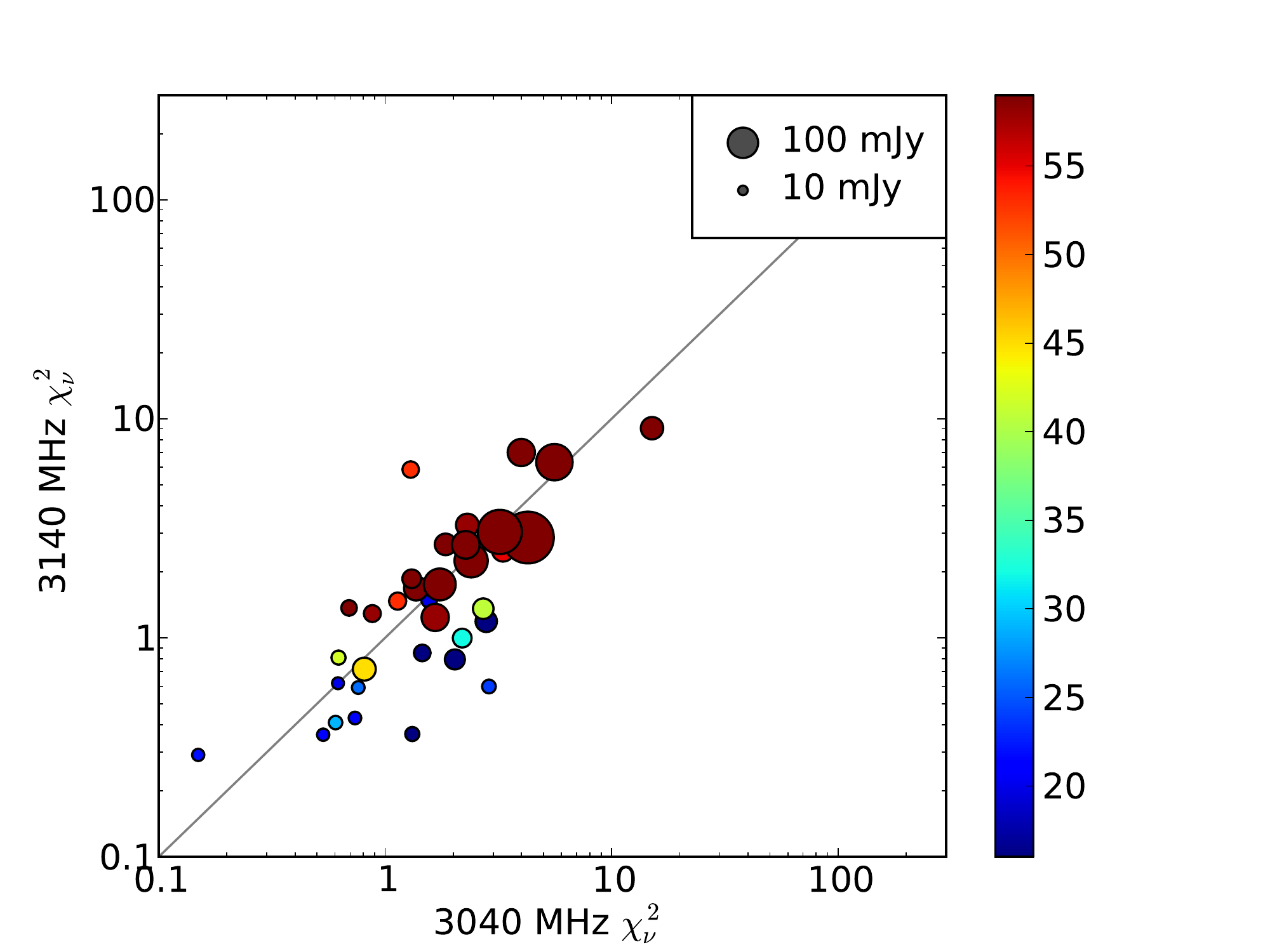}%
\includegraphics[width=0.49\linewidth,draft=false]{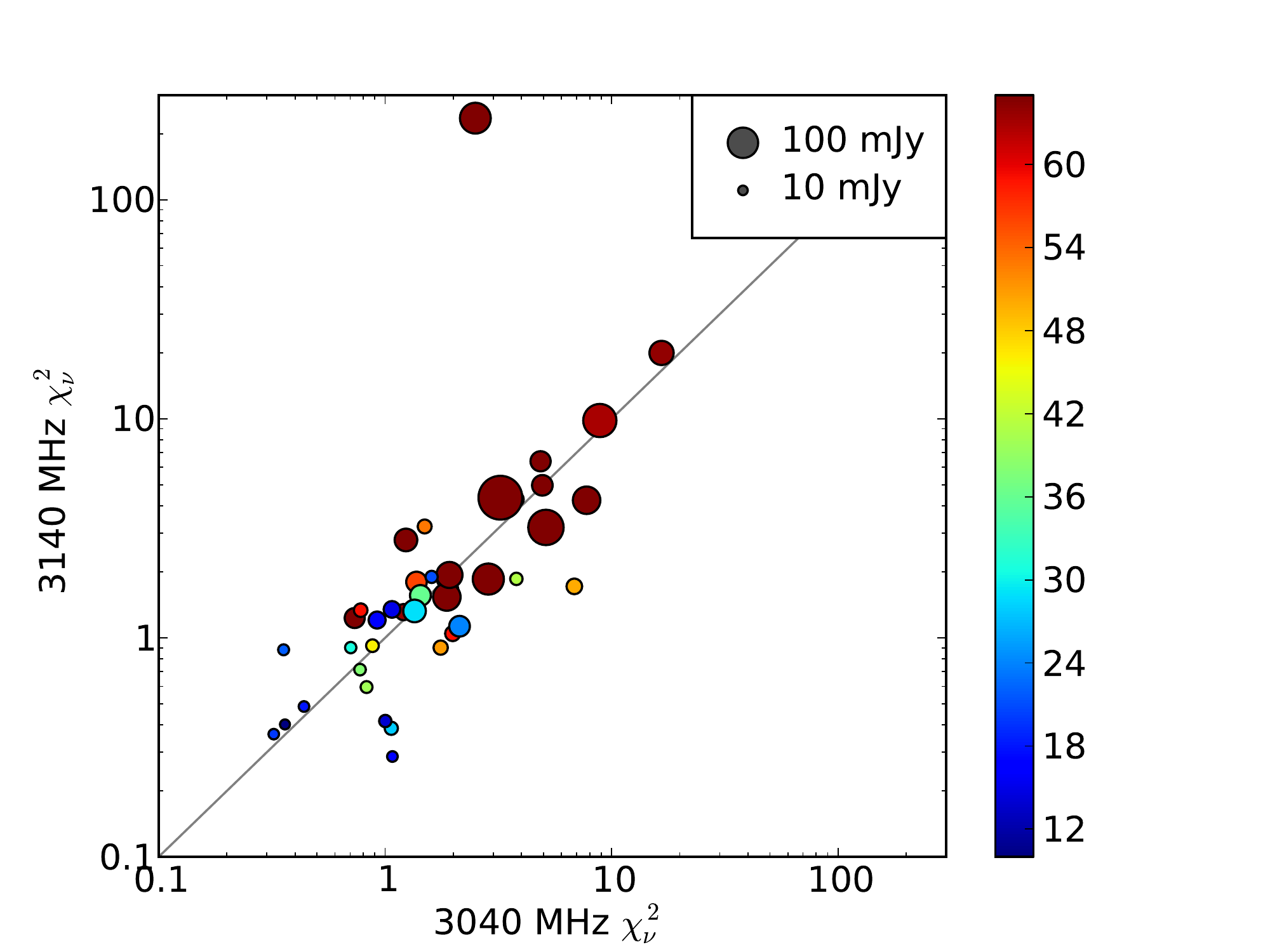}
\includegraphics[width=0.49\linewidth,draft=false]{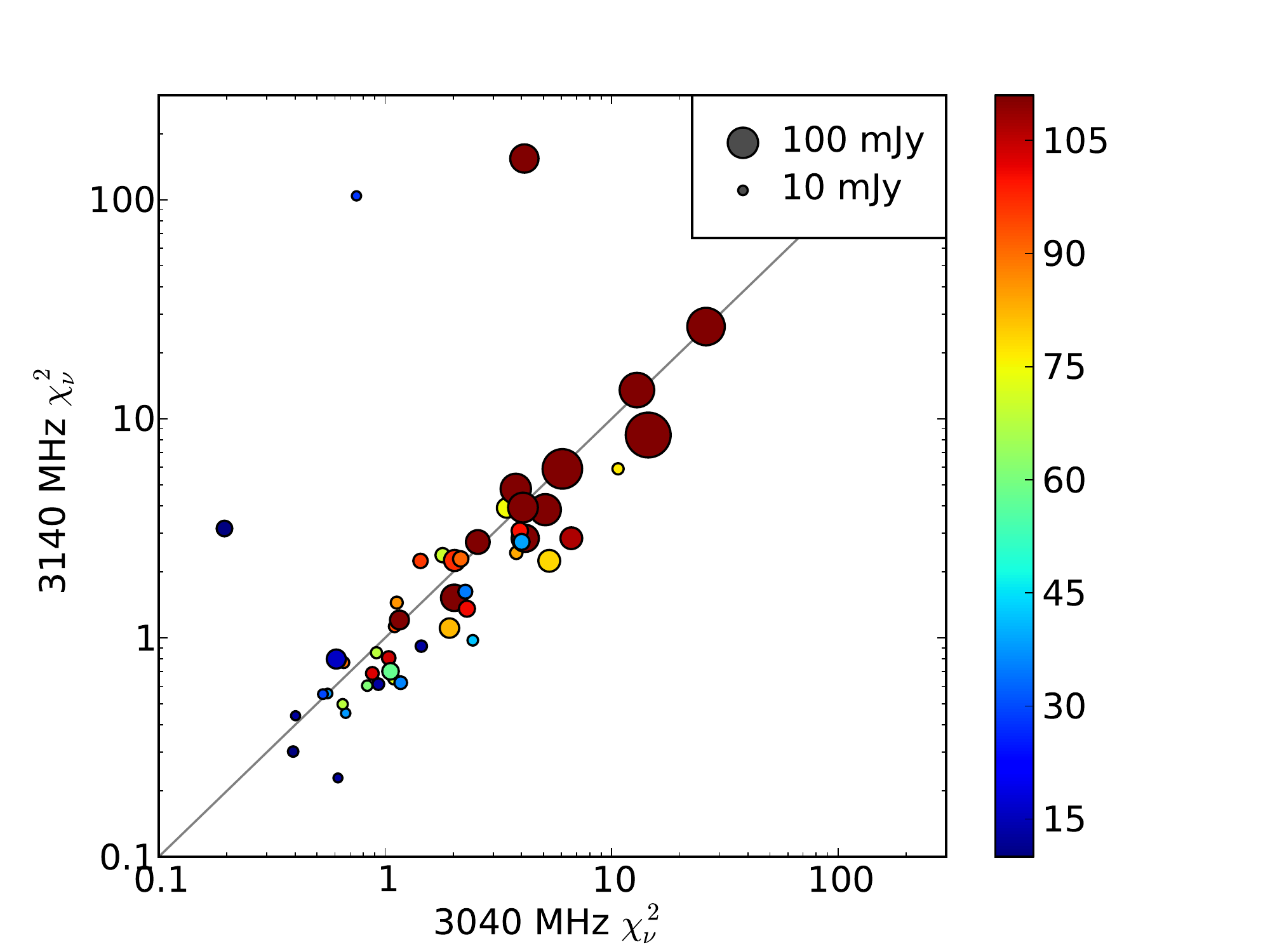}%
\includegraphics[width=0.49\linewidth,draft=false]{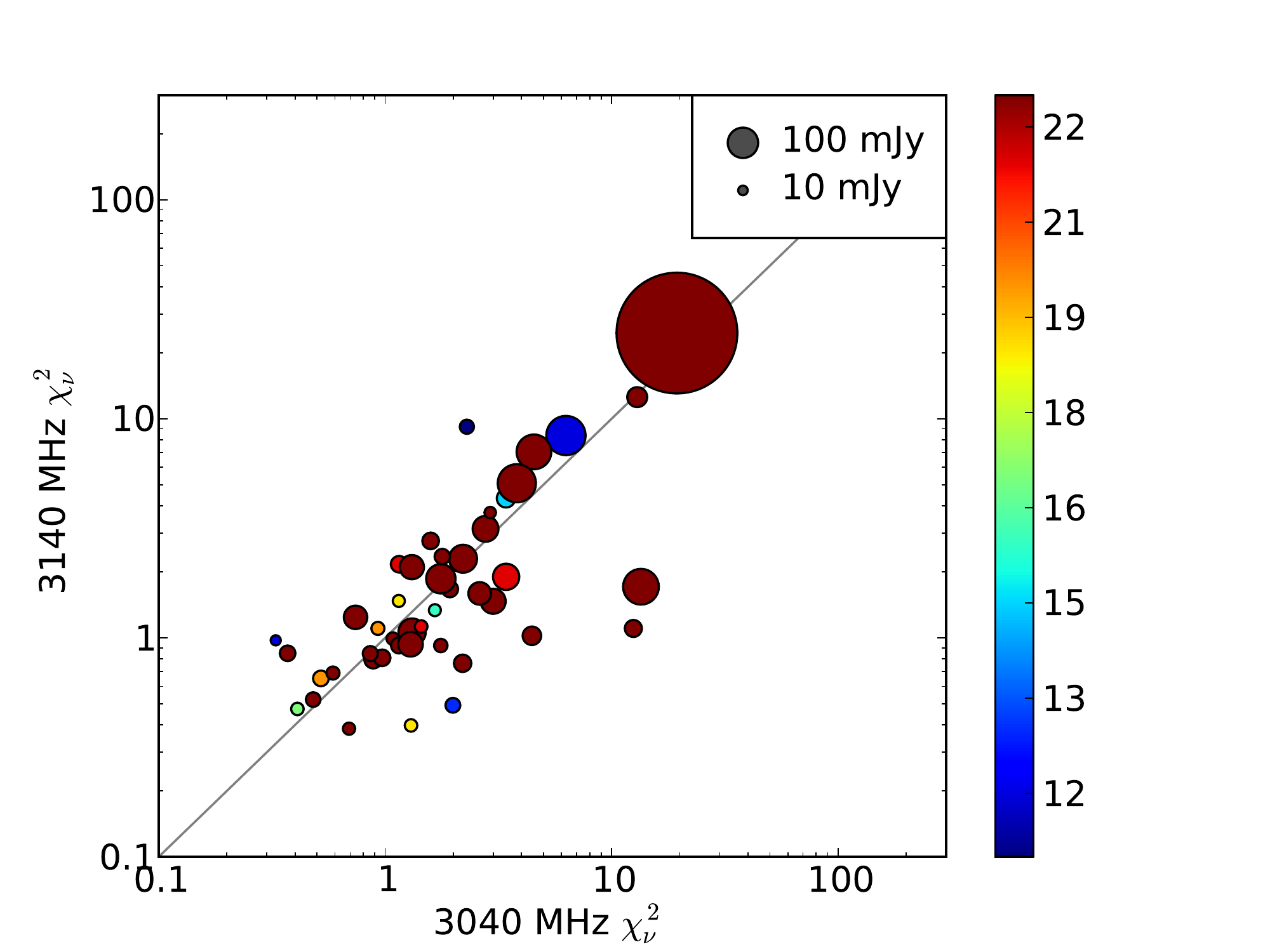}
\caption{\label{fig:chichi}
Reduced chi-squared, \rchisq, for the hypothesis that sources do not vary in individual epochs, computed for sources in (from top left to bottom right panel) the NDWFS, Lockman, ELAIS-N1 and Coma fields. We consider only epochs where sources are detected at both frequencies, and only sources with dual-frequency detections in at least 10 good epochs. Sources are color-coded according to the number of epochs in which they were detected. Circle sizes scale with mean flux density, $\bar{S}$. The grey line guides the eye for sources with equal \rchisq\ at both frequencies.
}
\end{figure*}

For a source detected in $n_{epoch}$ good epochs, we can compute a reduced chi-squared, $\rchisq = \chisq / (n_{epoch} - 1)$, enabling us to compare the significance of the detection of variability for sources detected in different numbers of epochs. We compute \rchisq\ independently for the 3040\,MHz and 3140\,MHz light curves for each source, considering only epochs where the source is detected at both frequencies (from the subsample of epochs that passed the criteria for our variability analysis). Fig.~\ref{fig:chichi} compares \rchisq\ at the two frequencies. For sources detected in a large number of epochs (which naturally tend to be the brighter sources), the two measurements of \rchisq\ tend to agree. These bright sources do not necessarily have larger fractional variability; the \chisq\ is larger because we can detect variability more easily for these well-characterized sources.

Some fainter sources, but also a handful of brighter sources, have discrepant measures of \rchisq, sometimes appearing variable at one frequency but not at the other. Again, this is typically due to erroneous outlying flux density measurements in small numbers of epochs. If we had considered only a single frequency, however, we might have concluded that large values of \rchisq\ generally indicate significant variability. We appear, though, to be in a similar regime as for transients (Section~\ref{sec:falsepos}), where the non-Gaussian tail of outliers is in danger of dominating truly interesting sources showing large intrinsic variability.

\begin{figure*}[htp]
\centering
\includegraphics[width=0.49\linewidth,draft=false]{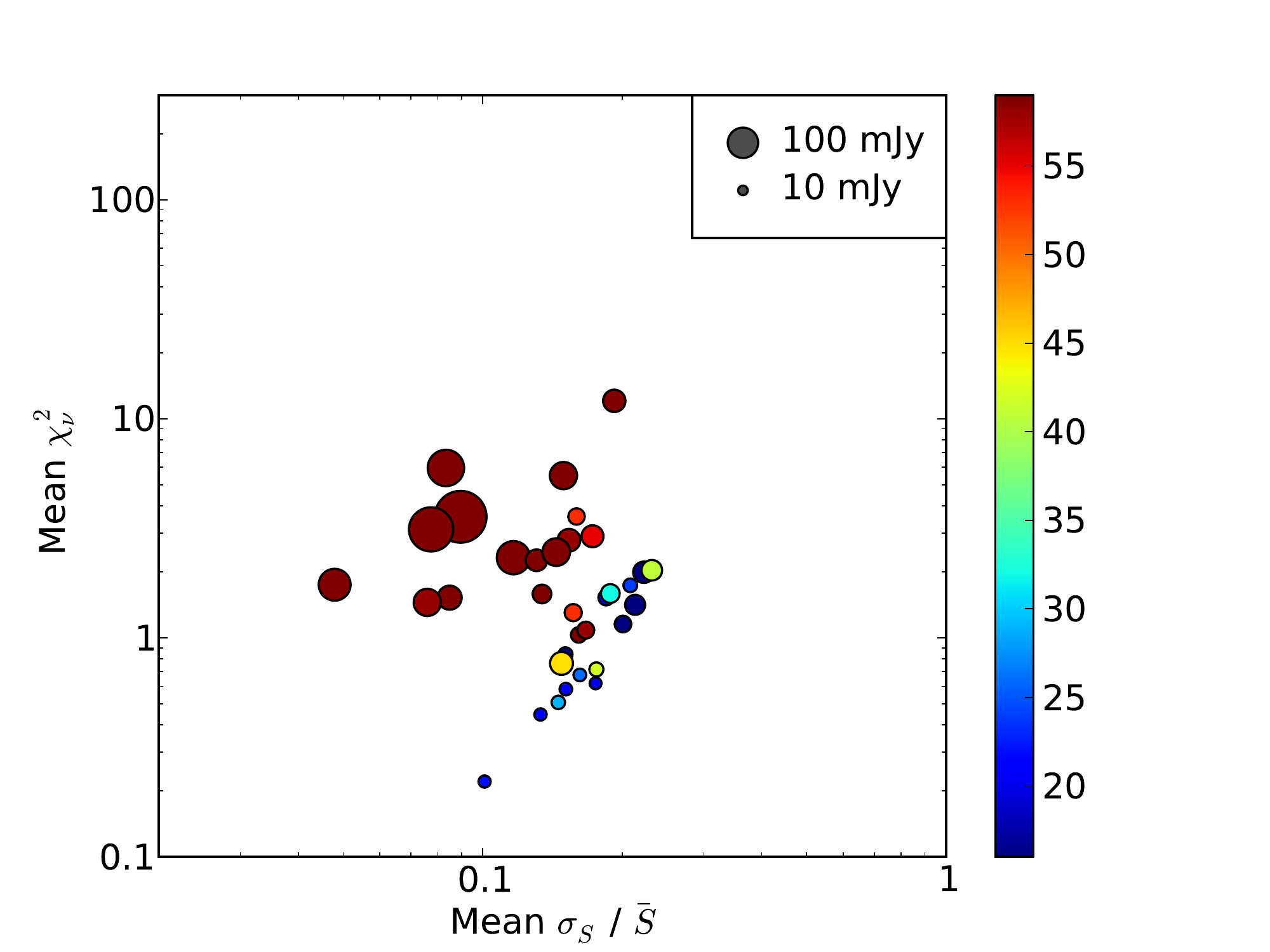}%
\includegraphics[width=0.49\linewidth,draft=false]{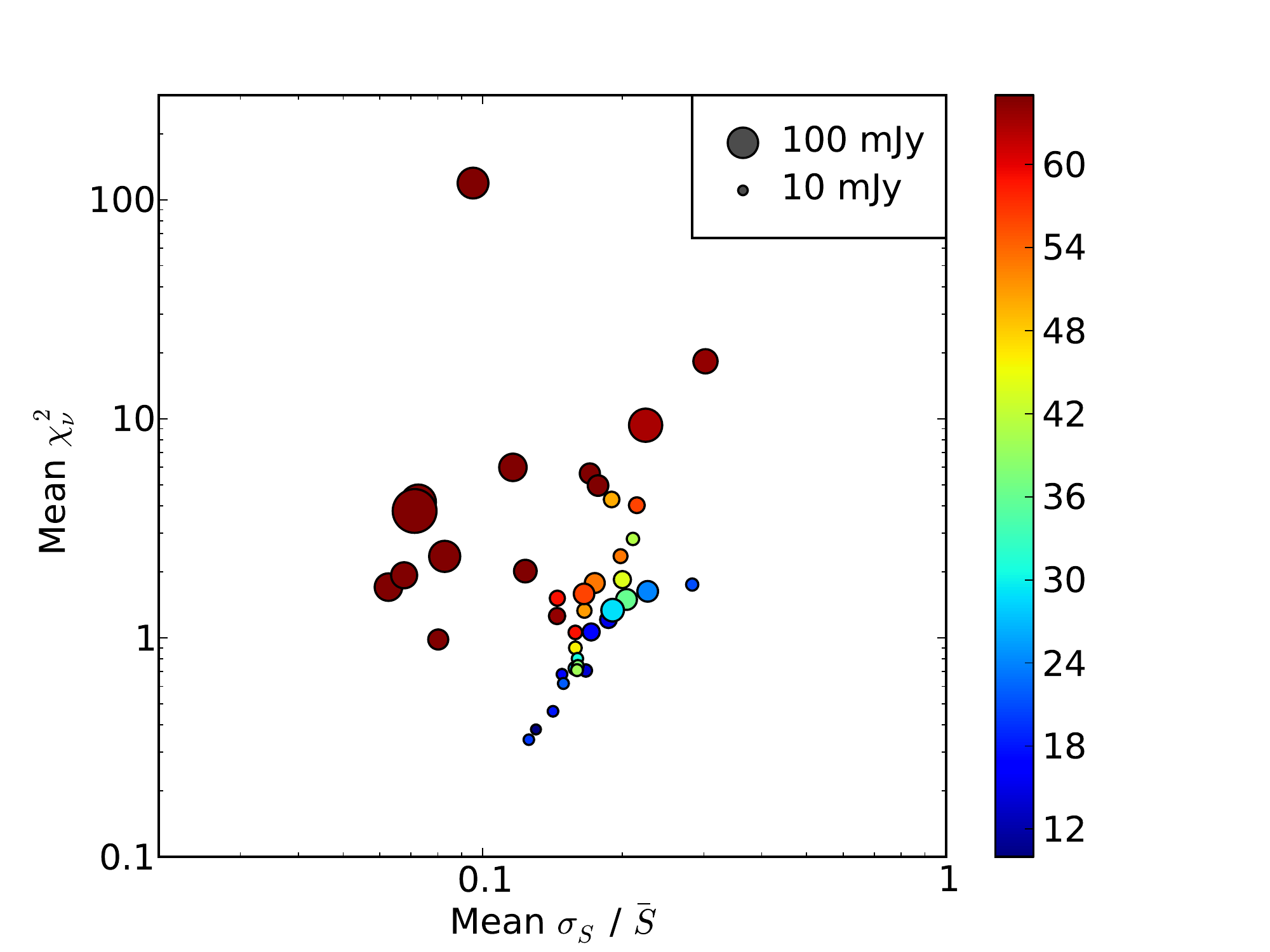}
\includegraphics[width=0.49\linewidth,draft=false]{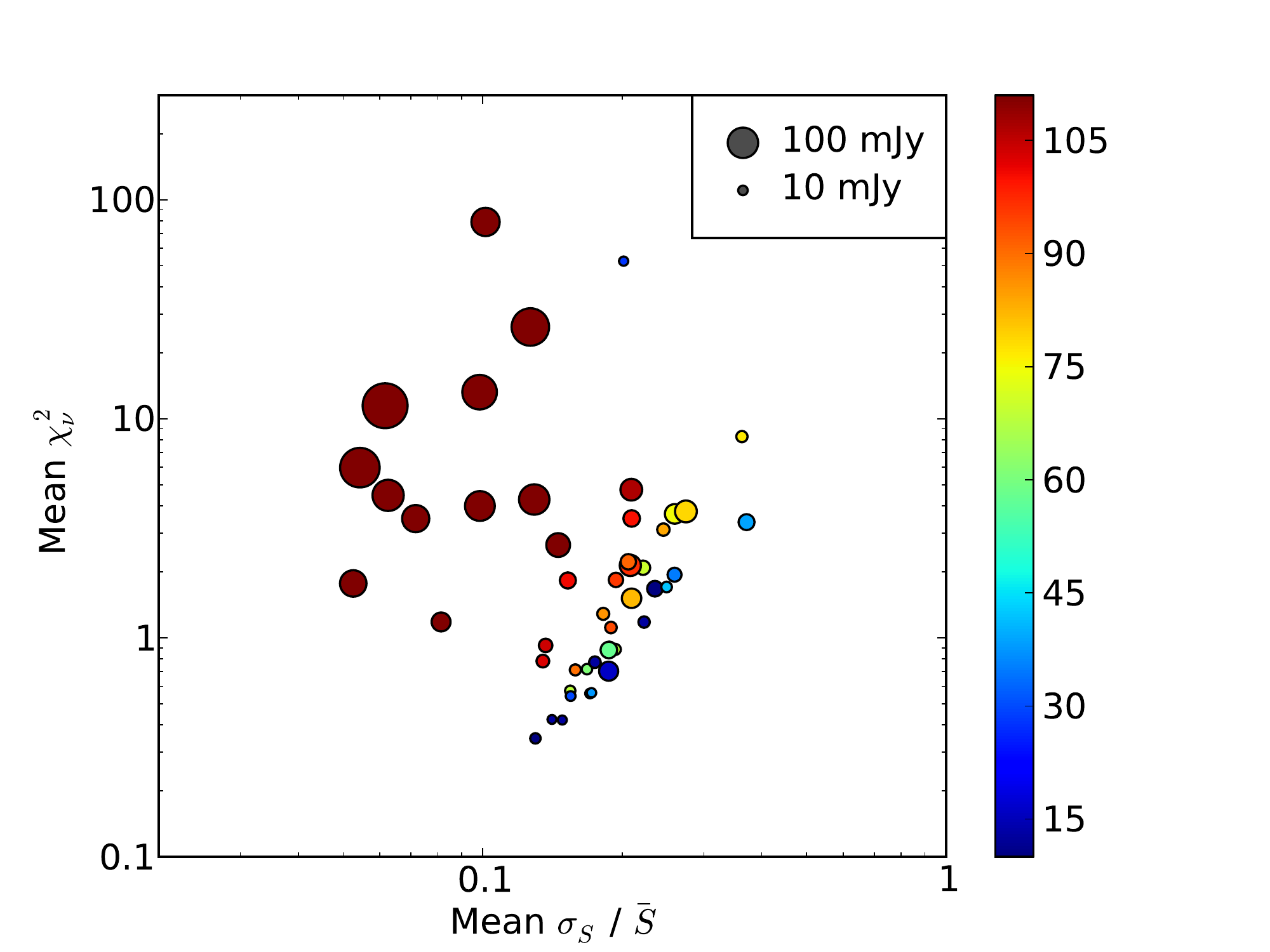}%
\includegraphics[width=0.49\linewidth,draft=false]{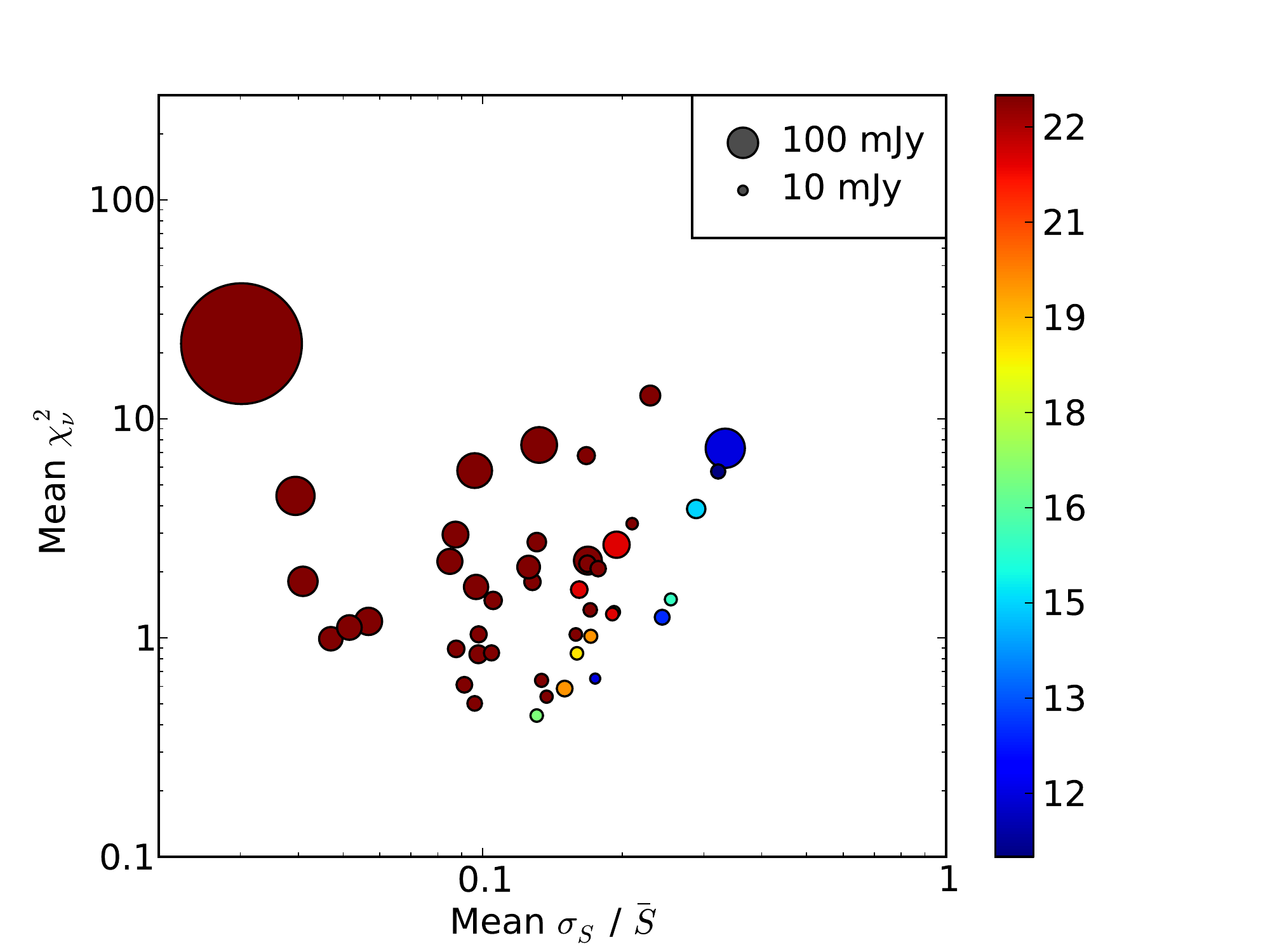}
\caption{\label{fig:chivar}
Reduced chi-squared, \rchisq, for the hypothesis that sources do not vary in individual epochs, computed for sources in (from top left to bottom right panel) the NDWFS, Lockman, ELAIS-N1 and Coma fields, plotted against the standard deviation divided by the mean flux density for the same sources. Values plotted are the mean of those computed independently at 3040\,MHz and 3140\,MHz for each source with dual-frequency detections in at least 10 good epochs. We consider only epochs where sources are detected at both frequencies. Sources are color-coded according to the number of epochs in which they were detected. Circle sizes scale with mean flux density, $\bar{S}$.
}
\end{figure*}

In Fig.~\ref{fig:chivar}, we plot \rchisq\ against mean $\sigma_S / \bar{S}$ for the same sources. The variability seen in faint sources is less well-characterized than for their brighter counterparts (because they are less well detected) and so they tend to have lower values of \rchisq. Faint sources with larger measured variability ($\sigma_S / \bar{S}$) have higher \rchisq\ because such variability becomes increasingly large relative to the measurement uncertainties on the flux densities. Bright sources tend to have larger \rchisq\ on average, but a wide range in both parameters, suggesting that some are well-characterized (as measured by \rchisq) as relatively non-variable, and others well-characterized as more variable. 

\subsection{Structure Function}

We calculate structure functions for our sources following \citet{lovell:08},

\begin{equation}
D_{\nu} \left( \tau \right) = \frac{1}{N_{\tau}} \sum_{j,k} \left(S_{\nu,j} - S_{\nu,k} \right)^2,
\end{equation} 

where $S_{\nu,j}$ and $S_{\nu,k}$ are pairs of flux density measurements normalized by the mean flux density of the source over all epochs (treating the two frequency bands, $\nu$, independently), and $N_{\tau}$ is the number of pairs of measurements with time lag $\tau$ in bins equally spaced in ${\rm log} (\tau)$. Most of the structure functions for sources with dual-frequency detections in at least 10 good epochs appear quite flat, with normalized amplitudes typically between 0.01 and 0.1, comparable to the results of \citet{lovell:08} on shorter timescales.

In Fig.~\ref{fig:strucfunc} we show structure functions, computed independently in our two frequency bands, for the 12 sources with $\rchisq > 5$ in both bands, \ie, those where variability is best detected (although not necessarily the sources with the largest measured variability; see Fig.~\ref{fig:chivar}). Again, most of the structure functions appear quite flat, although a few of the sources show signs of an upturn on long time lags, and a few also still show disagreement between the two frequency bands, indicative of remaining bad data, despite us using only data where sources are detected at both frequencies in at least 10 good epochs (defined as PIC gain within the good range) here. The flat structure functions likely indicate that intrinsic variability of these sources is rather low, although some may be variable on timescales shorter than measured here (for example, as intraday variables).

\begin{figure*}[htp]
\centering
\includegraphics[width=0.49\linewidth,draft=false]{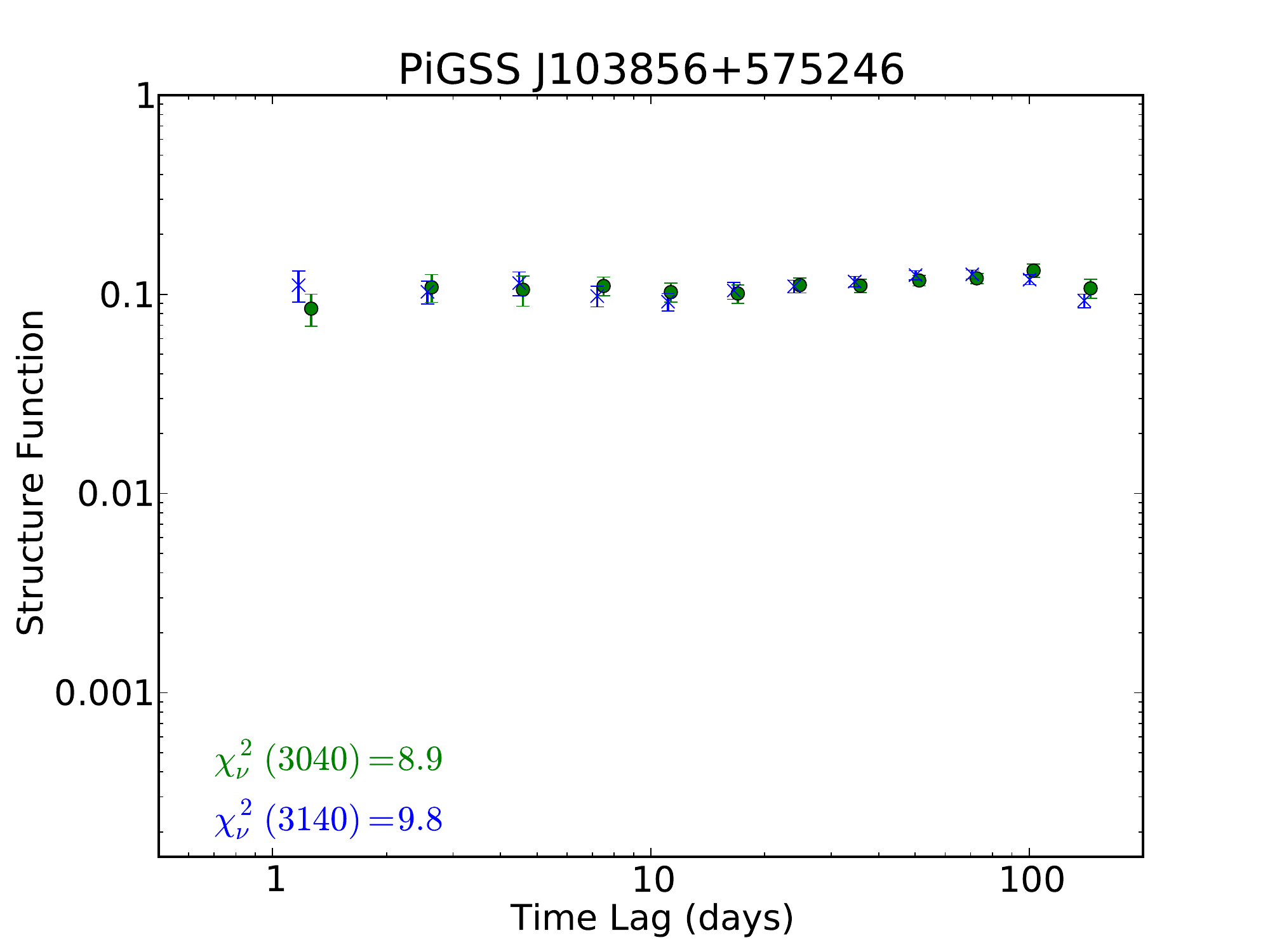}%
\includegraphics[width=0.49\linewidth,draft=false]{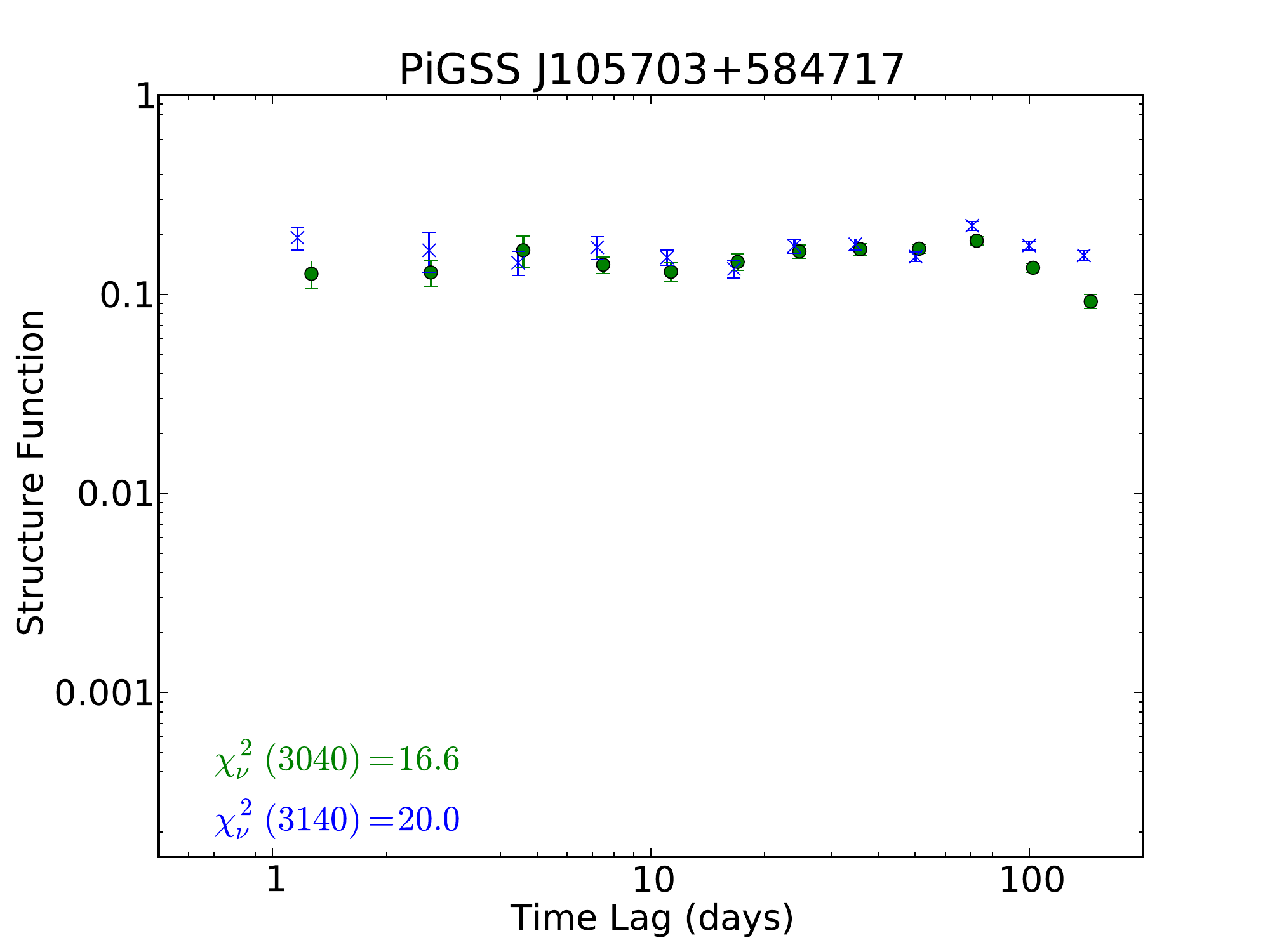}
\includegraphics[width=0.49\linewidth,draft=false]{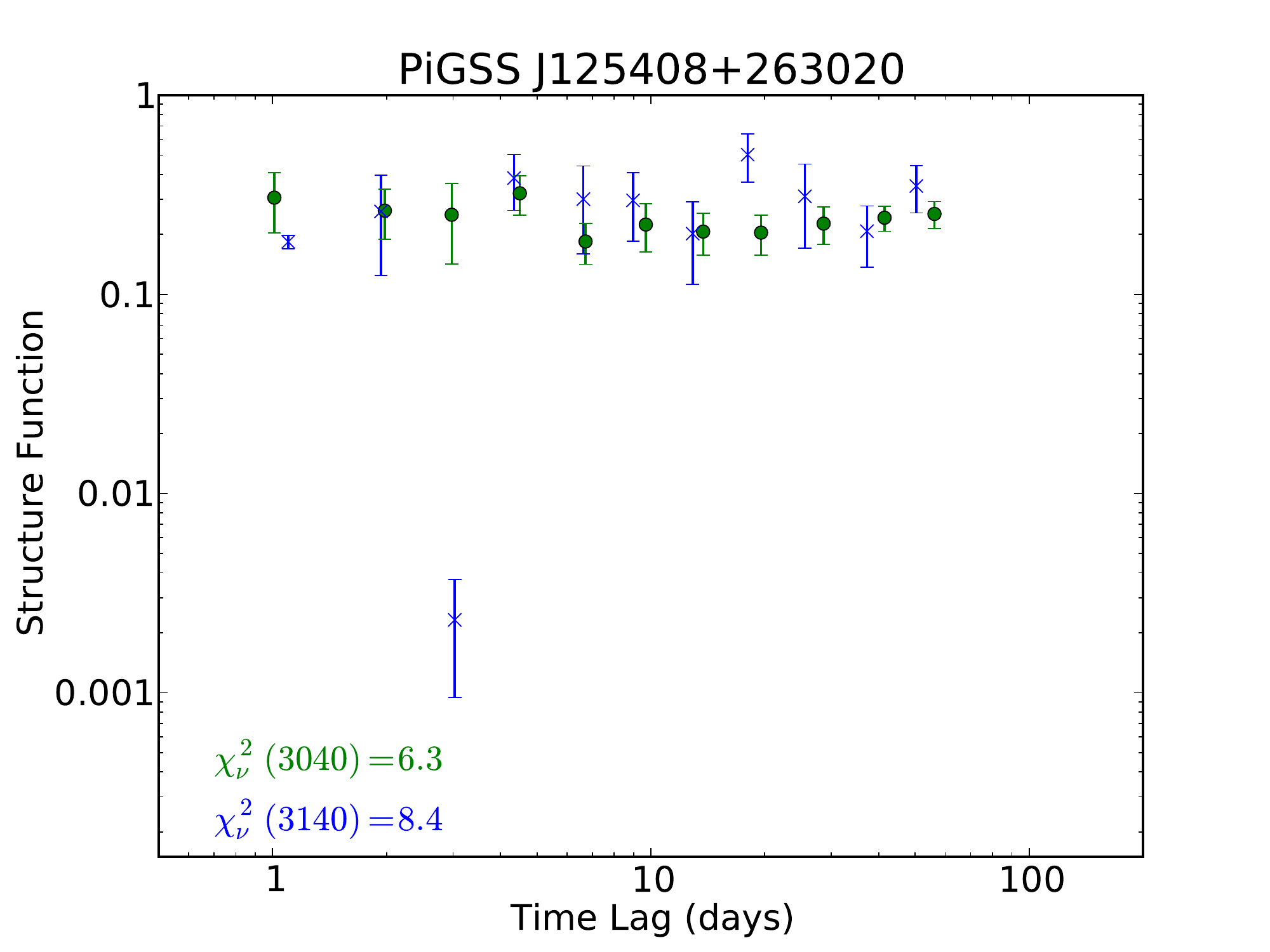}%
\includegraphics[width=0.49\linewidth,draft=false]{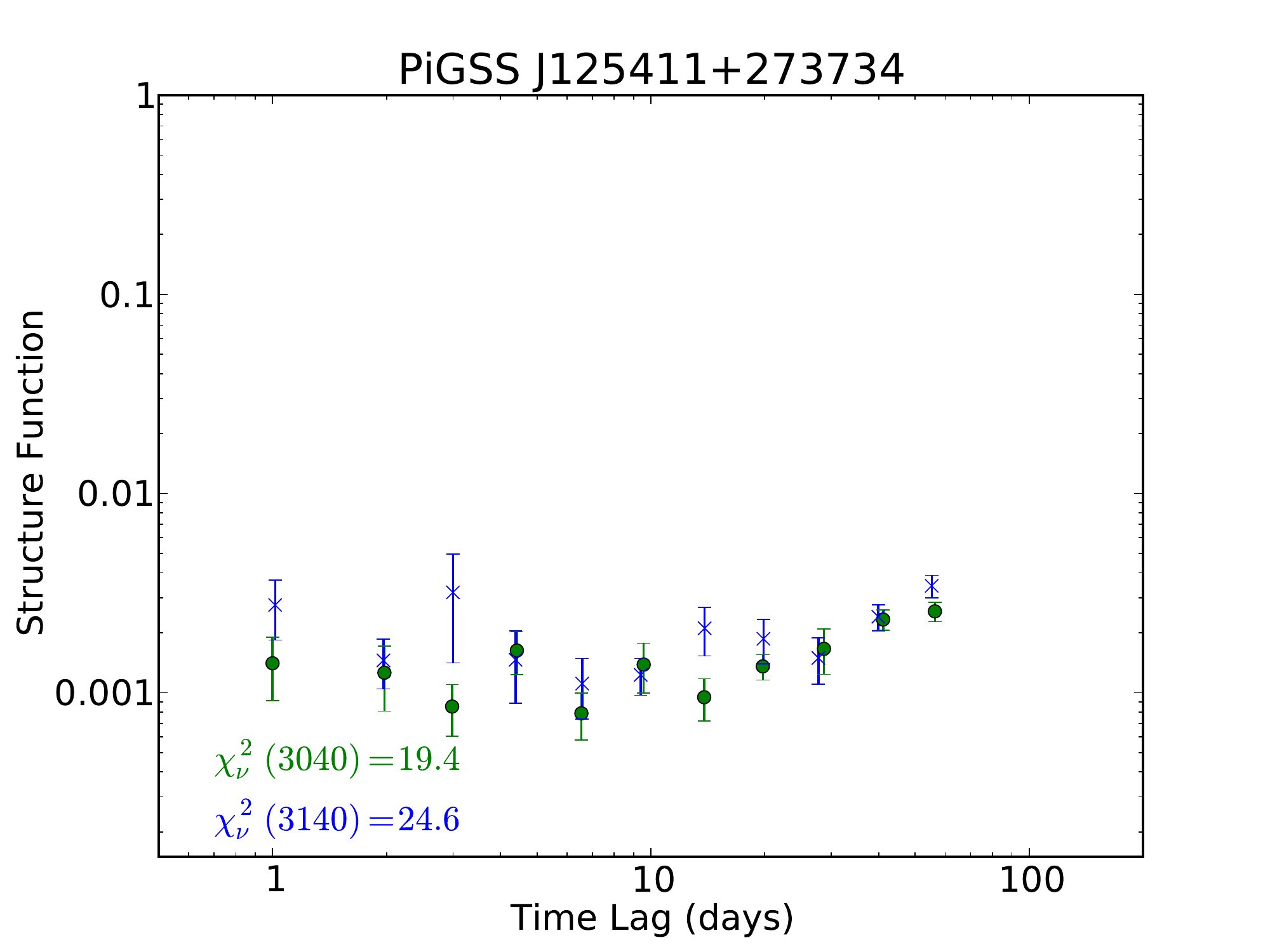}
\includegraphics[width=0.49\linewidth,draft=false]{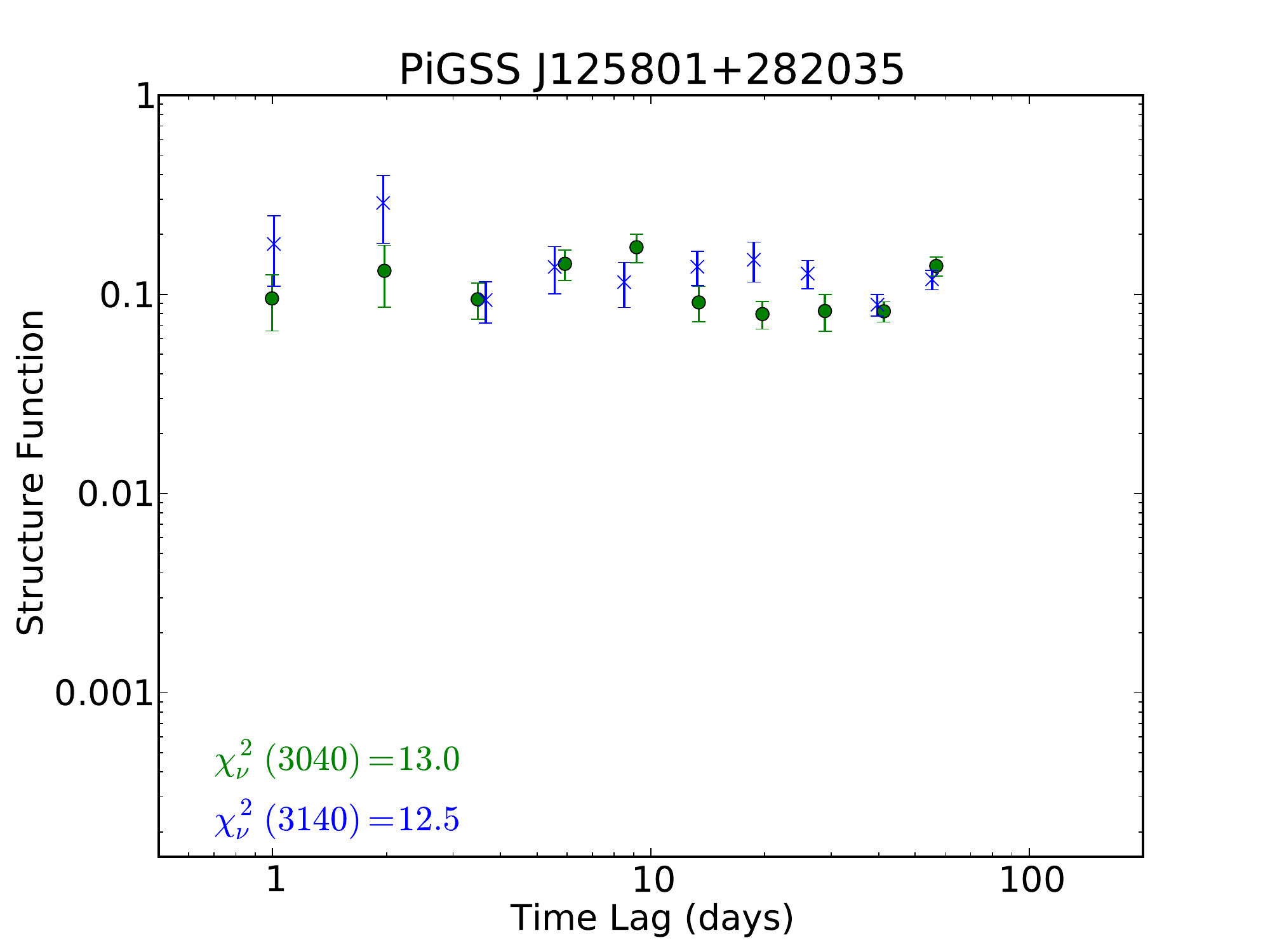}%
\includegraphics[width=0.49\linewidth,draft=false]{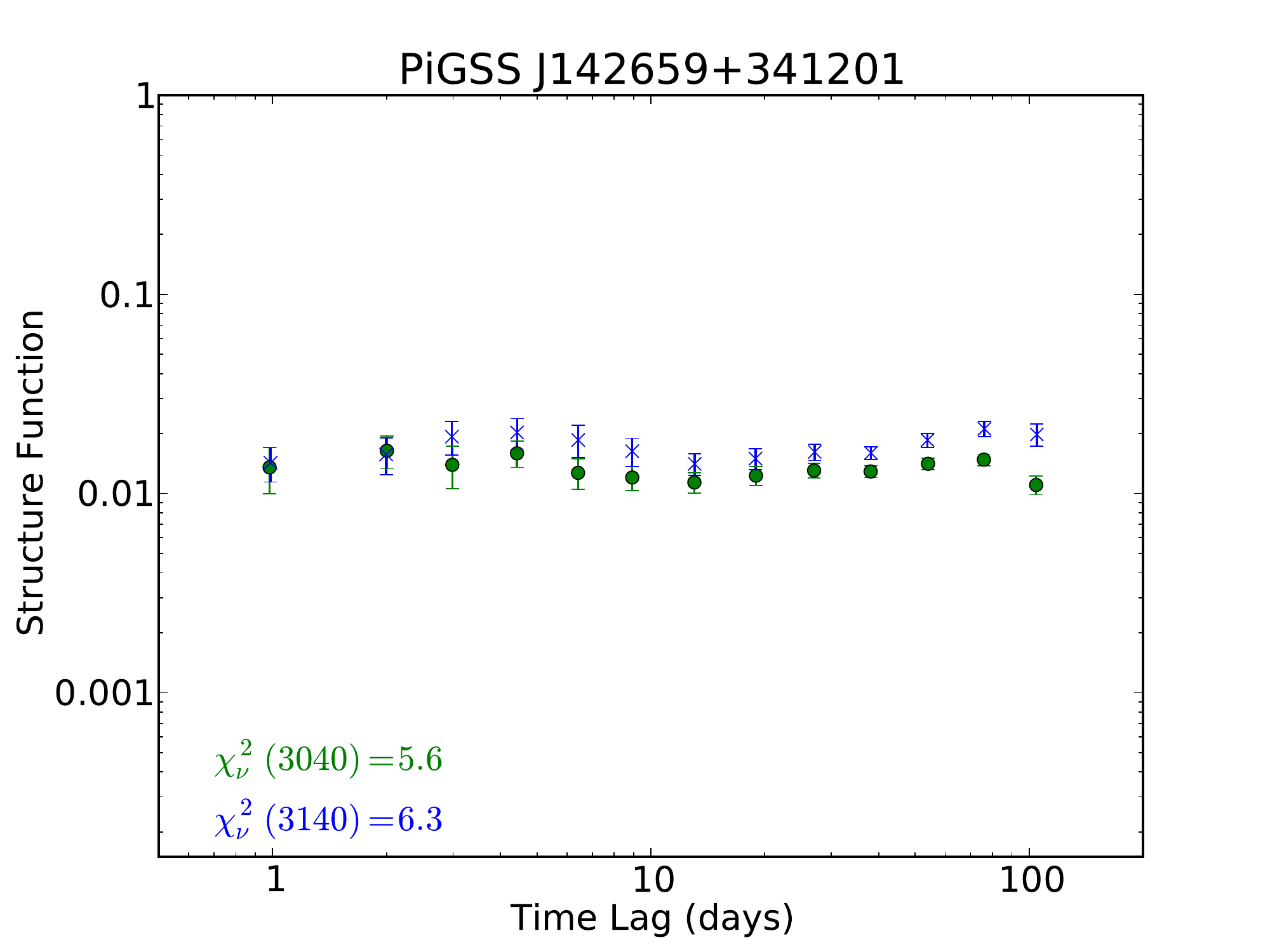}
\caption{\label{fig:strucfunc}
Structure functions (see text) for the twelve sources where variability is best detected ($\rchisq > 5$ in both bands). Data at 3040\,MHz (green) and 3140\,MHz (blue) are considered independently. (Figure continued on next page \ldots).
}
\end{figure*}

\begin{figure*}[htp]
\addtocounter{figure}{-1}
\centering
\includegraphics[width=0.49\linewidth,draft=false]{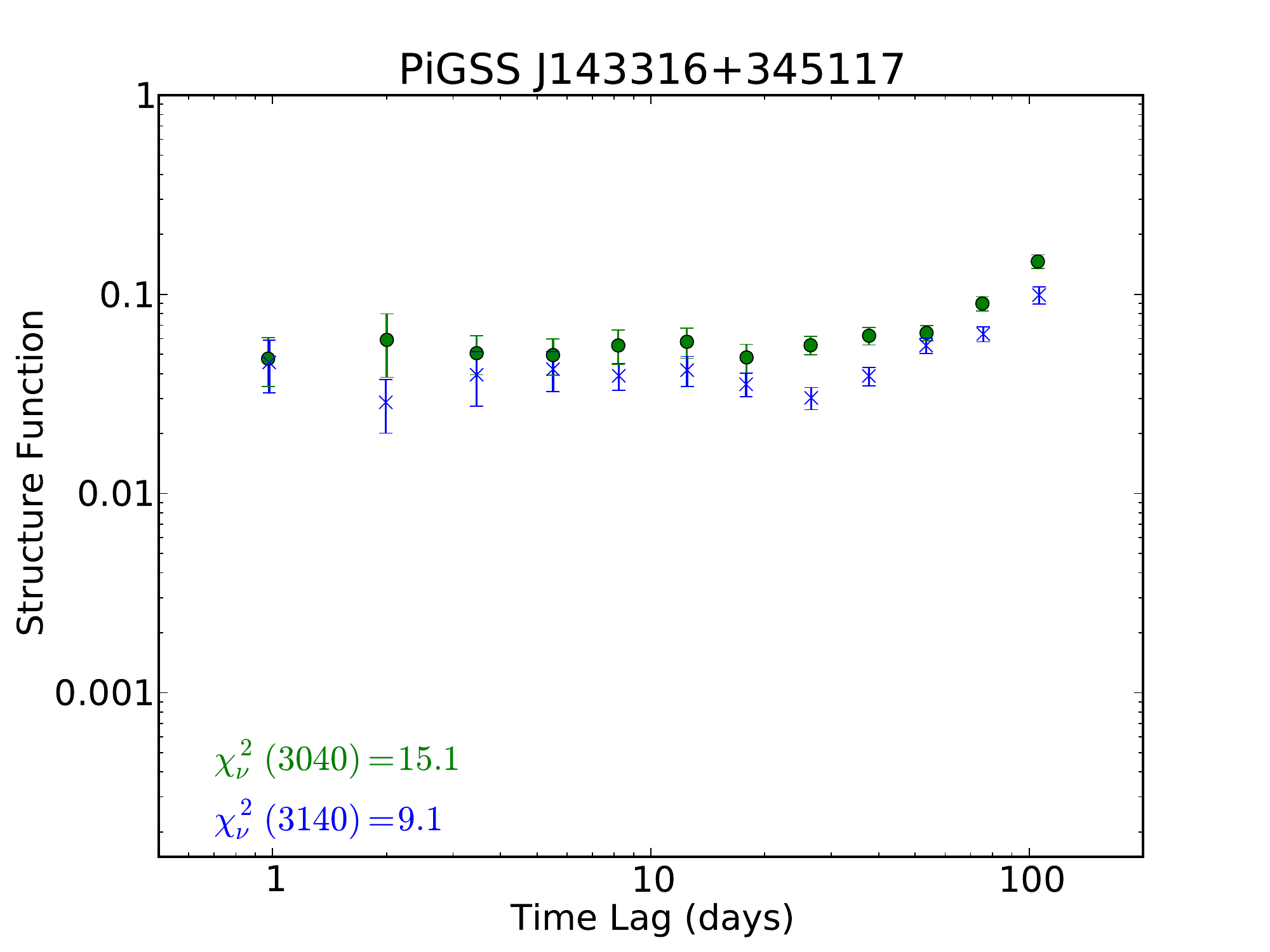}%
\includegraphics[width=0.49\linewidth,draft=false]{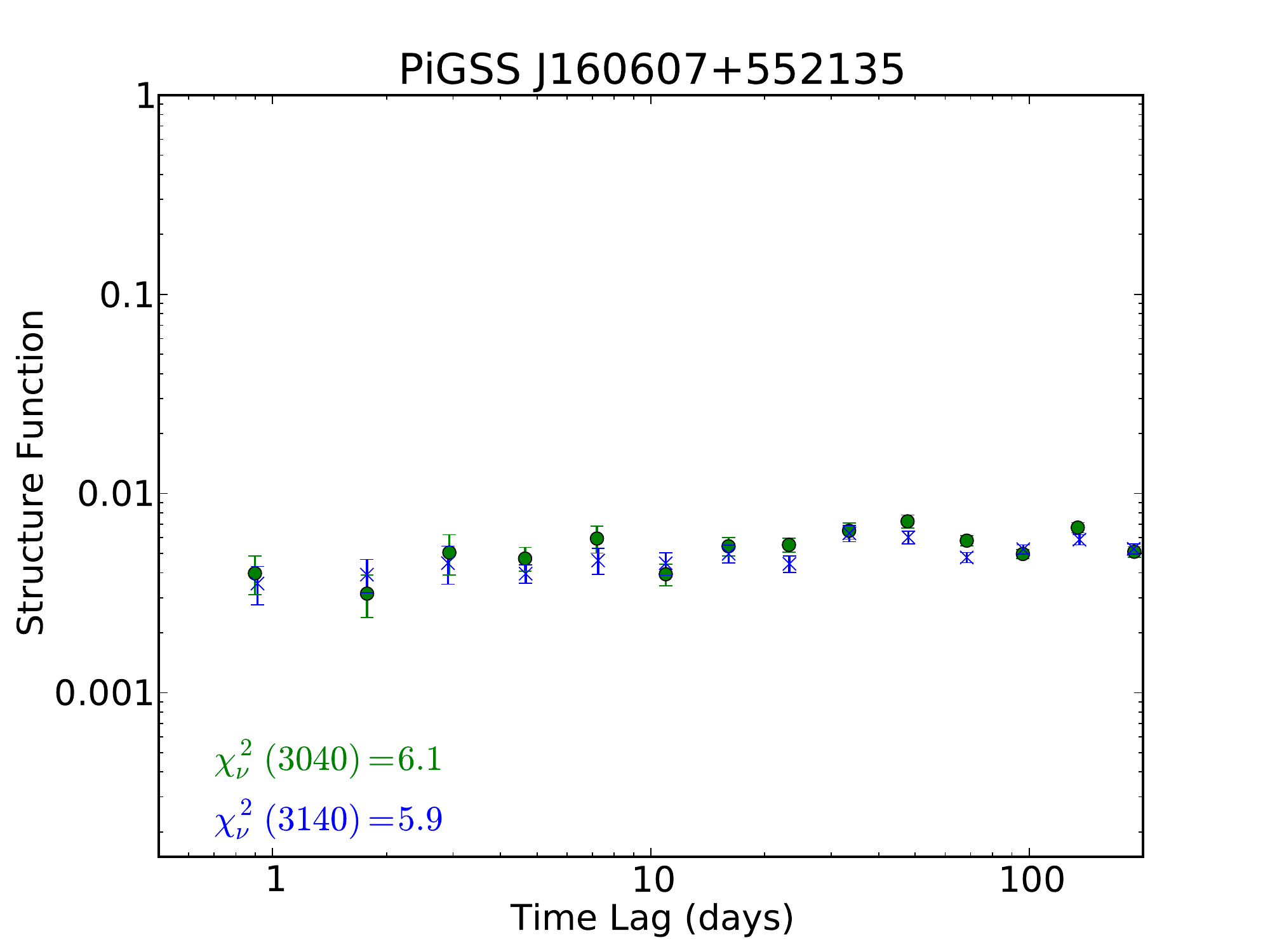}
\includegraphics[width=0.49\linewidth,draft=false]{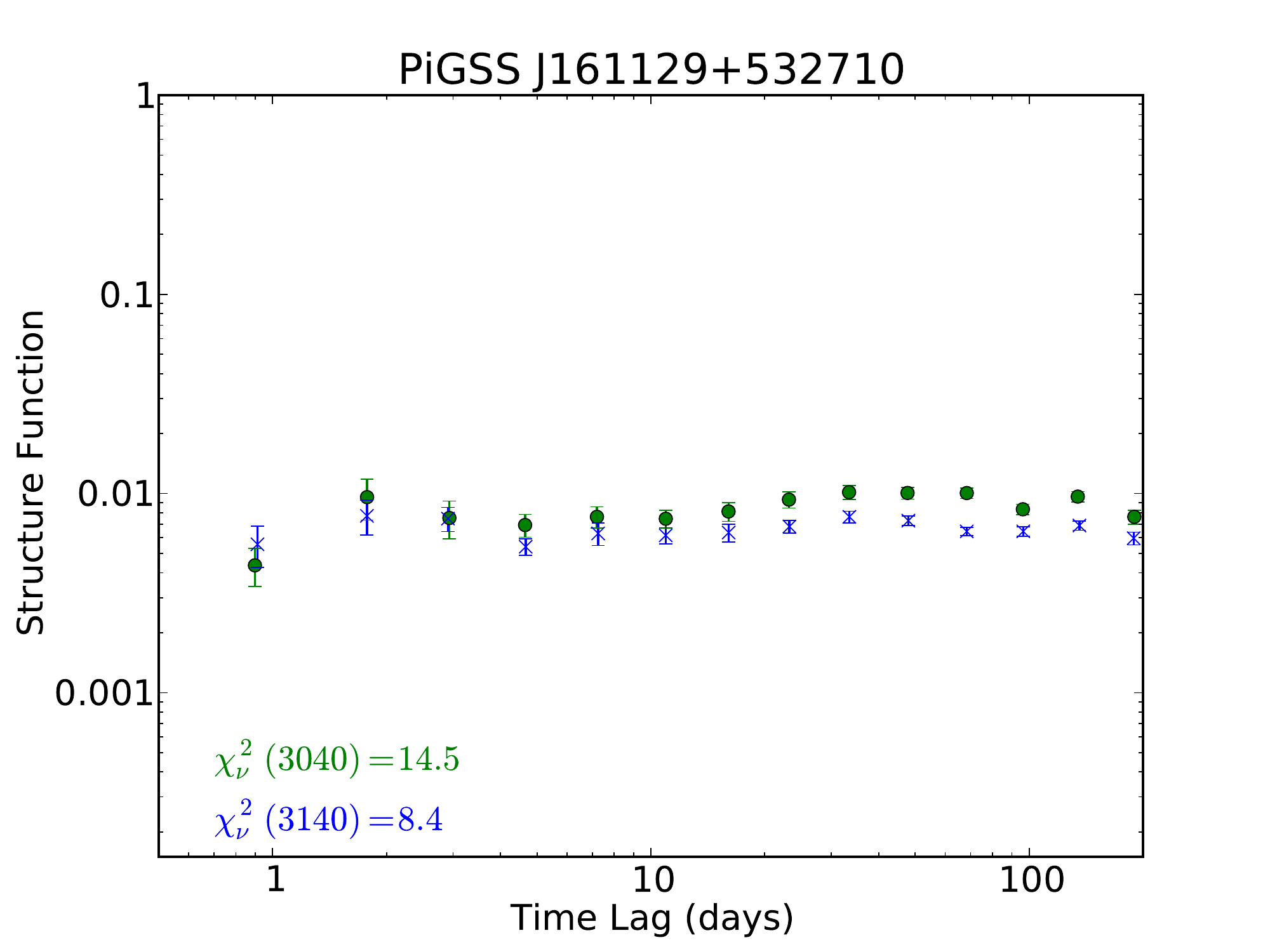}%
\includegraphics[width=0.49\linewidth,draft=false]{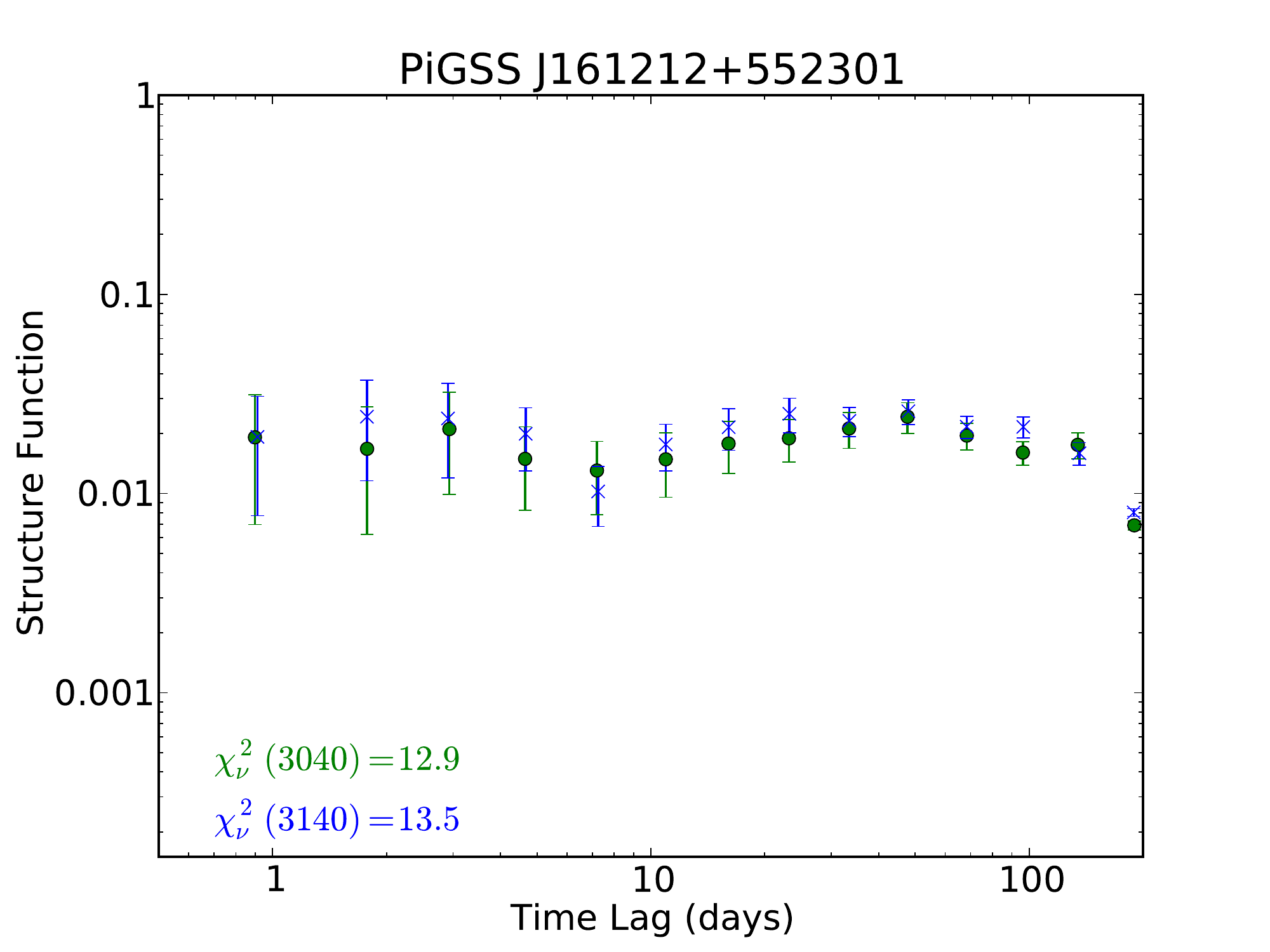}
\includegraphics[width=0.49\linewidth,draft=false]{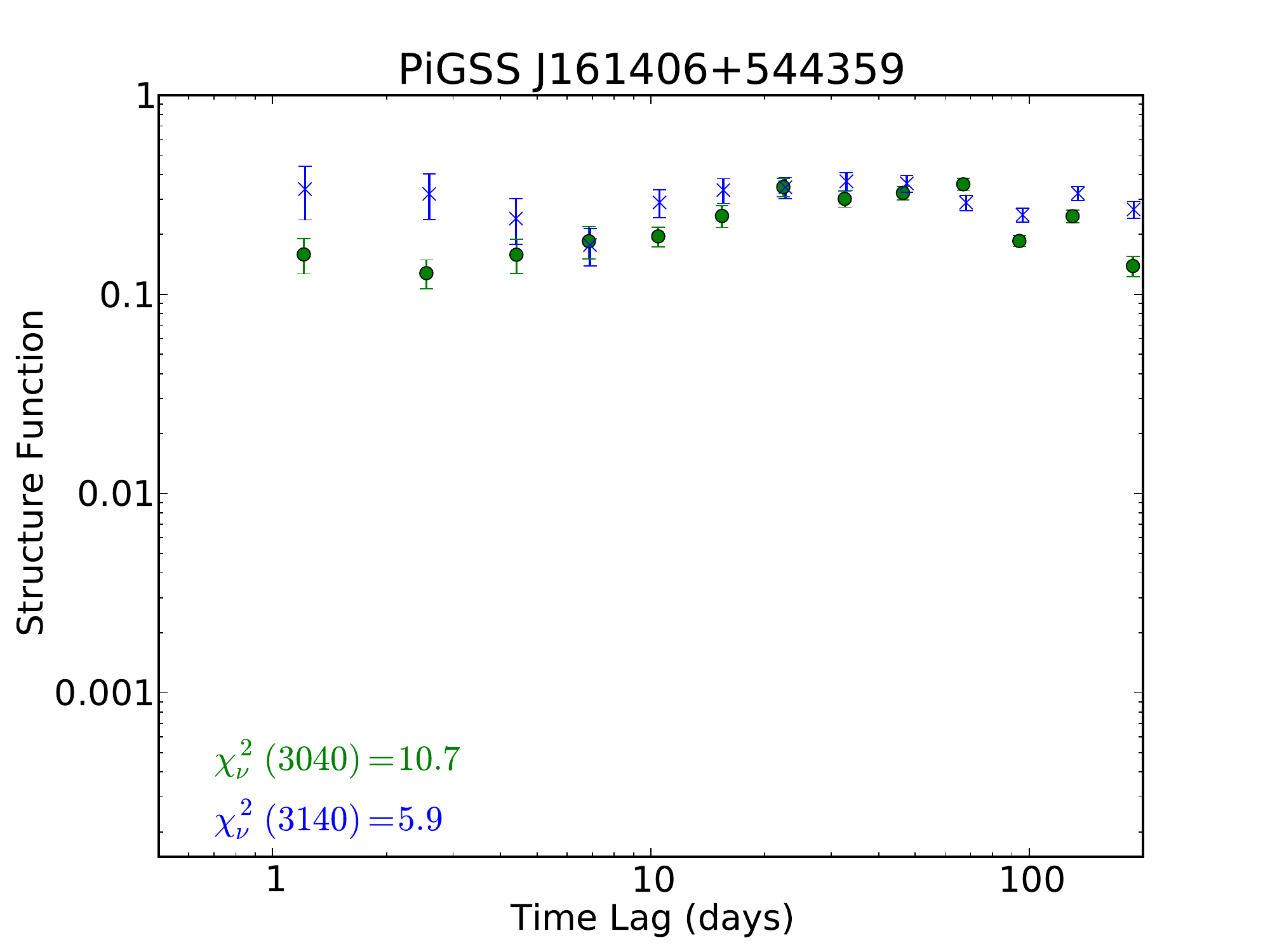}%
\includegraphics[width=0.49\linewidth,draft=false]{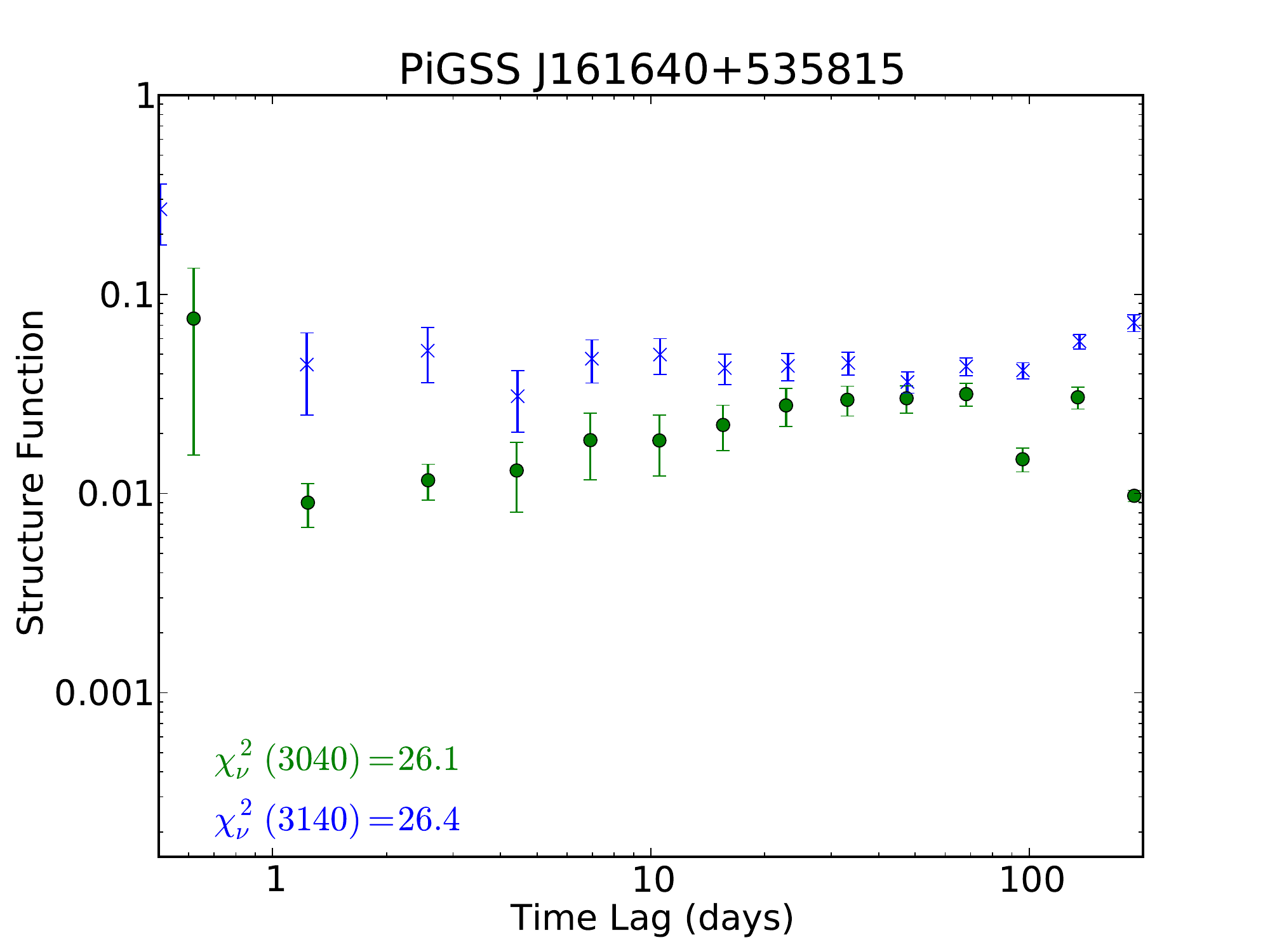}
\caption{
(continued)
}
\end{figure*}

\section{Conclusions} 

We report on an ATA 3.1\,GHz program that observed one of four 12 \sqdeg\ deep fields an average of a little less than once per day, as part of the PiGSS project (covering all four fields over the course of $\sim 2$\,yrs). We detect no transients in comparisons of images made from 379 individual epochs to deep images made from all usable observations of a particular field, or from comparison of the deep images to images made from monthly averages of our data or to legacy surveys from the literature. We place upper limits on the surface densities of transients on timescales of days to years, and flux densities brighter than a few mJy, that are still around an order of magnitude higher than predictions of rates from the literature, for example for tidal disruption events from \citet{frail:12}. Blind surveys covering more than an order of magnitude greater effective area ought to begin to detect such events if theoretical predictions are correct.

We discuss a number of caveats to be aware of in the search for transient and variable sources, and note in particular that the rejection of data from bad epochs (for example, by examining post-imaging calibration fits of the single-epoch flux densities to those from the deep comparison image, and by rejecting epochs with poor completeness) as well as rejection of image defects and other spurious sources which do not obey simple Gaussian statistics (in particular, by considering only sources that are detected in images made from neighboring frequency bands or from data split in some other way) are critically important in the regime where the potential false positive rate is large compared to the true rate of candidates of interest. We also note that some of the sources which appear most variable in our survey (and perhaps in others) appear variable because of discrepant flux density measurements in only one of our two frequencies, further reinforcing the dangers of relying on single flux density measurements without a simultaneous second confirming measurement.

A forthcoming paper will discuss the results from a two-epoch survey, also undertaken as part of PiGSS, which surveyed thousands of \sqdeg\ to similar sensitivity to the single-epoch images presented here.

\acknowledgments

The authors would like to acknowledge the generous support of the Paul
 G. Allen Family
 Foundation, which has provided major support for the design, construction,
 and operation of
 the ATA. Contributions from Nathan Myhrvold, Xilinx Corporation, Sun
 Microsystems,
 and other private donors have been instrumental in supporting the ATA.
 The ATA has been
 supported by contributions from the US Naval Observatory in addition
 to National Science
 Foundation grants AST-050690, AST-0838268 and AST-0909245. SC acknowledges support from the American Astronomical Society through the International Travel Grant program. 
  
 We are grateful for the support of the entire ATA team, and acknowledge in particular helpful discussions with Melvyn Wright, Peter Williams, Casey Law, and Garrett Keating, as well as the efforts of the HCRO onsite staff.
 
 We thank the anonymous referee for a very thorough reading of the manuscript and for many helpful comments and suggestions.
 
 \bibliographystyle{apj}

\end{document}